\documentclass[twocolumn]{aastex631}
\usepackage{graphicx}
\usepackage{epsfig}
\usepackage{natbib}
\usepackage{multirow}
\usepackage{amsmath}

\newcommand{\eg}{{\rm e.g.}}

\newcommand{\km}{{\rm\thinspace km}}
\newcommand{\g}{{\rm\thinspace g}}

\newcommand{\s}{{\rm\thinspace s}}

\newcommand{\kmps}{\hbox{$\km\s^{-1}\,$}}

\newcommand{\fesc}{$f_{\rm esc}$}

\newcommand{\fesclya}{$f_{{\rm esc,Ly}\alpha}$}
\newcommand{\fabs}{$f_{\rm abs}$}

\newcommand{\logten}{$\log_{\rm 10}$}

\newcommand{\lya}{Ly$\alpha$}

\newcommand{\G}{{\rm G}}
\newcommand{\K}{{\rm K}}
\newcommand{\hi}{H\thinspace{\sc i}}

\newcommand{\oiii}{[O\thinspace{\sc iii}]}

\newcommand{\oii}{[O\thinspace{\sc ii}]}
\newcommand{\oi}{[O\thinspace{\sc i}]}

\newcommand{\neiii}{[Ne\thinspace{\sc iii}]}

\newcommand{\sii}{[S\thinspace{\sc ii}]}

\newcommand{\sithree}{Si\thinspace{\sc iii}}

\newcommand{\sigsfr}{$\Sigma_{\rm SFR}$}

\newcommand{\Msol}{\hbox{\thinspace M$_{\sun}$}}

\shorttitle{Multivariate Prediction of Escape Fraction II}

\begin{document}

\title{Multivariate Predictors of LyC Escape II: Predicting LyC Escape Fractions for High-Redshift Galaxies\footnote{Based on observations made with the NASA/ESA Hubble Space Telescope, obtained at the Space Telescope Science Institute, which is operated by the Association of Universities for Research in Astronomy, Inc., under NASA contract NAS 5-26555. These observations are associated with programs GO-15626, GO-13744, GO-14635, GO-15341, and GO-15639.}}

\author[0000-0002-6790-5125]{Anne E. Jaskot}
\affiliation{Department of Astronomy, Williams College, Williamstown, MA 01267, USA}
\author{Anneliese C. Silveyra}
\affiliation{Department of Astronomy, Williams College, Williamstown, MA 01267, USA}
\affiliation{Department of Physics, University of Nevada, Reno, NV 89557, USA}
\author[0000-0001-6281-7727]{Anna Plantinga}
\affiliation{Department of Mathematics \&\ Statistics, Williams College, Williamstown, MA 01267, USA}
\author[0000-0002-0159-2613]{Sophia R. Flury}
\affiliation{Department of Astronomy, University of Massachusetts Amherst, Amherst, MA 01002, USA}
\author[0000-0001-8587-218X]{Matthew Hayes}
\affiliation{Department of Astronomy, Oskar Klein Centre, Stockholm University, AlbaNova, SE-10691 Stockholm, Sweden}
\author[0000-0002-0302-2577]{John Chisholm}
\affiliation{Department of Astronomy, University of Texas at Austin, Austin, TX 78712, USA}
\author[0000-0001-6670-6370]{Timothy Heckman}
\affiliation{Department of Physics and Astronomy, Johns Hopkins University, Baltimore, MD 21218, USA}
\author[0000-0001-8940-6768]{Laura Pentericci}
\affiliation{INAF - Osservatorio Astronomico di Roma, via Frascati 33, 00078, Monteporzio Catone, Italy}
\author[0000-0001-7144-7182]{Daniel Schaerer}
\affiliation{Observatoire de Gen{\`e}ve, Universit{\'e} de Gen{\`e}ve, Chemin Pegasi 51, 1290 Versoix, Switzerland}
\author[0000-0002-6849-5375]{Maxime Trebitsch}
\affiliation{Astronomy, Kapteyn Astronomical Institute, Landleven 12, 9747 AD Groningen, The Netherlands}
\author[0000-0002-2201-1865]{Anne Verhamme}
\affiliation{Observatoire de Gen{\`e}ve, Universit{\'e} de Gen{\`e}ve, Chemin Pegasi 51, 1290 Versoix, Switzerland}
\affiliation{Univ. Lyon, Univ. Lyon 1, ENS de Lyon, CNRS, Centre de Recherche Astrophysique de Lyon UMR5574, 69230 Saint-Genis-Laval, France}
\author[0000-0003-4166-2855]{Cody Carr}
\affiliation{Center for Cosmology and Computational Astrophysics, Institute for Advanced Study in Physics, Zhejiang University, Hangzhou 310058, China}
\affiliation{Institute of Astronomy, School of Physics, Zhejiang University, Hangzhou 310058, China}
\author[0000-0001-7113-2738]{Henry C. Ferguson}
\affiliation{Space Telescope Science Institute, 3700 San Martin Dr., Baltimore, MD 21218, USA}
\author[0000-0001-7673-2257]{Zhiyuan Ji}
\affiliation{Steward Observatory, University of Arizona, Tucson, AZ 85721, USA}
\author[0000-0002-7831-8751]{Mauro Giavalisco}
\affiliation{Department of Astronomy, University of Massachusetts Amherst, Amherst, MA 01002, USA}
\author[0000-0002-6586-4446]{Alaina Henry}
\affiliation{Space Telescope Science Institute, 3700 San Martin Dr., Baltimore, MD 21218, USA}
\author[0000-0001-8442-1846]{Rui Marques-Chaves}
\affiliation{Observatoire de Gen{\`e}ve, Universit{\'e} de Gen{\`e}ve, Chemin Pegasi 51, 1290 Versoix, Switzerland}
\author[0000-0002-3005-1349]{G{\"o}ran {\"O}stlin}
\affiliation{Department of Astronomy, Oskar Klein Centre, Stockholm University, AlbaNova, SE-10691 Stockholm, Sweden}
\author[0000-0001-8419-3062]{Alberto Saldana-Lopez}
\affiliation{Department of Astronomy, Oskar Klein Centre, Stockholm University, AlbaNova, SE-10691 Stockholm, Sweden}
\author[0000-0002-9136-8876]{Claudia Scarlata}
\affiliation{Minnesota Institute for Astrophysics, School of Physics and Astronomy, University of Minnesota, 316 Church St. SE, Minneapolis, MN 55455, USA}
\author[0000-0003-0960-3580]{G{\'a}bor Worseck} 
\affiliation{VDI/VDE Innovation+Technik, Berlin, Germany}
\author[0000-0002-9217-7051]{Xinfeng Xu}
\affiliation{Center for Interdisciplinary Exploration and Research in Astrophysics, Northwestern University, Evanston, IL 60201, USA}

\begin{abstract}
{\it JWST} is uncovering the properties of ever increasing numbers of galaxies at $z>6$, during the epoch of reionization. Connecting these observed populations to the process of reionization requires understanding how efficiently they produce Lyman continuum (LyC) photons and what fraction (\fesc) of these photons escape into the intergalactic medium. By applying the Cox proportional hazards model, a survival analysis technique, to the Low-redshift Lyman Continuum Survey (LzLCS), we develop new, empirical, multivariate predictions for \fesc. The models developed from the LzLCS reproduce the observed \fesc\ for $z\sim3$ samples, which suggests that LyC emitters may share similar properties at low and high redshift. Our best-performing models for the $z\sim3$ galaxies include information about dust attenuation, ionization, and/or morphology. We then apply these models to $z\gtrsim6$ galaxies. For large photometric samples, we find a median predicted \fesc=0.047-0.14. For smaller spectroscopic samples, which may include stronger emission line galaxies, we find that $\geq33$\%\ of the galaxies have \fesc\ $>0.2$, and we identify several candidate extreme leakers with \fesc\ $\geq0.5$. The current samples show no strong trend between predicted \fesc\ and UV magnitude, but limited spectroscopic information makes this result uncertain. Multivariate predictions can give significantly different results from single variable predictions, and the predicted \fesc\ for high-redshift galaxies can differ significantly depending on whether star formation rate surface density or radius is used as a measure of galaxy morphology. We provide all parameters necessary to predict \fesc\ for additional samples of high-redshift galaxies using these models.

\end{abstract}

\section{Introduction}
\label{sec:intro}

Reionization represents a fundamental transformation of the universe's hydrogen gas and dramatically illustrates the effect galaxies can have on their surroundings. The presence of \lya\ absorption in quasar spectra suggests that reionization ended sometime near $z\sim6$ \citep[\eg,][]{fan06, mcgreer15, robertson22}. The electron scattering optical depth of the cosmic microwave background is consistent with this general picture, constraining the midpoint of reionization to $z\sim7.8$ \citep{planck20}. However, the exact timeline of reionization remains uncertain, in part because of the inhomogeneous nature of the reionization process \citep[\eg,][]{eilers18, jung20, becker21}. 

The recently launched {\it James Webb Space Telescope (JWST)} is rapidly expanding our knowledge of galaxy properties during the epoch of reionization. Surveys are detecting more bright galaxies than anticipated, challenging our understanding of early galaxy evolution \citep[\eg,][]{castellano22, naidu22, finkelstein23, donnan23, harikane23, mcleod24}. {\it JWST} images suggest that galaxies in the reionization era are morphologically compact \citep[\eg,][]{robertson23, ormerod24, morishita24}, and spectroscopic observations are tracing the evolution of nebular ionization and metallicity to redshifts well above 7 \citep[\eg,][]{schaerer22b, sanders23, tang23, fujimoto23, curti23, backhaus24}. Measurements of nebular emission lines in early galaxies are also constraining their ionizing, Lyman continuum (LyC) photon production rate, $\xi_{\rm ion}$, a key quantity needed to understand reionization. Such observations suggest that faint galaxies with bursty star formation histories may have elevated $\xi_{\rm ion}$ \citep[\eg,][]{atek24, simmonds24, saxena24}. 

Several complementary approaches can give insights as to which galaxy populations dominated the reionization process. A galaxy's contribution to reionization depends on its $\xi_{\rm ion}$ and the fraction \fesc\ of these LyC photons that escape into the intergalactic medium (IGM). Using constraints on galaxy number densities at $z>6$, some studies explore different distributions of $\xi_{\rm ion}$ and \fesc\ among the galaxy population and seek to match the observed constraints on the reionization timeline. Depending on the model assumptions, either moderately bright or very faint galaxies may provide the majority of the ionizing photons \citep[\eg,][]{finkelstein19, naidu20}.

Cosmological hydrodynamical simulations can also model the progress of reionization over time and can explore how and why LyC photons escape simulated galaxies. Simulations consistently find that feedback plays an essential role in clearing out obscuring material \citep[\eg,][]{wise09, cen15, paardekooper15, trebitsch17}. LyC input may peak shortly following a burst of star formation after supernovae explode \citep[\eg,][]{ma15, trebitsch17}, although other factors such as binary star evolution or an extended star formation history may affect this timeline \citep[\eg,][]{ma15, barrow20, katz23}. The geometry and mechanism of LyC escape may also vary across the galaxy population \citep[\eg,][]{katz23, bremer23}. Despite general agreement on the importance of feedback, however, simulation predictions have yet to be confirmed observationally and the galaxies that are the primary drivers of reionization have not yet been conclusively identified.

Directly detecting LyC and constraining \fesc\ observationally becomes difficult above $z>4$ \citep[\eg,][]{inoue14} due to high IGM attenuation. Fortunately, LyC observations at lower redshifts can help test simulation predictions regarding which galaxy properties regulate LyC escape. Over the past decade, the number of LyC detections has grown rapidly, with dozens of LyC Emitters (LCEs) now known at $z\sim2-3$ \citep[\eg,][]{mostardi15, shapley16, vanzella16, vanzella18, bassett19, fletcher19, riverathorsen19, ji20, saxena22}, multiple detections of LyC emission in stacked samples at $z\sim2-4$ \citep[\eg,][]{marchi18, steidel18, bian20, nakajima20}, and more than 50 known LCEs at $z<0.5$ \citep[\eg,][]{leitet11, leitet13, borthakur14, izotov16b, izotov18b, izotov21, leitherer16, wang19, flury22a}. At both low and intermediate redshift, LCEs appear deficient in absorbing gas and dust \citep[\eg,][]{gazagnes18, chisholm18, saldana22, steidel18, ji20}, morphologically compact \citep[\eg,][]{borthakur14, izotov18b, flury22b, vanzella16, marchi18, riverathorsen19}, and bright in higher ionization emission lines such as \oiii~$\lambda$5007 \citep[\eg,][]{izotov18b, flury22b, vanzella16, fletcher19, nakajima20}. 

Building on these observational and theoretical efforts, several studies have proposed diagnostics to predict \fesc\ at $z>6$ based on observable properties. Using simulated galaxies from the SPHINX cosmological radiation hydrodynamics simulation, \citet{choustikov24} develop a method to predict \fesc\ from a linear combination of observables, including the UV slope, E(B-V), H$\beta$ luminosity, UV magnitude, and nebular line ratios. Most other studies have taken an empirical approach, constructing \fesc\ diagnostics based on the large observational sample of LCEs at $z\sim0.3$ \citep[\eg,][]{verhamme17, wang21, flury22b, xu23}. The largest of these $z\sim0.3$ samples is the Low-redshift Lyman Continuum Survey (LzLCS; \citealt{flury22a}), a set of 66 galaxies with LyC measurements from the {\it Hubble Space Telescope (HST)} and ancillary ultraviolet and optical data from {\it HST} and the Sloan Digital Sky Survey (SDSS; \citealt{blanton17}). Combined with archival datasets \citep{izotov16a, izotov16b, izotov18a, izotov18b, izotov21, wang19}, this $z\sim0.3$ sample, hereafter the LzLCS+, consists of 89 galaxies with measured LyC or stringent upper limits. Based on an analysis of the LzLCS+, \citet{chisholm22} find that the UV slope $\beta_{\rm 1550}$ shows one of the strongest correlations with \fesc\ and propose that this single variable can serve as a predictor of \fesc\ at high redshift. Other recent studies of the LzLCS+ consider \fesc\ diagnostics that incorporate information from multiple variables. \citet{saldana22} generate an equation to predict \fesc\ from E(B-V) and low-ionization UV absorption lines. By tracing the gas and dust that destroy LyC photons, these parameters closely track \fesc. Employing this method at high redshift may be a challenge, however, as measuring weak absorption lines requires high signal-to-noise observations of galaxy continua (but see \citealt{saldana23}). Emission lines and photometry offer a simpler, if less direct, means of predicting \fesc. \citet{mascia23} propose a new \fesc\ diagnostic using $\beta_{\rm 1550}$, \oiii~$\lambda$5007/\oii~$\lambda$3727=O32, and half-light radius, three of the variables that correlate strongly with \fesc\ in the LzLCS+. By applying this diagnostic at high redshift using {\it JWST} observations from the GLASS and CEERS surveys, they find predicted \fesc\ of $\sim0.1$ for galaxies at $z>6$ \citep{mascia23, mascia24}. With a similar combination of variables ($\beta_{\rm 1550}$, O32, and UV magnitude), \citet{lin24} develop a regression model for the probability of LyC escape. Their model suggests that \fesc\ may be high ($\sim0.2$) in brighter galaxies in the epoch of reionization \citep{lin24}. These studies all agree that plausible LCE candidates exist at high redshift, although they differ in their \fesc\ prediction methods.

In Jaskot et al. (2024), hereafter Paper I, we use the LzLCS+ to develop new empirical multivariate models for predicting \fesc. Because the LzLCS+ contains both LyC detections and upper limits, we adopt the statistical techniques of survival analysis, which are suitable for such censored data. Specifically, we employ the semiparametric Cox proportional hazards model to generate \fesc\ predictions based on a desired set of input observables. We show that a model limited to observables accessible at $z>6$ can reproduce the observed \fesc\ in the LzLCS+ with a root-mean-square scatter of 0.46 dex. Of these observables, three variables (O32, $\beta_{\rm 1550}$, and the star formation rate surface density) are statistically significant in the fit, and a model limited to these three input observables predicts \fesc\ as well as the full model. The Cox model technique can be customized to include any combination of variables that are available for most of the LzLCS+ galaxies and hence offers a flexible tool for predicting \fesc\ at high redshift.

In this paper, we apply Cox models for \fesc\ to samples of high-redshift galaxies. Following the techniques in Paper I, which we summarize in \S\ref{sec:methods}, we generate new models for the sets of variables in published high-redshift samples. In \S\ref{sec:z3}, we first test the models' performance at $z\sim3$ using samples with published LyC measurements. With published samples at $z\gtrsim6$, we then generate \fesc\ predictions for galaxies in the epoch of reionization in \S\ref{sec:z6}. In \S\ref{sec:compare}, we compare our models with alternative proposed \fesc\ prediction methods from the literature, and we discuss the implications of our results for studies of reionization in \S\ref{sec:discuss}. We summarize our conclusions in \S\ref{sec:conclusions}. In the Appendix, we provide parameters for all models as well as examples on how to apply these models to future samples. We adopt a cosmology of  $H_0 = 70$ km s$^{-1}$ Mpc$^{-1}$, $\Omega_m = 0.3$, and $\Omega_\Lambda = 0.7$.

\section{Methods}
\label{sec:methods}
\subsection{Sample: The Low-redshift Lyman Continuum Survey}
\label{sec:lzlcs}

We derive our empirical predictions of \fesc\ from the LzLCS+, a combined, homogeneously processed dataset consisting of the Low-redshift Lyman Continuum Survey \citep{flury22a} and archival samples with {\it HST} Cosmic Origins Spectrograph (COS) LyC observations \citep{izotov16a, izotov16b, izotov18a, izotov18b, izotov21, wang19}. A full description of the data processing and UV and optical measurements appears in \citet{flury22a}, \citet{saldana22}, and Paper I, but we summarize key points here. 

The LzLCS+ contains 89 galaxies at $z\sim0.3$, a distance where the LyC redshifts into a sensitive wavelength range for the COS detector. The LzLCS targets were selected on properties proposed to indicate LyC escape: high O32 ratios (O32 $\geq 3$), high star formation rate surface densities (\sigsfr $> 0.1$ \Msol\ yr$^{-1}$ kpc$^{-2}$ ), and/or blue UV slopes ($\beta < -2$). The LzLCS+ covers a wide range of parameter space, spanning $\sim2$ dex in O32, $\sim2$ dex in \sigsfr, and an observed (not dust-corrected) UV absolute magnitude range of $M_{\rm 1500}=-18.3$ to -21.5. In Paper I and this study, we exclude one galaxy from the LzLCS+ sample: J1333+6246 \citep{izotov16b}. This galaxy has visibly truncated emission lines in its SDSS spectrum and unphysical Balmer line ratios, which together suggest that its nebular line flux measurements may be inaccurate. For this work, our total sample therefore includes 88 $z\sim0.3$ galaxies, 49 of which have detected LyC. 

To measure the LyC, we use COS G140L observations, processed using the {\tt calcos} pipeline (v3.3.9) and the \textsc{FaintCOS} software routines \citep{worseck16, makan21}. We measure the LyC in a 20~\AA-wide wavelength bin near rest-frame 900 \AA, while excluding any wavelengths above 1180~\AA\ in the observed frame because of telluric emission. We follow the definitions of LyC detections and upper limits from \citet{flury22a}, where detections are observations with a probability $< 0.02275$ of originating from background counts and the upper limit for non-detections represents the the 84th percentile of the background count distribution. To correct for Milky Way attenuation, we adopt the \citet{green18} dust maps and \citet{fitzpatrick99} attenuation law. 

\citet{flury22a} investigate several different measures of \fesc. Here, as in Paper I, we adopt the absolute \fesc, with \textsc{Starburst99} \citep{leitherer11a, leitherer14} spectral energy distribution (SED) fits to the UV continuum providing the estimate of the initial intrinsic LyC \citep{chisholm19, saldana22}. Alternative estimates of the intrinsic LyC from H$\beta$ neglect LyC photons absorbed by dust, assume an isotropic geometry, and require an assumed simplistic star formation history \citep[\eg,][]{flury22a}. Another alternative measure of LyC escape is the $F_{\lambda {\rm LyC}}/F_{\lambda {\rm 1100}}$ flux ratio. As a ratio of two observed fluxes, this method requires no model assumptions. However, we found in Paper I that this quantity was more difficult to predict from easily accessible observables with the survival analysis models. Hence, for the \fesc\ models in this paper, we proceed with the absolute \fesc\ estimates from the UV SED fits for the LzLCS+.

Optical and UV observations supply information about numerous other properties of the LzLCS+ galaxies (see \citealt{flury22a} and \citet{saldana22} for details). We obtain stellar mass ($M_*$) estimates from \textsc{Prospector} \citep{leja17, johnson19} fits to the SDSS and {\it Galaxy Evolution Explorer} ({\it GALEX}; \citealt{martin03}) photometry (\citealt{flury22a}, Ji et al.\ in prep.). With multi-Gaussian fits to nebular lines, we measure nebular line fluxes and equivalent widths (EWs), and we estimate the nebular dust attenuation, E(B-V)$_{\rm neb}$, from Balmer line ratios. We calculate the oxygen abundance using the direct method and the {\tt pyneb} package \citep{luridiana15}, adopting $n_e=100$ cm$^{-3}$ and the estimated \oiii~$\lambda$4363 flux from the \citet{pilyugin06} ``ff-relation" in cases where the \sii~$\lambda\lambda$6716,6731 doublet or \oiii~$\lambda$4363 auroral line are undetected. We estimate star formation rates (SFRs) from the \citet{kennicutt12} SFR calibration using the dust-corrected H$\beta$ luminosities and Case B H$\alpha$/H$\beta$ ratio \citep{storey95}. The COS near-UV acquisition images allow us to measure the UV half-light radius $r_{\rm 50,NUV}$, which we then use to calculate \sigsfr\ as
\begin{equation}
\Sigma_{\rm SFR} = \frac{\rm SFR}{2 \pi r^2_{\rm 50,NUV}}.
\end{equation}
In addition to \fesc, we derive several other parameters from the {\it HST} COS UV spectra. We obtain estimates of the dust attenuation E(B-V) from the UV spectrum \textsc{Starburst99} SED fits; we label this parameter as E(B-V)$_{\rm UV}$ to distinguish it from the nebular dust attenuation E(B-V)$_{\rm neb}$ derived from the Balmer lines. Although the G140L spectra do not extend to rest-frame 1500 \AA, we can estimate the ``observed" (non-extinction corrected) absolute magnitude at 1500 \AA\ ($M_{1500}$) and the power law index slope at 1550 \AA\ ($\beta_{1550}$) by extrapolating the SED fits to longer wavelengths \citep{saldana22}. The inferred $\beta_{\rm 1550}$ values do match the observed values for the few galaxies with existing longer wavelength UV spectra \citep{chisholm22}. We also measure \lya\ properties from the COS spectra. We derive \lya\ fluxes and EWs by linearly fitting the continuum within 100 \AA\ of \lya, excluding regions affected by nebular or stellar features, and integrating all emission above the continuum. We then derive the \lya\ escape fraction (\fesclya) using the dust-corrected H$\beta$ flux and intrinsic Case B \lya/H$\beta$ ratio \citep{storey95} appropriate for the galaxies' measured electron temperatures and densities. We note that the \lya\ measurements represent the net sum of both underlying absorption along the line of sight and scattered \lya\ emission within the aperture. Nine galaxies have detectable \lya\ absorption troughs that overlap with the \sithree~$\lambda$1206 absorption feature. We increase our uncertainties to account for the change in flux from omitting wavelengths within 500 km s$^{-1}$ of the \sithree\ line. The inclusion or exclusion of this region has only minor effects, changing the \lya\ EW by $<3$\AA\ and with typical changes to \fesclya\ of only 0.001.

\subsection{Multivariate Survival Analysis: The Cox Proportional Hazards Model}
\label{sec:cox}

In order to generate multivariate diagnostics for \fesc, we need to incorporate information from both the \fesc\ detections and the upper limits. Within the field of statistics, survival analysis techniques are appropriate for censored datasets that contain limits. One such survival analysis method is the Cox proportional hazards model (\citealt{cox72}; see \citealt{clark03} and \citealt{bradburn03} for reviews and \citealt{feigelson85} and \citealt{isobe86} for examples in astronomy). Here we describe the basic form of the Cox model and its assumptions. We refer the reader to Paper I for a more thorough discussion of this model and its application to the LzLCS+. To implement the Cox model, we use the CoxPHFitter routine in the {\tt lifelines} python package \citep{davidson19}.

The Cox proportional hazards model predicts the probability of a particular \fesc\ value given a set of input variables. Like many other survival analysis techniques, implementations of the Cox model typically assume the dataset contains measurements and lower limits (so-called ``right-censored" data). In contrast, our dependent variable data consist of \fesc\ measurements and associated upper limits. Consequently, we transform our \fesc\ values to the absorbed fraction of LyC, \fabs\ = 1 - \fesc, for use in the Cox model. However, for ease of interpretation, we put the \fabs\ results back into the form of \fesc\ in all figures and in the tabulated results in the Appendix.

As applied to the LzLCS+, the Cox proportional hazards regression model fits for the probability of a LyC detection in an infinitesimally small increment of \fabs, for a given set of independent variables and assuming no detection at a lower value of \fabs\ (higher value of \fesc). The Cox model assumes a particular functional form for this \fesc\ probability, which is known as the ``hazard function":
\begin{equation}
h(f_{\rm abs}|x) = h_0(f_{\rm abs})\exp[\sum_{i=1}^{n} b_i(x_i - {\bar x_i})].
\label{eqn:coxhazard}
\end{equation}
In this equation, $b_i$ are the best-fit coefficients for each input variable $x_i$, and $\bar x_i$ is the mean value of each input variable in the reference LzLCS+ dataset. The term $h_0$(\fabs) is the baseline hazard function, the probability of having \fabs\ in the case where all input variables match their average values within the LZLCS+. The Cox model is semi-parametric, in that the dependence on the input variables has a fixed exponential functional form, but the baseline hazard function, $h_0$, is estimated non-parametrically. Despite its fixed functional form for the input variable dependence, the Cox model allows the input variables to take any form. For instance, we could define an input variable $x_i$ as a measured value $j$, the logarithm of that measurement \logten($j$), the square of that measurement $j^2$, or any other functional form of our choosing. In our case, we opt to use logarithmic forms of our input variables where possible (e.g, \logten(O32), $M_{\rm 1500}$, \logten($M_*$)). In this logarithmic form, most variables have a similar order of magnitude and scale in the same manner, and an order of magnitude increase in a particular input variable simply translates into increasing the probability $h(f_{\rm abs}|x)$ by a factor of $e^{b_i}$. In this paper, the only variables that we do not use in a logarithmic fashion are EW(\lya) and \fesclya, because they range from negative (\lya\ absorption) to positive (\lya\ emission) within the LzLCS+.

Although Equation~\ref{eqn:coxhazard} gives the {\it probability} of \fabs, we would like to predict the expected {\it values} of \fabs\ and \fesc\ and their associated uncertainty. We adopt the median of the probability distribution as this expected value \citep[\eg,][]{bradburn03, davidson19}; the model predicts that \fesc\ will be above this value 50\%\ of the time and below it 50\%\ of the time. To determine this median \fabs\ mathematically, we first calculate the survival function, $S$(\fabs), the probability that we do not detect LyC at $f_{\rm abs,detect} < $ \fabs\ (i.e., at $f_{\rm esc,detect} > $ \fesc). In the Cox model, the survival function is
\begin{equation}
S(f_{\rm abs}) = \exp[-{\rm HF}_0(f_{\rm abs})\cdot{\rm ph}(x)],
\end{equation}
where HF$_0$ is the baseline cumulative hazard function
\begin{equation}
{\rm HF}_0(f_{\rm abs}) = \int_0^{f_{\rm abs}} h_0(f) df
\end{equation}
and ph($x$) is the partial hazards function, which describes how the probability scales with the set of input variables $x$:
\begin{equation}
{\rm ph}(x) = \exp[\sum_{i=1}^{n} b_i(x_i - {\bar x_i})].
\end{equation}
\citep[\eg,][]{cox72, bradburn03, davidson19, mclernon23}. The median \fabs, which corresponds to our predicted \fesc\ value, is the \fabs\ value where $S$(\fabs)$ = 0.5$. In some cases, $S$(\fabs)$ > 0.5$, even for our largest tabulated value of \fabs, which implies that \fabs\ is closer to 1 than we can determine and \fesc\ $\sim0$. In this circumstance, we infer that \fesc\ is arbitrarily small and report a predicted \fesc\ $=0$. 

The {\tt lifelines} CoxPhFitter returns the best-fit coefficients $b_i$ and the cumulative baseline hazard HF$_0$(\fabs) for each value of \fabs\ corresponding to a LyC detection in the LzLCS+ (see Paper I for more details about the {\tt lifelines} methodology). In the Appendix, we provide these best-fit parameters for the models in Paper I and the models in this paper, and we give examples of how to use these models to calculate the predicted \fesc\ for a set of observed input variables. The coefficients and hazards can be used to predict the expected \fesc\ for any galaxy, as long as it has estimated values for each of the independent variables used in the model. 

\subsubsection{Uncertainty in the Predicted \fesc}
\label{sec:uncertainty}
Because the survival function represents a probability distribution, we can also use it to calculate the expected uncertainty in our predicted \fesc\ estimate. The \fabs\ values where $S$(\fabs) = 0.159 and $S$(\fabs) = 0.841 represent the range in the predicted \fesc\ corresponding to the Normal theory 1-$\sigma$ uncertainty. This uncertainty range reflects the inherent scatter of the relationships between \fesc\ and the input variables, where this scatter can come both from measurement uncertainties and from genuine variation among the galaxy population. In Paper I, we tested the effect of measurement uncertainty by performing a Monte Carlo (MC) resampling of each independent and dependent variable according to its observational uncertainty. We found that the distribution of predicted \fesc\ from the resampled inputs was nearly always smaller than the 1-$\sigma$ bounds inferred from the survival function. In other words, the inherent scatter in the correlations, not the measurement error, is the dominant source of uncertainty in the predicted \fesc, and the survival function bounds serve as a reasonable estimate of the uncertainty in the \fesc\ predictions.

\subsubsection{Goodness-of-Fit Metrics}
\label{sec:metrics}
As in Paper I, we evaluate the Cox models' performance using several complementary metrics. The concordance index, $C$, is particularly useful, as it is appropriate for censored data. The concordance index assesses whether the model \fesc\ predictions correctly sort the dataset in the order of its observed \fesc. To evaluate this sorting, the concordance index calculation compares each possible pair of data points. Concordant pairs are those where the galaxy with higher observed \fesc\ also has higher predicted \fesc, discordant pairs are the opposite, and tied pairs have identical predicted \fesc. Some pairs with upper limits in \fesc\ lead to ambiguous rankings and do not appear in the concordance index calculation. With all pairs evaluated, the concordance index is calculated as
\begin{equation}
C = \frac{n_c+0.5n_t}{n_c+n_d+n_t},
\end{equation}
where $n_c$, $n_d$, and $n_t$ are the number of concordant, discordant, and tied pairs, respectively. A value of $C=1.0$ indicates a perfect rank ordering, 0.5 is perfectly random ordering, and 0 is perfect disagreement. The major advantage of $C$ is that it includes both LyC detections and non-detections. However, it only assesses the relative order of the predicted \fesc\ values, not their accuracy. To measure the latter, we turn to alternative quantities. One such metric is the $R^2$ statistic
\begin{equation}
R^2=1-\frac{\sum_i(y_i-f_i)^2}{\sum_i(y_i-{\bar y})^2},
\label{eqn:r2}
\end{equation}
where $y_i$ are the observed values of \logten(\fesc), ${\bar y}$ is their mean value, and $f_i$ are the predicted \fesc\ values from the Cox model. As in Paper I and \citet{maji22}, we also calculate a variant of $R^2$, the adjusted $R^2$, which accounts for the number of free parameters $p$ in the model and number of data points $n$: 
\begin{equation}
R^2_{\rm adj} = 1-(1-R^2)\frac{n-1}{n-p-1}.
\end{equation}
When increasing the number of variables, $R^2_{\rm adj}$ increases only if a variable improves the predictions more than expected by chance. Finally, we report the root-mean-square (RMS) dispersion
\begin{equation}
{\rm RMS}=\sqrt{\frac{\sum_i(y_i-f_i)^2}{n}}.
\end{equation}
As explained in Paper I, we evaluate these three quantities ($R^2$, $R^2_{\rm adj}$, and RMS) using \logten(\fesc). The scatter in the \logten(\fesc) predictions is relatively consistent across the full range of observed \fesc, whereas the scatter in the linear \fesc\ changes systematically across this range. We can only calculate the $R^2$, $R^2_{\rm adj}$, and RMS metrics for the LzLCS+ galaxies with both detected LyC and non-zero predicted \fesc, since these metrics require observed values of \logten(\fesc) and finite predicted \logten(\fesc). Hence, these metrics only indicate the accuracy of the model predictions for LCEs. However, the most successful models according to the $R^2$, $R^2_{\rm adj}$, and/or RMS metrics also tend to have high $C$ and vice versa (see Paper I). 

\subsubsection{Choice of Input Variables}
\label{sec:variable_choice}
In generating Cox models from the LzLCS+, we consider a limited set of input variables. Although the dependent variable in the Cox model can contain measurement limits, the independent variables cannot. We therefore only use independent variables that have measurements for nearly all the LzLCS+ galaxies, which requires us to exclude fainter emission line measurements (e.g., \oi~$\lambda$6300 and \sii~$\lambda\lambda$6716,6731) or measurements not widely available for the full sample (e.g., \lya\ velocity peak separation). In this paper, all 88 galaxies in the LzLCS+ have measurements for our chosen input variables with one exception. The non-leaker J1046+5827 does not have a reported \sigsfr\ or $r_{\rm 50, NUV}$, and we do not include it in deriving Cox models that use these variables. The Cox model may also fail to converge if we include highly collinear variables, which trace nearly identical properties. Given this limitation, we choose to use only one measure of UV dust attenuation (E(B-V)$_{\rm UV}$ or $\beta_{\rm 1550}$) per model, and we do not include both \sigsfr\ and $r_{\rm 50, NUV}$ in a single model. 

We explored a variety of variable combinations in Paper I, and our best model attained $R^2=0.69$, $R^2_{\rm adj}=0.60$, RMS $= 0.31$ dex, and $C=0.91$. We provide the parameters for this model in the Appendix. However, this model included the average EW of Lyman-series absorption lines (EW(\hi,abs)) and \fesclya, both of which will be affected by the IGM at $z>6$. Models in Paper I without UV absorption lines or \lya\ showed higher scatter but an overall ability to identify LCEs, with $R^2=0.29-0.40$, $R^2_{\rm adj}=0.14-0.35$, RMS $= 0.44-0.47$ dex, and $C = 0.83$ (Paper I; see Appendix for these model parameters). In this paper, we are specifically concerned with variables that are easily observable at $z>6$ or measured in large samples at $z\sim3$. Hence, we likewise omit UV absorption line measurements, and we avoid \lya\ measurements for most of the models tailored to $z>6$ galaxies. 

\section{Testing the Cox Models at $z\sim3$}
\label{sec:z3}

We apply the Cox models developed on the LzLCS+ sample to galaxies at $z\sim3$ and $z\gtrsim6$. We use the $z\sim3$ galaxies to test whether the models based on the low-redshift LzLCS+ galaxies can successfully predict \fesc\ for galaxies at high redshift. We then apply the Cox models to $z\gtrsim6$ samples to predict \fesc\ for galaxies in the epoch of reionization. 

\subsection{High-redshift Datasets and Models}
\label{sec:highzmodels}
First, we compile samples of $z\sim3$ galaxies with reported global absolute LyC \fesc\ and at least three of the input variables from Paper I, which include stellar mass $M_*$, $M_{\rm 1500}$, nebular EWs, metallicity, optical nebular line ratios, \sigsfr, $r_{\rm 50,NUV}$, E(B-V)$_{\rm UV}$, $\beta_{\rm 1550}$, and \lya\ measurements. The $z\sim3$ LyC measurements include individual detections, reported LyC upper limits, and stacked samples that average over variations in IGM attenuation \citep{hainline09, vasei16, james18, debarros16, vanzella16, steidel18, pahl21, bassett19, fletcher19, nakajima20, bian20, marqueschaves21, marqueschaves22, liu23, kerutt24}. We exclude one AGN from the \citet{fletcher19} sample and one shock-dominated galaxy from the \citet{bassett19} sample. Some well-known $z\sim3$ LCEs do not appear in our sample, because they lack published global absolute \fesc\ measurements (e.g., Ion3 and the Sunburst Arc). We also note that some of the high-redshift samples fall outside of the parameter space probed by the LzLCS+, such that our model predictions will extrapolate for these galaxies. The $z\sim3$ data has limitations as well; the compiled $z\sim3$ samples differ in the methods used for \fesc\ and input variable measurements, and individual $z\sim3$ \fesc\ measurements have high uncertainty due to unknown variations in IGM attenuation. We discuss the limitations of the $z\sim3$ comparison further in the following sections.

In Paper I, we introduced a JWST Cox Model, which could be applied to an ideal $z>6$ sample, with eight relevant variables from the LzLCS+ included. We also found that a ``limited JWST Model", fit using only the three top-ranked variables ($\beta_{\rm 1550}$, \logten(O32), and \logten(\sigsfr)) performed equally well for predicting \fesc\ in the LzLCS+ sample. Unfortunately, many of the required variables for both the full and limited JWST models have not been measured for large samples of $z\sim3$ LCEs or $z>6$ galaxies. Consequently, we run new models limited to the variables available for the $z\sim3$ and for $z\gtrsim6$ samples. In Table~\ref{table:highz:models} and Table~\ref{table:highz:samples}, we list these models, the variables they include, and the samples to which they apply. 

For the $z\gtrsim6$ samples, we prioritize models that can apply to large samples \citep[\eg,][]{endsley21, endsley23, morishita24}, models that can apply to faint galaxies \citep{atek24}, and models that have at least two of the most statistically significant variables in Paper I (a measure of dust attenuation plus a measurement of ionization or morphology). The ``TopThree" model includes only the three top-ranked variables from Paper I: dust, O32, and \sigsfr. In the TopThree model, we use E(B-V)$_{\rm UV}$ rather than $\beta_{\rm 1550}$, since it enables us to compare with a larger $z\sim3$ sample. As noted in \citet{chisholm22}, the $\beta_{\rm 1550}$ and E(B-V)$_{\rm UV}$ measurements for the LzLCS+ track each other almost perfectly and provide equivalent information. Most models in Table~\ref{table:highz:models} use E(B-V)$_{\rm UV}$; models with ``$\beta$" in their name use $\beta_{\rm 1550}$ instead, and one model, LAE-O32-nodust, has no dust attenuation measurement. All models except the TopThree model include $M_{\rm 1500}$, and most models include $M_*$ as well. The only models without $M_*$ are the LAE, LAE-O32-nodust, and ELG-O32-$\beta$ models.

We provide the best-fit coefficients and cumulative baseline hazards for each model in the Appendix, which can be used to derive the median predicted \fesc\ for a given galaxy and the uncertainty from the 16-84th percentiles of the \fesc\ probability distribution. As with the LzLCS+ galaxies, we use an MC method to sample the variable uncertainties for the $z\sim3$ and $z\gtrsim6$ observations and re-generate the predicted \fesc. We again find that the uncertainty estimated from the Cox model survival function dominates over the uncertainty from sampling the input variables in nearly all cases. 

\begin{deluxetable*}{llllll}
\tablecaption{Cox Models for High-Redshift Predictions}
\label{table:highz:models}
\tablehead{
\colhead{Model} & \multicolumn{5}{c}{Variables} \\
\cline{2-6}
\colhead{} & \colhead{Dust} & \colhead{\lya} & \colhead{Nebular} & \colhead{Luminosity} &  \colhead{Morphology}}
\startdata
TopThree & E(B-V)$_{\rm UV}$ & --- & \logten(O32) & --- & \logten(\sigsfr) \\ 
LAE & E(B-V)$_{\rm UV}$ & EW(\lya) & --- & $M_{\rm 1500}$ & --- \\ 
LAE-O32 & E(B-V)$_{\rm UV}$ & EW(\lya) & \logten(O32) & $M_{\rm 1500}$, \logten($M_*$) & --- \\ 
LAE-O32-nodust &--- & EW(\lya) & \logten(O32) & $M_{\rm 1500}$ & --- \\ 
ELG-EW & E(B-V)$_{\rm UV}$ & --- &  \logten(EW([O III]+H$\beta$)) & $M_{\rm 1500}$, \logten($M_*$) & ---\\ 
ELG-O32 & E(B-V)$_{\rm UV}$ & --- &  \logten(O32) & $M_{\rm 1500}$, \logten($M_*$) & ---\\ 
ELG-O32-$\beta$ & $\beta_{\rm 1550}$ &--- & \logten(O32) & $M_{\rm 1500}$ &--- \\ 
ELG-O32-$\beta$-Ly$\alpha$ & $\beta_{\rm 1550}$ & \fesclya\ & \logten(O32) & $M_{\rm 1500}$, \logten($M_*$) & --- \\ 
R50-$\beta$ & $\beta_{\rm 1550}$ &--- & ---& $M_{\rm 1500}$, \logten($M_*$) &  \logten($r_{\rm 50,NUV}$)) \\
$\beta$-Metals & $\beta_{\rm 1550}$ &--- & 12+\logten(O/H) & $M_{\rm 1500}$, \logten($M_*$) &--- \\ 
\enddata
\end{deluxetable*}

\movetabledown=2.5in
\begin{rotatetable*}
\begin{deluxetable*}{llll}
\tablecaption{Cox Model High-Redshift Samples}
\label{table:highz:samples}
\tablehead{
\colhead{Model} & \multicolumn{2}{l}{$z\sim3$ Samples} & \colhead{$z\sim6$ Samples}}
\startdata
TopThree & 1 Detection: & Ion2\tablenotemark{a} & \\
& 1 Non-Detection: & Cosmic Horseshoe\tablenotemark{b} & \\
& {\it 1 with Input Limits:} & J1316+2614\tablenotemark{c}  & \\
\hline
LAE & 28 Detections+13 Stacks: & Ion2\tablenotemark{a}, \citet{bassett19}, \citet{fletcher19}\tablenotemark{d}, \citet{liu23}\tablenotemark{e} & \citet{saxena23} \\
& &  \citet{kerutt24}, J0121+0025\tablenotemark{f}, J1316+2614\tablenotemark{c}, \citet{steidel18}\tablenotemark{g} & \\
& 6 Non-Detections+5 Stacks: & \citet{bassett19}, \citet{steidel18}\tablenotemark{g}, \citet{bian20} & \\
& {\it 3 with Input Limits:} &  \citet{fletcher19}\tablenotemark{d} & \\
\hline
LAE-O32 & 4 Detections & Ion2\tablenotemark{a}, \citet{bassett19},  \citet{fletcher19}\tablenotemark{d} &  \citet{saxena23} \\
& 5 Non-Detections: & \citet{bassett19} & \\
& {\it 5 with Input Limits:} & \citet{bassett19},  \citet{fletcher19}\tablenotemark{d}, J1316+2614\tablenotemark{c} & \\
\hline
LAE-O32-nodust & 5 Detections+1 Stack: & Ion2\tablenotemark{a}, \citet{bassett19},  \citet{nakajima20}, J1316+2614\tablenotemark{c} &  \citet{saxena23} \\
& 14 Non-Detections+2 Stacks: &  \citet{bassett19},  \citet{nakajima20} & \\
& {\it 19 with Input Limits:} & \citet{bassett19},  \citet{nakajima20} & \\
\hline
ELG-EW & 4 Detections: & Ion2\tablenotemark{a}, \citet{fletcher19}\tablenotemark{d} & \citet{endsley21, endsley23}, \\
& {\it 4 with Input Limits: } & \citet{fletcher19}\tablenotemark{d}, J1316+2614\tablenotemark{c} & \citet{bouwens23}, \citet{tang23}\tablenotemark{h}, \\
& & & \citet{fujimoto23}, \citet{saxena23}\\
\hline
ELG-O32 & 4 Detections: & Ion2\tablenotemark{a}, \citet{bassett19},  \citet{fletcher19}\tablenotemark{d} & \citet{williams23}, \citet{tang23}\tablenotemark{h}, \\
& 5 Non-Detections:  & \citet{bassett19} &  \citet{fujimoto23}, \citet{saxena23} \\
& {\it 5 with Input Limits:} & \citet{bassett19},  \citet{fletcher19}\tablenotemark{d}, J1316+2614\tablenotemark{c} & \\
\hline
ELG-O32-$\beta$ & 2 Detections: &  Ion2\tablenotemark{a}, J1316+2614\tablenotemark{c} &  \citet{williams23}, \citet{schaerer22b}\tablenotemark{i}, \\
& & & \citet{mascia23}, \citet{saxena23, saxena24} \\ 
\hline
ELG-O32-$\beta$-\lya\ & & & \citet{saxena23} \\
\hline
R50-$\beta$ & 1 Detection: & Ion2\tablenotemark{a} & \citet{mascia23}, \citet{morishita24}\tablenotemark{j} \\
\hline
$\beta$-Metals & 1 Detection: & Ion2\tablenotemark{a} & \citet{schaerer22b}\tablenotemark{i}, \\
 & {\it 1 with Input Limits:}  & J1316+2614\tablenotemark{c} & \citet{williams23}, \citet{atek24}\tablenotemark{k}  \\
\enddata
\tablecomments{{\it (a)} From \citet{debarros16, vanzella16, vanzella20}. {\it (b)} From \citet{hainline09, vasei16, james18}. We adopt the average dust-corrected O32 ratio of the two similar-flux regions in \citet{hainline09}. The E(B-V)$_{\rm UV}$ and \sigsfr\ values are flux-weighted averages of the regions in \citet{james18}. {\it (c)} From \citet{marqueschaves22}.{\it (d)} From \citet{fletcher19, nakajima20}.{\it (e)} We adopt the \fesc\ determined from the SED-fitting method in \citet{liu23}.{\it (f)} From \citet{marqueschaves21}.{\it (g)} From \citet{steidel18, pahl21}.{\it (h)} We adopt $M_*$ calculated using the bursty non-parametric star formation history, but include the alternative values in the uncertainties. We use the direct method metallicities for the two galaxies with secure \oiii~$\lambda$4363 detections and the modeled metallicities for the others.{\it (i)} We correct $M_*$ for magnification, using the values in \citet{schaerer22b}. For the source without a direct method metallicity, we adopt the strong-line method 12+\logten(O/H).{\it (j)} We exclude sources flagged as having significant residuals in the morphological fits.{\it (k)} Metallicities calculated using strong-line methods.}
\end{deluxetable*}
\end{rotatetable*}

\begin{deluxetable*}{lllllll}
\tablecaption{Metrics for High-Redshift Cox Models}
\label{table:highz:metrics}
\tablehead{}
\startdata
\multicolumn{7}{l}{LzLCS+ Sample} \\
Model & $N_{\rm Gal}$ & $N_{\rm Detect}$ & $R^2$ & $R^2_{\rm adj}$ & RMS & $C$ \\
\hline
TopThree & 87 & 45 & 0.38 & 0.34 & 0.44 & 0.82 \\ 
LAE & 88 & 46 & 0.21 & 0.15 & 0.50 & 0.81 \\ 
LAE-O32 & 88 & 44 & 0.37 & 0.29 & 0.45 & 0.82 \\ 
LAE-O32-nodust & 88 & 43 & 0.02 & -0.06 & 0.59 & 0.77 \\ 
ELG-EW & 88 & 43 & 0.14 & 0.05 & 0.53 & 0.79 \\ 
ELG-O32 & 88 & 42 & 0.42 & 0.36 & 0.44 &  0.79 \\ 
ELG-O32-$\beta$ & 88 & 42 & 0.30 & 0.25 & 0.48 & 0.79 \\ 
ELG-O32-$\beta$-\lya\ & 88 & 43 & 0.40 & 0.32 & 0.45 & 0.81 \\ 
R50-$\beta$ & 87 & 45 & 0.36 & 0.30 & 0.44 & 0.84 \\ 
$\beta$-Metals & 88 & 44 & 0.11 & 0.02 & 0.55 & 0.77 \\ 
\hline
\hline
\multicolumn{7}{l}{} \\
\multicolumn{7}{l}{$z\sim3$ Sample} \\
Model & $N_{\rm Gal}$ & $N_{\rm Detect}$ & $R^2$ & $R^2_{\rm adj}$ & RMS & $C$ \\
\hline
TopThree & 2 & 1 & --- & --- & 0.15 & 1.00 \\ 
LAE & 52 & 35 & -4.18 & --- & 0.92 & 0.49 \\ 
LAE-O32 & 9 &  4 & -0.35 & --- & 0.32 & 0.71 \\ 
LAE-O32-nodust & 22 & 6 & -0.12  & --- & 0.31  & 0.80 \\ 
ELG-EW & 4 & 4 & -4.81 & --- & 0.65 & 0.50 \\ 
ELG-O32 & 9 & 4 & 0.09 & --- & 0.27 & 0.71 \\ 
ELG-O32-$\beta$ & 2 & 2 & -0.91 & --- & 0.04 & 1.00 \\ 
ELG-O32-$\beta$-\lya\  & --- & --- & --- & --- & ---\\ 
R50-$\beta$ & 1 & 1 & --- & --- & 0.09 & --- \\ 
$\beta$-Metals & 1 & 1 & --- & --- & 0.43 & --- \\  
\hline
\hline
\multicolumn{7}{l}{} \\
\multicolumn{7}{l}{Combined Sample} \\
Model & $N_{\rm Gal}$ & $N_{\rm Detect}$ & $R^2$ & $R^2_{\rm adj}$ & RMS & $C$ \\
\hline
TopThree & 89 & 46 & 0.43 & 0.39 & 0.44 & 0.83 \\ 
LAE & 140 & 81 & -0.47 & -0.53 & 0.71 & 0.70 \\ 
LAE-O32 &  97 & 48 & 0.42 & 0.35 & 0.44 & 0.82 \\ 
LAE-O32-nodust & 110 & 49 & 0.19 & 0.13 & 0.56 & 0.78 \\ 
ELG-EW & 92 & 47 & 0.16 & 0.08 & 0.54 & 0.79 \\ 
ELG-O32 & 97 & 46 & 0.48 & 0.42 & 0.43 & 0.80 \\ 
ELG-O32-$\beta$ & 90 & 44 & 0.40 & 0.36 & 0.47 & 0.80 \\ 
ELG-O32-$\beta$-\lya\  & --- & --- &  --- & --- &--- & ---\\ 
R50-$\beta$ & 88 & 46 & 0.41 & 0.35 & 0.43 & 0.84 \\ 
$\beta$-Metals & 89 & 45 & 0.17 & 0.08 & 0.55 & 0.77\\ 
\enddata
\tablecomments{$N_{\rm Gal}$ is the number of galaxies assessed in the fit and does not include any galaxies with limits for input variables. $N_{\rm Detect}$ is the number of galaxies with LyC detections, finite \fesc\ predictions, and no upper or lower limits for input variables; the $R^2$ and RMS metrics use only these galaxies. The $R^2$ statistic measures how well the predictive model explains the observed variance in the \logten(\fesc) data, with higher $R^2$ values corresponding to more accurate models. The $R^2_{\rm adj}$ metric  accounts for the number of parameters used in the model and increases only if a variable improves the fit more than expected by chance. RMS is the root-mean-square dispersion of the predicted vs.\ observed \logten(\fesc) for the LyC detections. Higher $C$ values indicate that the model more accurately sorts the observations in the correct order of increasing \fesc. $C$ includes both detections and galaxies with \fesc\ upper limits. See \S\ref{sec:metrics} for a full description of these statistics.}
\end{deluxetable*}

We first test the Cox model predictions on $z\sim3$ samples to see whether the high-redshift LCEs behave similarly to the LzLCS+ sample. We derive Cox models for the variable sets in Table~\ref{table:highz:models} using only the LzLCS+ data; the high-redshift samples are not included in the fitting process. We then use the models to predict  \fesc\ for each set of $z\sim3$ measurements, and we calculate goodness-of-fit metrics for the LzLCS+ sample alone, the high-redshift sample alone, and for the combined sample of low- and high-redshift galaxies. We list these goodness-of-fit metrics in Table~\ref{table:highz:metrics}.

\subsection{Model Performance}
\label{sec:highzmetrics}
\begin{figure*}
\gridline{\fig{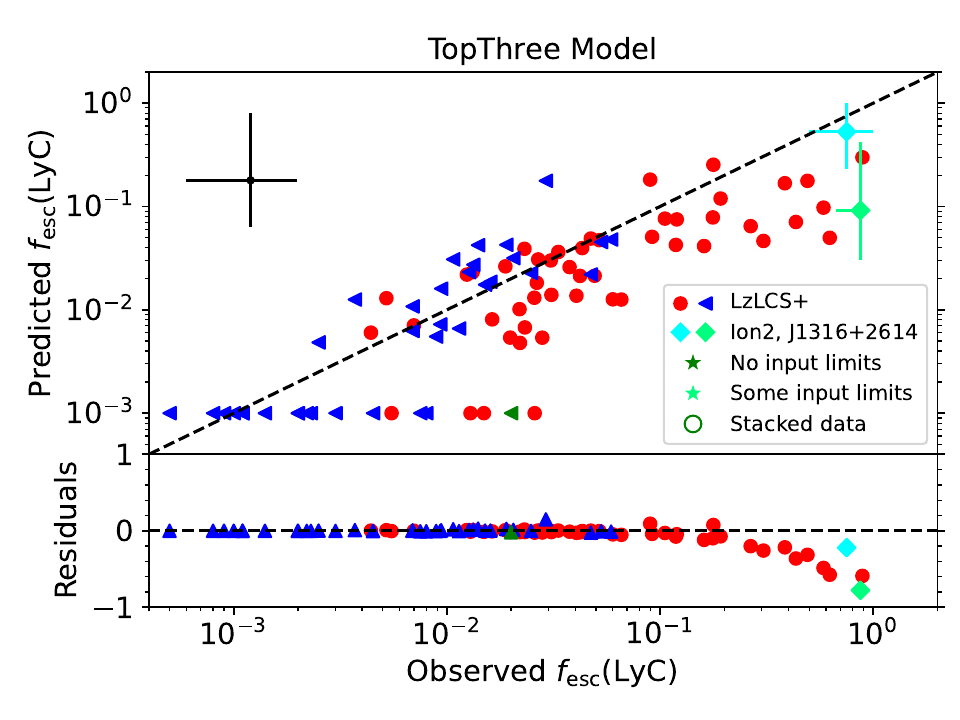}{0.5\textwidth}{(a)}
	\fig{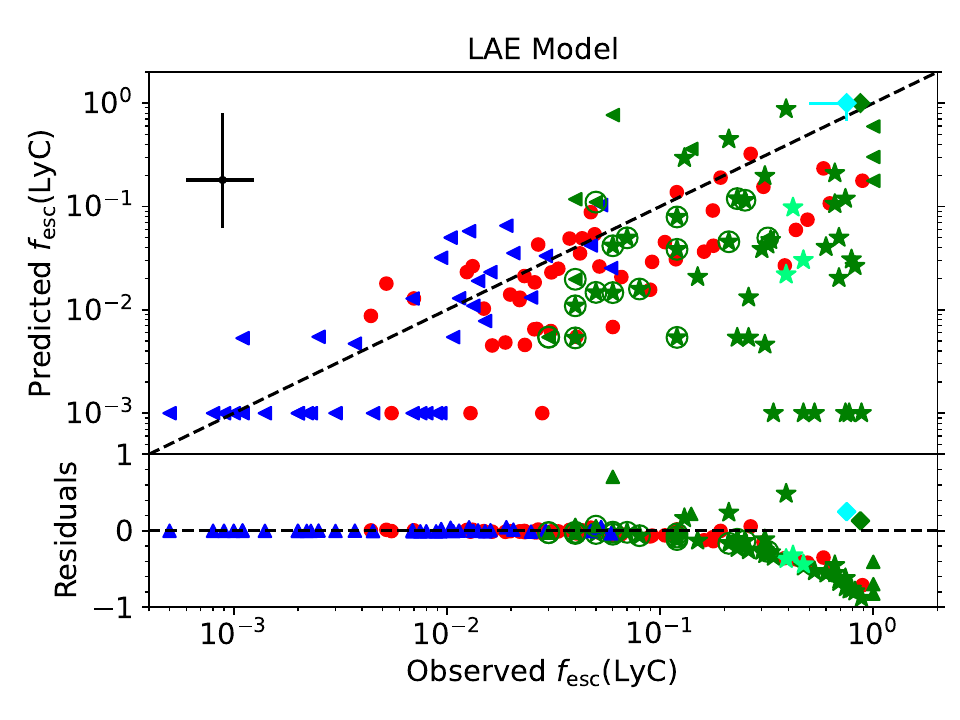}{0.5\textwidth}{(b)}
	}
\gridline{\fig{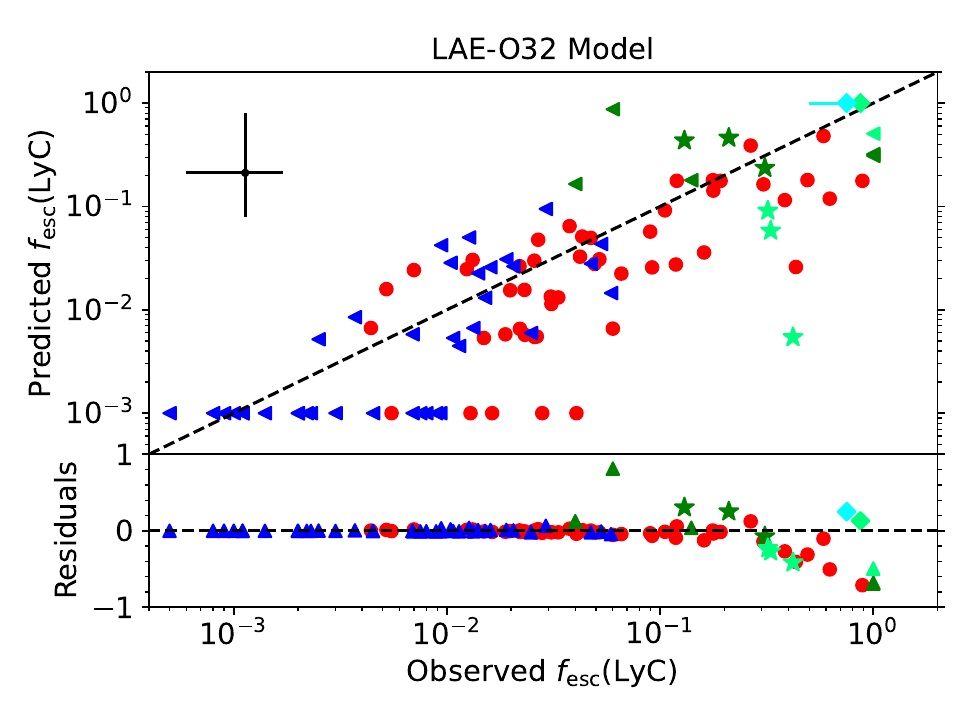}{0.5\textwidth}{(c)}
	\fig{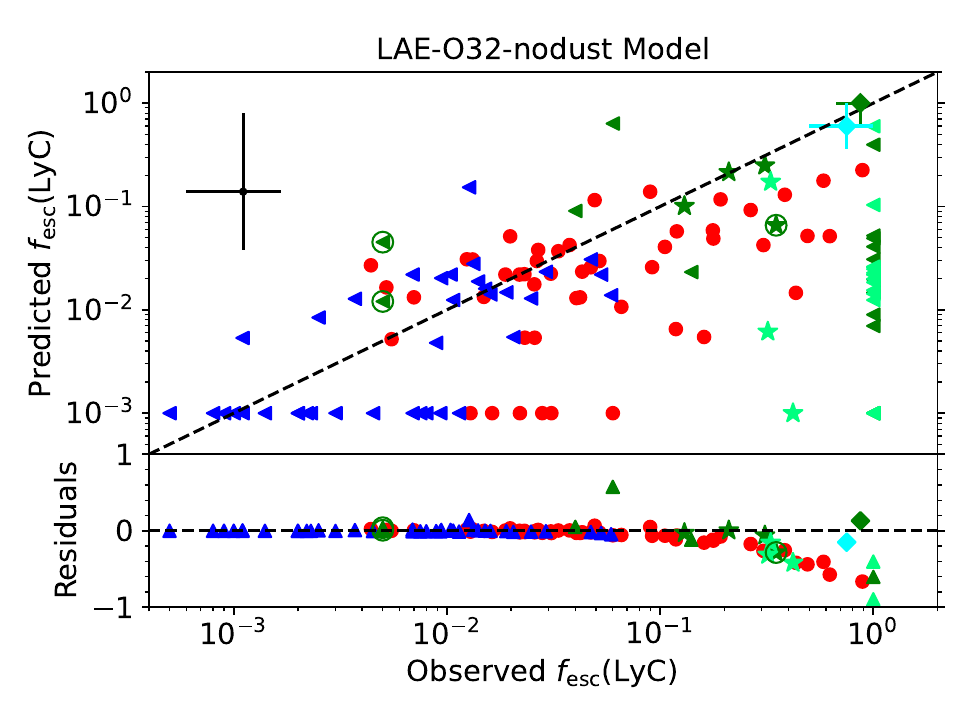}{0.5\textwidth}{(d)}
	}
\caption{The \fesc\ predictions from the TopThree (a), LAE (b), LAE-O32 (c), and LAE-O32-nodust (d) models. See Table~\ref{table:highz:models} for model descriptions. Red circles represent LzLCS+ LyC detections, and blue triangles represent upper limits. We plot $z\sim3$ galaxies in green, with stars denoting LyC detections and triangles denoting upper limits. We identify the two strongest high-redshift LCEs, Ion2 and J1316+2614, by a teal and a green diamond, respectively. Light green symbols indicate high-redshift galaxies that have a limit for one or more input variables. We draw circles around data points representing high-redshift galaxy stacks, which are less subject to uncertainty in IGM attenuation. The error bar in the upper left corner indicates the median size of the uncertainties in the observed and predicted \fesc\ for the combined sample of low- and high-redshift galaxies. The dashed line shows a one-to-one correspondence. Several of the Cox models predict \fesc\ for both low- and high-redshift galaxies with comparable accuracy.
\label{fig:highz_a}}
\end{figure*}

\begin{figure*}
\gridline{\fig{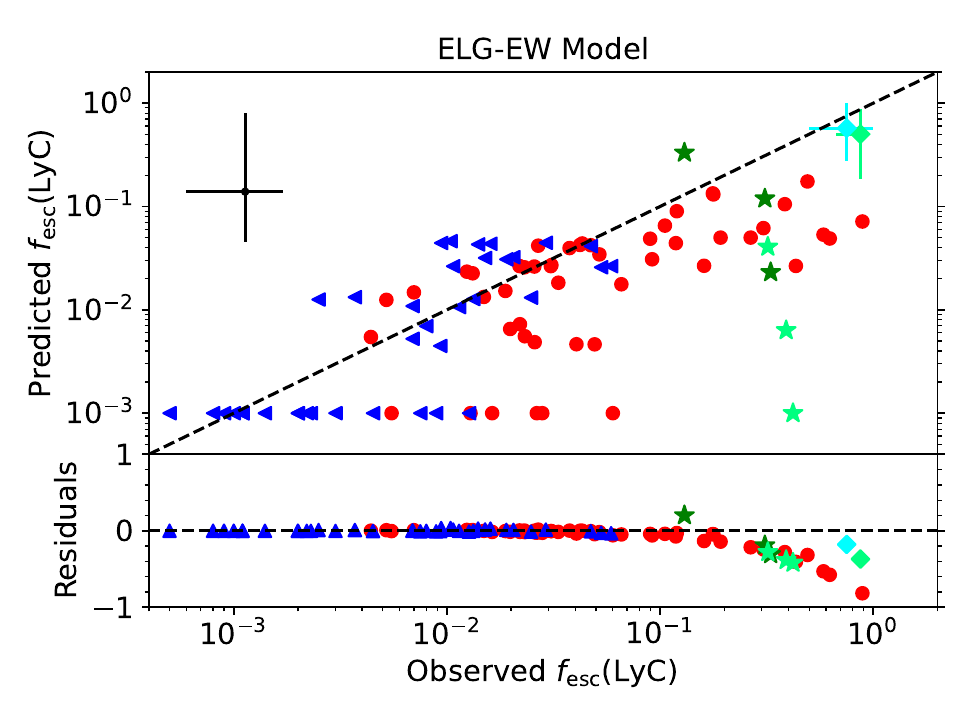}{0.5\textwidth}{(a)}
	\fig{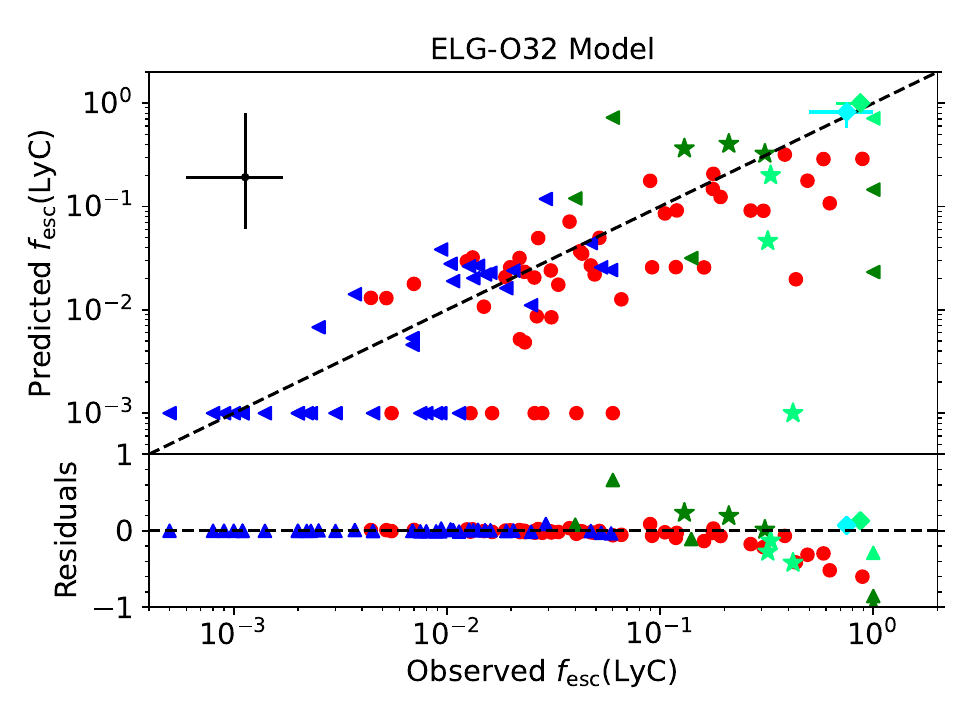}{0.5\textwidth}{(b)}
	}
\gridline{\fig{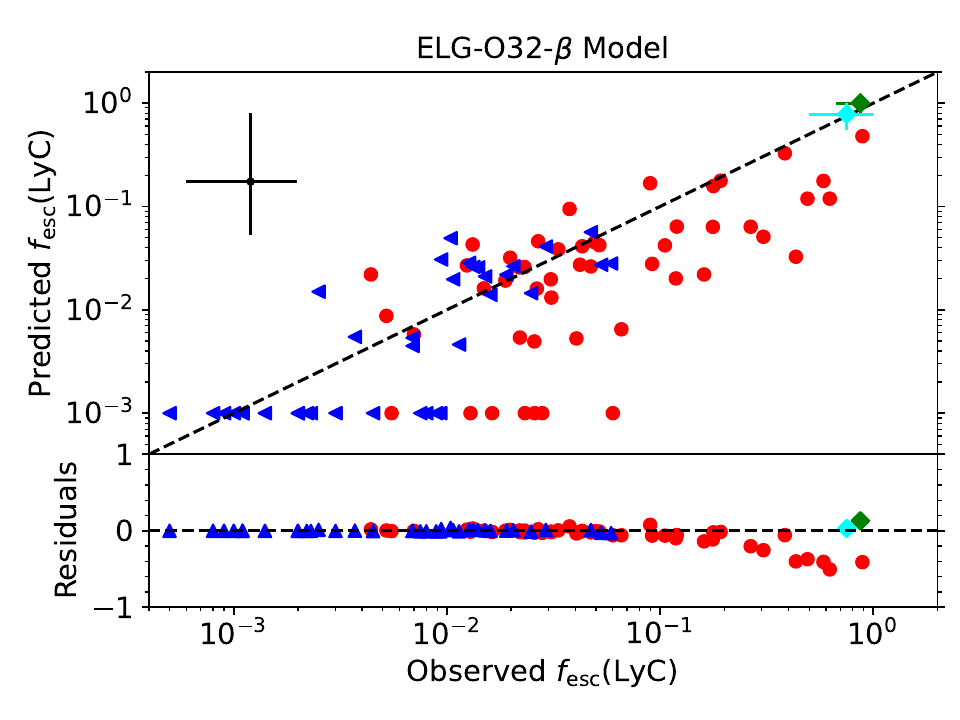}{0.5\textwidth}{(c)}
	\fig{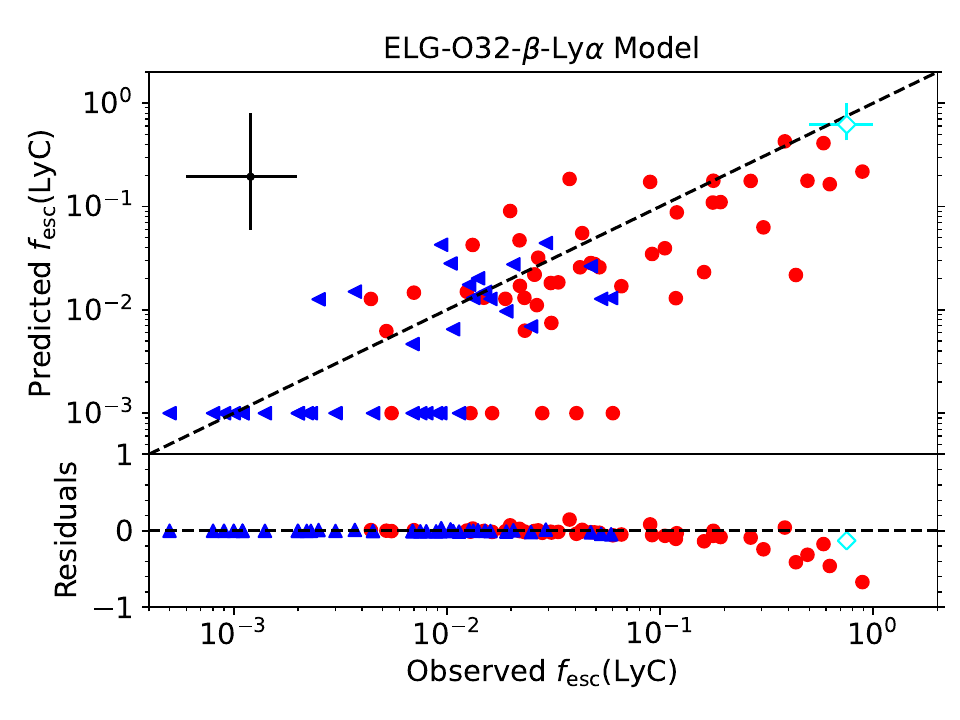}{0.5\textwidth}{(d)}
	}
\gridline{\fig{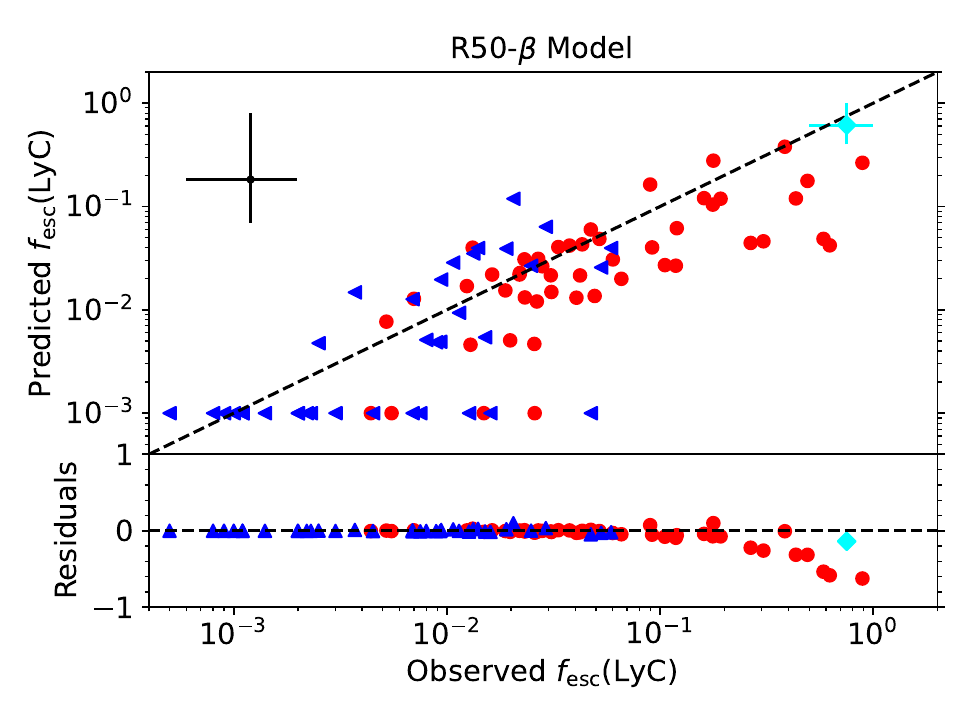}{0.5\textwidth}{(e)}
	\fig{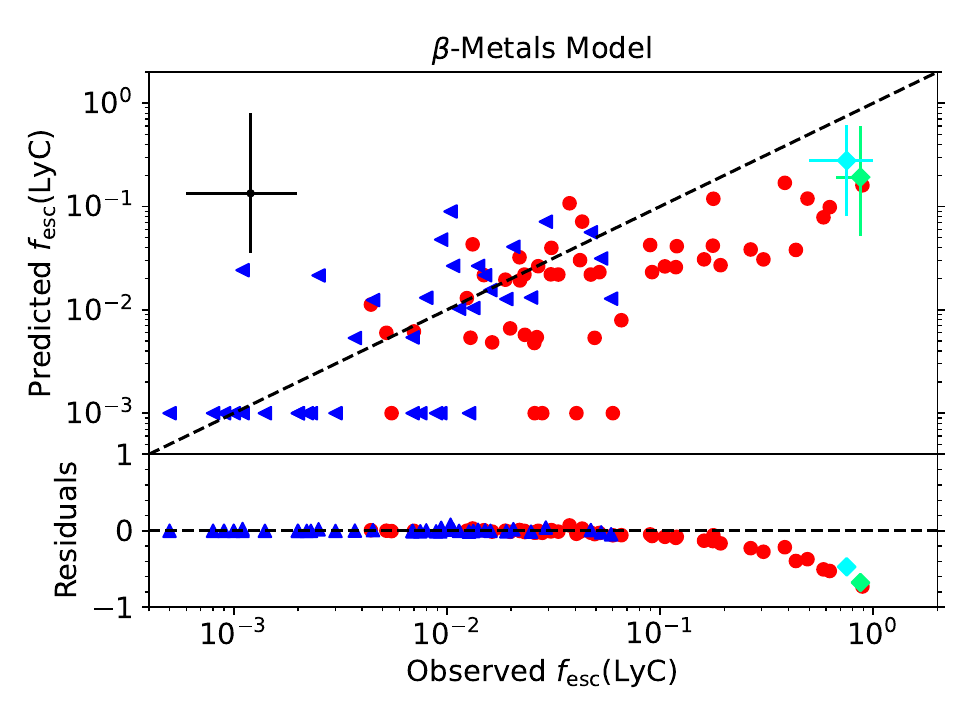}{0.5\textwidth}{(f)}
	}
\caption{The \fesc\ predictions from the ELG-EW (a), ELG-O32 (b), ELG-O32-$\beta$ (c), ELG-O32-$\beta$-Ly$\alpha$ (d), R50-$\beta$ (e), and $\beta$-Metals (f) models. See Table~\ref{table:highz:models} for model descriptions. Ion2 is shown by an open symbol in panel d, because its prediction is based on a limit in \fesclya. Other symbols are the same as in Figure~\ref{fig:highz_a}.
\label{fig:highz_b}}
\end{figure*}

Figures~\ref{fig:highz_a} and \ref{fig:highz_b} show the predicted vs.\ observed \fesc\ for the LzLCS+ and $z\sim3$ samples. As seen from the plots and the goodness-of-fit statistics in Table~\ref{table:highz:metrics}, these models do not perform as well as the fiducial model from Paper I, but several (TopThree, LAE-O32, ELG-O32, ELG-O32-$\beta$, ELG-O32-$\beta$-\lya, R50-$\beta$) are comparable to or better than the Paper I JWST model (see Tables~\ref{atab:U}-\ref{atab:JWSTbeta} for the list of input variables used in the Paper I models). The metrics for the LzLCS+ sample are $R^2=0.02-0.42$ (vs.\ 0.59 for the fiducial model and 0.29 for JWST), $R^2_{\rm adj}=-0.06$ to 0.36 (vs.\ 0.48 for the fiducial model and 0.14 for JWST), RMS$=0.44-0.59$ (vs.\ 0.37 for the fiducial model and 0.47 for JWST), and $C=0.77$ to 0.84 (vs.\ 0.88 for the fiducial model and 0.83 for JWST).

The models developed on the LzLCS+ sample generally reproduce the observed \fesc\ values of the $z\sim3$ LCEs. In fact, the high-redshift samples often have a lower RMS scatter than the predictions for the LzLCS+, and in several models, the $R^2$ and $C$ values for the combined high- and low-redshift sample are comparable to or higher than the $R^2$ and $C$ values for the LzLCS+ sample alone. Notably, predictions for the two strongest high-redshift LCEs, Ion2 and J1316+2614, match the observations nearly perfectly in most cases (Figures~\ref{fig:highz_a} and \ref{fig:highz_b}), demonstrating the success of the model in identifying extreme LCEs. 

The goodness-of-fit metrics give clues as to which variables are most important in predicting \fesc. As seen in Table~\ref{table:highz:metrics}, the models with the highest $R^2$ and lowest RMS for the LzLCS+ sample are the TopThree, LAE-O32, ELG-O32 and variants, and the R50-$\beta$ models. These same models are the only ones that include both a measure of UV dust attenuation and either O32 or radius as variables. Turning to the $C$ metric, which includes non-detections, the best-performing models for the LzLCS+ sample are the R50-$\beta$, TopThree, LAE-O32, ELG-O32-$\beta$-\lya, and LAE models. To properly estimate \fesc\ for weak and non-LCEs, UV dust attenuation again appears important, but each of the best-performing models also includes either morphological information or \lya\ measurements. These additional variables may help distinguish non-LCEs from LCEs. 

By all metrics, the LAE-O32-nodust and $\beta$-Metals models perform the worst for the LzLCS+ galaxies. The LAE-O32-nodust model is the only one that lacks E(B-V)$_{\rm UV}$ or $\beta_{\rm 1550}$, which highlights the crucial role of dust extinction in LyC escape. Conversely, the $\beta$-Metals model relies almost exclusively on dust attenuation to infer \fesc\ and lacks information on O32, \lya, or morphology, which may better constrain the LyC absorption due to \hi.

\subsubsection{Model Performance for $z\sim3$ Samples}
\label{sec:z3perform}
Comparing the models' performance for the high-redshift and combined low- and high-redshift samples is difficult, because each high-redshift model applies to a different set of galaxies. For example, the ELG-O32-$\beta$ model has the lowest RMS for the high-redshift galaxies, but this RMS is based on a single galaxy: Ion2. Ion2 is the only galaxy that is included in most of the models (all but ELG-O32-$\beta$-\lya, where it has an input limit). For Ion2, we adopt an observed absolute \fesc=0.75, in the middle of the 0.5-1 bounds reported by \citet{vanzella16}. Using this \fesc, Ion2's LyC escape is predicted most accurately by the ELG-O32-$\beta$ (RMS=0.02 dex) and ELG-O32 (RMS=0.04 dex) models. For Ion2, dust attenuation, luminosity, and ionization appear to be key factors in predicting its \fesc, although most other models are still consistent with its reported range of \fesc\ (RMS=0.09-0.15 dex for the next six models). The $\beta$-Metals model, which lacks information about O32 and morphology, is the only inaccurate model (RMS=0.43 dex). More surprisingly, however, the TopThree model gives the lowest predicted \fesc=0.53 (RMS=0.15 dex), which highlights the role of UV magnitude in the \fesc\ predictions, a point we discuss further in \S\ref{sec:z6}.

Nine galaxies, including four LCEs, appear in each of the LAE, LAE-O32, LAE-O32-nodust, and ELG-O32 model test samples. Although the LAE-O32-nodust model performed worst for the LzLCS+, for this common subsample of nine high-redshift galaxies, the LAE-O32-nodust model has the lowest RMS (0.09 dex) and highest $C$ (0.76), compared to RMS values of 0.27-0.32 dex and $C$ values of 0.65-0.71 for these galaxies in the other three models. Given the generally successful performance of all four of these models, models with O32 and/or EW(\lya) appear able to predict \fesc\ for high-redshift galaxies equally well. Hence, while models with E(B-V)$_{\rm UV}$ and O32 or morphology work best for the LzLCS+ sample, we cannot rule out the possibility that EW(\lya) may also be important in predicting \fesc\ at $z\sim3$. 

Nevertheless, using EW(\lya) as the sole nebular measurement is not sufficient to accurately predict \fesc. The LAE model (Figure~\ref{fig:highz_a}b), which applies to the largest set of $z\sim3$ LyC measurements, is one of the least successful models at predicting \fesc\ in the high-redshift galaxies. As seen in Table~\ref{table:highz:metrics}, this model has the largest RMS scatter for the $z\sim3$ sample (0.9 dex), lowest $C$ (0.49), and the second worst $R^2$ (-4.2). The other model with comparably poor $R^2$ and $C$ for $z\sim3$ galaxies is the ELG-EW model, which is the only other model that exclusively relies on an emission line EW for its input nebular information. Both models also have some of the lowest $R^2$ values for the LzLCS+ sample, which reflects the fact that these models substantially under-predict the \fesc\ of the strongest LCEs in the LzLCS+. The LAE model shows this same tendency towards under-prediction for the $z\sim3$ LCEs as well; 13 $z\sim3$ galaxies with observed \fesc\ $\geq0.2$ are predicted to have \fesc\ of only $\leq0.03$ by the model. 

The poor performance of the LAE model may result from limitations in both the model itself and in the $z\sim3$ data. The fact that both the LAE and ELG-EW models fare poorly for the LzLCS+ and for the high-redshift galaxies, especially for strong LCEs, suggests that nebular EW is a flawed tracer of \fesc. Indeed, \oiii\ EW and \lya\ EW should be low both for the weakest and for the strongest LCEs. In non-leakers, a high optical depth and corresponding lack of \lya\ escape should lead to a low \lya\ EW. Low \oiii\ and \lya\ EWs could also indicate a weak current burst of star formation, without significant feedback or LyC production, which might result in undetected LyC. At lower optical depths and moderate \fesc, \lya\ EW may increase due to enhanced escape, and \oiii\ EWs may likewise increase if these lower optical depths are preferentially associated with stronger starbursts. However, the \oiii\ and \lya\ EWs will decrease again in the strongest LCEs, as a general lack of nebular gas results in the limited production of nebular emission lines \citep[\eg,][]{zackrisson13, nakajima14}. Thus, EW may be an ambiguous indicator of LyC escape. 

The $z\sim3$ data also have limitations. The strong LCEs that are most severely under-predicted by the LAE model all come from the \citet{liu23} and \citet{kerutt24} samples of individual LAEs with LyC detections. Both papers note that their LyC detections represent the extreme of the population and may not be representative of the average galaxy population with these parameters, whose \fesc\ may be much lower. In Paper I, we found a similar result for the strongest LCEs in the LzLCS+, whose only distinguishing feature compared to more moderate LCEs is their low line-of-sight \hi\ content, suggestive of a favorable orientation. Another limitation of the high-redshift data is the uncertainty in the IGM attenuation. At $z\sim3$, this attenuation is significant and varies along different lines of sight \citep[\eg,][]{rudie13, inoue14, vanzella16, steidel18}. This unknown sightline dependence leads to additional uncertainty in the reported \fesc\ for individual galaxies and could cause some genuine LCEs to appear as non-detections or to have overestimated \fesc. 

The stacked samples \citep{steidel18, nakajima20, bian20} shown circled in Figure~\ref{fig:highz_a}, average over these variations and are less subject to this uncertainty. In the LAE model in Figure~\ref{fig:highz_a}b, we see that the stacked samples do indeed show less scatter than the individual high-redshift detections. However, the model does not perfectly predict \fesc\ for the stacked samples, and their scatter is still comparable to the scatter in the individual LzLCS+ galaxies. In addition to averaging over IGM attenuation variations, the stacks also average over any galaxy-to-galaxy variation in physical properties. Hence, the input variables may not represent the true set of properties of an individual system, leading to some uncertainty in the predicted \fesc. 

Unfortunately, the small sample sizes in most models and the lack of a common high-redshift sample across the models make it difficult to discern which model parameters are most important for predicting \fesc\ at high redshift. Our results suggest that if high-redshift galaxies behave like their lower redshift counterparts in the LzLCS+ sample, E(B-V)$_{\rm UV}$ or $\beta_{\rm 1550}$, O32, and \sigsfr\ or half-light radius are essential variables to include. For the existing $z\sim3$ samples, models with at least two of these parameters do successfully predict the \fesc\ of the available high-redshift galaxies to within $\sim$0.3 dex. 

This performance also shows that the LzLCS+ galaxies may indeed be reasonable analogs for high-redshift galaxies and that the same observable and physical properties may correlate with LyC escape at both low and high redshift \citep[\eg,][]{saldana23, schaerer22b, mascia23}. This agreement is not trivial, as in principle, the $z\sim0.3$ and $z\sim3$ galaxies could have different star formation histories, dust properties, morphologies, or other properties, any of which could affect \fesc. The LzLCS+ parameter space does cover the properties of the $z\sim3$ samples, with some exceptions. Seventeen $z\sim3$ targets (8 LCEs and 9 non-LCEs) are brighter than the LzLCS+ UV magnitude range by 0.1-3.18 magnitudes, including the strong LCEs Ion2 and J1316+2614. An additional two LCEs in the LAE model are fainter than the LzLCS+ by 0.5-0.7 magnitudes. The predictions for most models, all those with $M_{\rm 1500}$ as a variable, therefore extrapolate to an unobserved part of parameter space, yet still perform well. 

The data in Figures~\ref{fig:highz_a} and \ref{fig:highz_b} and metrics in Table~\ref{table:highz:metrics} include $z\sim3$ galaxies that have measured values for all of the input independent variables in the models\footnote{The reported E(B-V) for Ion2 is a stringent upper limit of E(B-V)$<0.04$. For consistency with Ion2's other reported measurements, we follow \citet{vanzella16} in adopting E(B-V)=0.0, and we choose to include Ion2 in our goodness-of-fit calculations.}. However, additional galaxies in these samples have upper or lower limits for some independent variables, which can still provide potentially useful constraints on \fesc. For these galaxies, we adopt the limit as the value of the independent variable and predict their \fesc\ using the LzLCS+ models. We plot these approximate estimates in Figures~\ref{fig:highz_a} and \ref{fig:highz_b} as light green symbols and an open symbol for Ion2, but we do not include these galaxies in the goodness-of-fit metrics. 

The galaxies with limits have one or more of the following: upper limits in $M_*$, lower limits in \sigsfr, lower limits in EW(\oiii+H$\beta$), lower limits in O32, lower limits in EW(\lya), lower limits in $M_{\rm 1500}$, and lower limits in \fesclya. One galaxy has an upper limit in EW(\oiii+H$\beta$) rather than a lower limit. Given the coefficients for these variables in the models, the mass and magnitude limits would cause the model to overestimate \fesc, while the lower limits in the nebular emission lines should generally cause the model to underestimate \fesc. For example, a galaxy with a lower limit of EW(\lya) $>100$\AA\ is an even stronger \lya\ emitter than we assume, and its predicted \fesc\ will be an underestimate. 

Most of the $z\sim3$ LCEs with input limits in the LAE-O32 and LAE-O32-nodust models only have lower limits in O32, such that their predicted \fesc\ should be an underestimate. For galaxies with lower limits in both a nebular property and $M_{\rm 1500}$, we cannot easily interpret the predicted \fesc\ as a lower or upper limit. However, the best-fit coefficients in the Appendix show that the models have a steeper dependence on the nebular lines than they do on $M_{\rm 1500}$. Figures~\ref{fig:highz_a} and \ref{fig:highz_b} show that for most of the galaxies with input limits, the measured limits still constrain \fesc\ as accurately as for the rest of the sample. However, a few galaxies (90675, 101846, 105937 from \citealt{fletcher19}) have strongly under-predicted \fesc, which suggests that their nebular lines may be much stronger than the reported limit.

Finally, we note that our model predictions do not account for systematic uncertainties, including differences in methodology. We adopt the published values for all data. However, each paper makes different assumptions, which could affect the tabulated values. For instance, papers differ in their adopted IGM transmission models, the models used in SED fits, and adopted dust attenuation laws. These assumptions could lead to systematic differences in properties like \fesc, $M_*$, and E(B-V)$_{\rm UV}$ among the different samples. In addition, LyC measurements may be photometric or spectroscopic and may cover different wavelength ranges. Despite these systematics, the models work well in predicting \fesc. Adopting consistent methodologies could potentially result in better predictions; however, as shown by the scatter in the LzLCS+ sample, which has a consistent methodology, the inherent uncertainty in the model itself also limits the possible accuracy of predictions.

\subsubsection{Summary of $z\sim3$ Results}
\label{sec:z3summary}
In conclusion, Cox models derived using the $z\sim0.3$ LzLCS+ sample can successfully predict \fesc\ in $z\sim3$ LCEs. This agreement suggests that LCEs may have similar physical properties at both low and high redshift \citep[\eg][]{saxena23}. We find that the most accurate models include E(B-V)$_{\rm UV}$ and O32 or morphology measurements as variables. However, larger high-redshift samples with a full suite of measurements are required to test this result. Future observations of $z\sim3$ LCEs with {\it JWST} will further clarify whether the relationship between \fesc\ and physical properties evolves with redshift or remains constant.  Preliminary {\it JWST} observations suggest that $z\gtrsim6$ galaxies may indeed share numerous physical properties with low-redshift analog samples like the LzLCS \citep[\eg,][]{schaerer22b, endsley23, mascia23, lin24}. The models developed on the LzLCS+ sample and tested on $z\sim3$ LCEs may therefore apply equally well at the reionization epoch.

\section{\fesc\ Predictions at $z\gtrsim6$}
\label{sec:z6}
\subsection{Model Predictions for $z\gtrsim6$ Samples}
\label{sec:z6samples}

Given the success of our models in predicting \fesc\ at $z\sim3$, we now use the Cox models to estimate \fesc\ for galaxies at $z\gtrsim6$, where LyC is not detectable because of the IGM opacity. We consider several models: the ``ELG" models, R50-$\beta$, and $\beta$-Metals. The $\beta$-Metals model is the most limited, as it only includes measurements of luminosity, mass, UV slope, and metallicity and has no information on nebular line strength or morphology. Based on the LzLCS+ galaxies, these models have an RMS scatter of 0.44-0.55 dex in \fesc\ (see Section~\ref{sec:z3} and Table~\ref{table:highz:metrics}), with the ELG-O32 or R50-$\beta$ models giving the most accurate predictions, depending on the metric considered. 

We can apply these models to several samples of reionization-era galaxies, which span $z\sim6-14$: photometric samples from \citet{endsley21}, \citet{endsley23}, \citet{bouwens23}, and \citet{morishita24} and spectroscopic samples from \citet{williams23}, \citet{schaerer22b}, \citet{tang23}, \citet{fujimoto23}, \citet{saxena23, saxena24}, \citet{mascia23}, and \citet{atek24}. The \citet{endsley21} sample is based on {\it Spitzer}, {\it HST}, and ground-based observations; all other samples incorporate {\it JWST} NIRCam photometry and several \citep{williams23, schaerer22b, tang23, fujimoto23, saxena23, saxena24, mascia23, atek24} use {\it JWST} NIRSpec spectroscopy as well. We select all galaxies at $z>5.9$ from these samples to investigate \fesc\ in the epoch of reionization.

\subsubsection{Photometric Samples}
\label{sec:z6:photometric}
We list the predicted \fesc\ values and their associated uncertainties for the $z\gtrsim6$ galaxies in Tables~\ref{table:predictendsley}-\ref{table:predictatek}. The ELG-EW and R50-$\beta$ models (Tables~\ref{table:predictendsley} and \ref{table:predictmorishita}) apply to the largest compilations of $z\gtrsim6$ galaxies, with 183 measurements corresponding to 180 unique galaxies for the ELG-EW samples and 278 measurements at $z>5.9$ for the R50-$\beta$ samples, although most of the galaxies for both models only have photometric redshifts. Most of the galaxies in the ELG-EW samples are from \citet{endsley21, endsley23}, and most of the R50-$\beta$ sample galaxies come from \citet{morishita24}. 

\begin{figure*}
\gridline{\fig{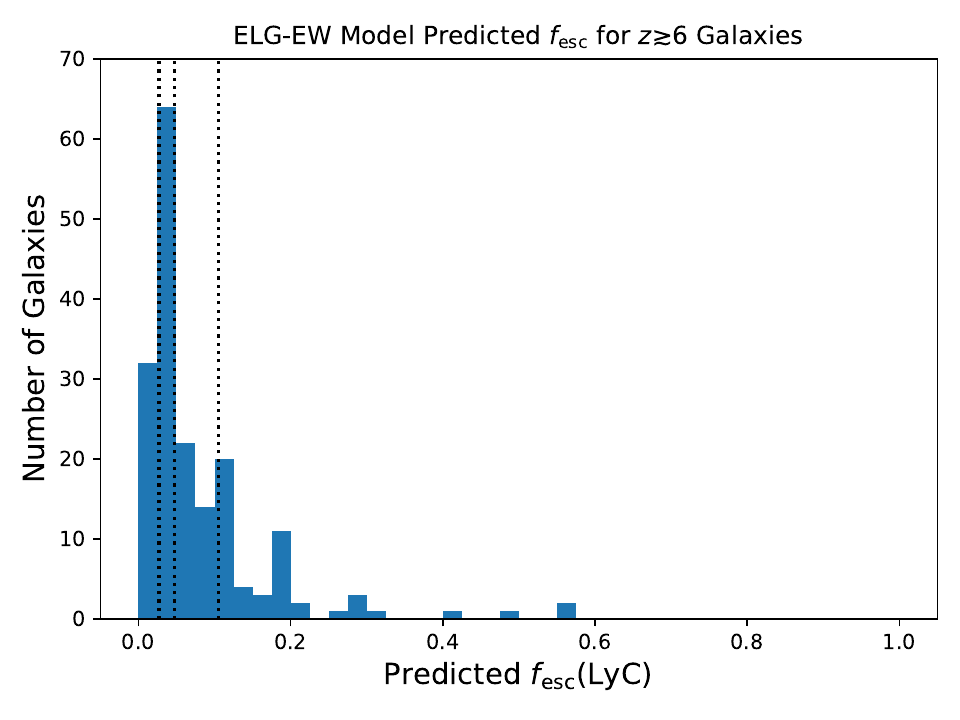}{0.5\textwidth}{(a)}
	\fig{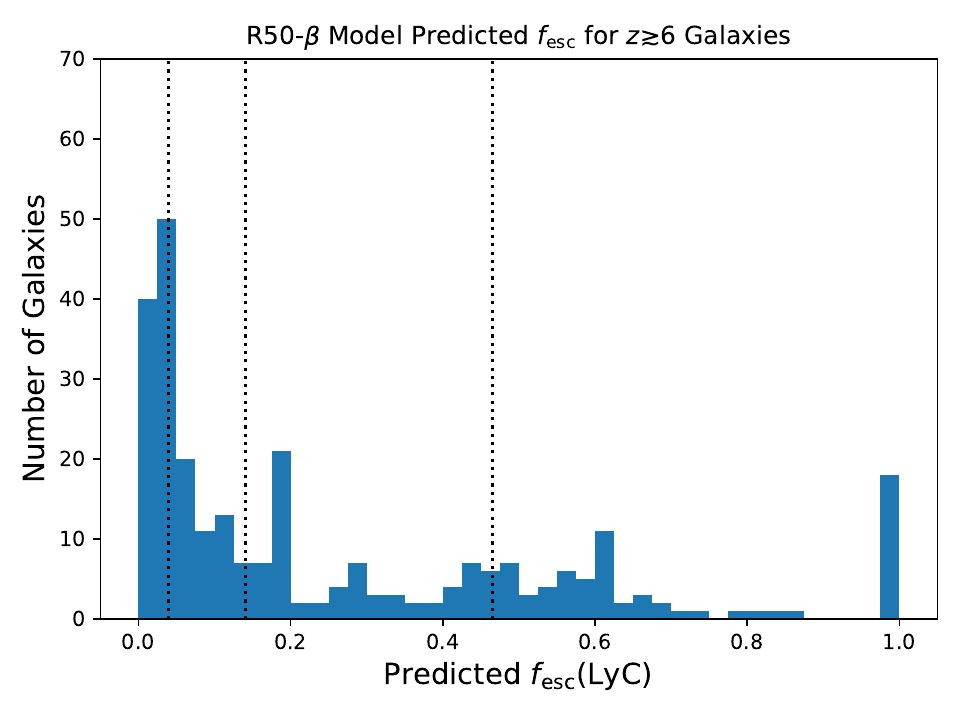}{0.5\textwidth}{(b)}}
\caption{The distribution of predicted \fesc\ from the ELG-EW model (a) for $z\gtrsim6$ galaxies from \citet{endsley21, endsley23}, \citet{bouwens23}, \citet{tang23}, \citet{fujimoto23}, and \citet{saxena23} and from the R50-$\beta$ model (b) for $z\gtrsim6$ galaxies from \citet{morishita24} and \citet{mascia23}. The vertical dotted lines show the median and quartiles of the distributions (\fesc\ $=0.027, 0.047, 0.105$ for ELG-EW and \fesc\ $=0.039, 0.141, 0.467$ for R50-$\beta$). Both models have some limitations but suggest that many, but not all, high-redshift galaxies may have \fesc\ $>0.1$.
\label{fig:endsleyhist}}
\end{figure*}

We show histograms of the ELG-EW and R50-$\beta$ model \fesc\ predictions in Figure~\ref{fig:endsleyhist}. The median of the \fesc\ predictions for the ELG-EW model samples is fairly low, \fesc\ $=0.047$, which is lower than the average \fesc\ value of 0.1-0.2 required to reionize the universe assuming a canonical ionizing photon production efficiency \logten($\xi_{\rm ion}$)$\sim25.3$\citep[\eg,][]{finkelstein15, finkelstein19, robertson15, naidu20}. Of these $z\gtrsim6$ galaxies, 27\%\ have \fesc\ $\geq 0.1$ and only 6\%\ have \fesc\ $\geq 0.2$.  However, as shown in Figure~\ref{fig:highz_b}a, the ELG-EW model is one of the less accurate models, with a greater tendency to underpredict the true \fesc\ of the LzLCS+ sample. This underprediction implies that the ELG-EW variable set, which includes EW(\oiii+H$\beta$) but not O32, does not distinguish strong LCEs from weaker LCEs. The more accurate R50-$\beta$ model (see Table~\ref{table:highz:metrics} and Figure~\ref{fig:highz_b}) implies a higher fraction of LCEs, with a median value of \fesc\ $=0.14$, and 56\%\ and 39\%\ of the galaxies with \fesc\ $>0.1$ and \fesc\ $>0.2$, respectively. Both models find a substantial fraction of weak or non-leakers, and 25\% of the galaxies in both models have \fesc\ $<0.03-0.04$. Taken at face value, our preliminary results would suggest that \fesc\ values $>0.1$ are common but not ubiquitous among moderate to bright galaxies ($M_{\rm 1500} < -18$) in the epoch of reionization (see also \citealt{mascia23, mascia24}).

\begin{deluxetable*}{lllllll}
\tablecaption{Predictions for $z\gtrsim6$ Galaxies from the ELG-EW Model}
\label{table:predictendsley}
\tablehead{
\colhead{Source ID} & \colhead{$z_{\rm phot}$} & \colhead{$z_{\rm spec}$} & \colhead{\fesc} & \colhead{$f_{\rm esc, min}$} & \colhead{$f_{\rm esc, max}$} & \colhead{Reference}}
\startdata
COS-83688 & 6.70 & --- & 0.190 & 0.056 & 0.602 & \citet{endsley21} \\
COS-87259 & 6.66 & --- & 0.024 & 0.005 & 0.119 & \citet{endsley21} \\
COS-237729 & 6.87 & ---& 0.119 & 0.033 & 0.488 & \citet{endsley21} \\
COS-312533 & 6.85 & ---& 0.043 & 0.016 & 0.265 & \citet{endsley21} \\
COS-400019 & 6.88 & ---& 0.109 & 0.031 & 0.463 & \citet{endsley21} \\
\enddata
\tablecomments{Predicted \fesc\ values from the ELG-EW model for the $z\gtrsim6$ galaxies from \citet{endsley21, endsley23}, \citet{bouwens23}, \citet{tang23}, \citet{fujimoto23}, and \citet{saxena23}. $z_{\rm phot}$ and $z_{\rm spec}$ are the photometric and spectroscopic redshifts. $f_{\rm esc, min}$ and $f_{\rm esc, max}$ represent the 15.9 and 84.1 percentiles of the model \fesc\ predictions. For galaxies with upper limits on EW(\oiii+H$\beta$), the \fesc\ predictions are also upper limits and are marked accordingly. The Reference column lists the publication used for the model input variables. The full, machine-readable version of this table is available online. We show the first five rows as an example here.} 
\end{deluxetable*}

\begin{deluxetable*}{lllllll}
\tablecaption{Predictions for $z\gtrsim6$ Galaxies from the R50-$\beta$ Model}
\label{table:predictmorishita}
\tablehead{
\colhead{Source ID} &\colhead{$z_{\rm phot}$} & \colhead{$z_{\rm spec}$} & \colhead{\fesc} & \colhead{$f_{\rm esc, min}$} & \colhead{$f_{\rm esc, max}$} & \colhead{Reference}}
\startdata
J1235-15534 & 5.9 & --- & 0.09 & 0.03 & 0.41 & \citet{morishita24} \\
JADESGDS-6734 & --- & 5.92 & 0.11 & 0.03 & 0.44 & \citet{morishita24} \\
JADESGDS-11449 & --- & 5.94  & 0.51 & 0.20 & 0.88 & \citet{morishita24} \\
JADESGDS-18169 & --- & 5.94 & 1.00 & 0.61 & 1.00 & \citet{morishita24} \\
JADESGDS-33803 & --- & 5.97 & 0.18 & 0.05 & 0.56 & \citet{morishita24} \\
\enddata
\tablecomments{Predicted \fesc\ values from the R50-$\beta$ model. $z_{\rm phot}$ and $z_{\rm spec}$ are the photometric and spectroscopic redshifts. $f_{\rm esc, min}$ and $f_{\rm esc, max}$ represent the 15.9 and 84.1 percentiles of the model \fesc\ predictions. The Reference column lists the publication used for the model input variables. The full, machine-readable version of this table is available online. We show the first five rows as an example here.} 
\end{deluxetable*}

We caution that some of the $z\gtrsim6$ galaxies in the ELG-EW sample and many in the R50-$\beta$ sample fall outside of the parameter space probed by the LzLCS+, although a similar extrapolation did not seem to adversely affect the \fesc\ predictions for the $z\sim3$ galaxies (Section~\ref{sec:z3}). Of the 183 measurements in the ELG-EW model sample, 14 galaxies have brighter UV magnitudes than the LzLCS+ galaxies, one is fainter, and 17 have higher EW(\oiii+H$\beta$). Likewise, the R50-$\beta$ sample contains 8 galaxies brighter than and 30 galaxies fainter than the LzLCS+ sample. More concerningly, 114 of the 278 galaxies in the R50-$\beta$ compilation are more compact than the LzLCS+ galaxies, which could mean that their high inferred \fesc\ values result from an incorrect extrapolation into this compact regime. 

\subsubsection{Spectroscopic Samples}
\label{sec:z6:spectroscopic}

The ELG-O32 and ELG-O32-$\beta$ model predictions (Tables~\ref{table:predictfujimoto} and \ref{table:predictsaxenab}) apply to fewer $z\gtrsim6$ galaxies, only 17 and 27 galaxies, respectively, but all of these galaxies have spectroscopic redshifts. Half the galaxies in the ELG-O32 model have predicted \fesc\ $<0.05$, although most of these are only lower limits in \fesc, and 25\% of the ELG-O32-$\beta$ sample galaxies have \fesc\ $\leq0.01$. Like the R50-$\beta$ model predictions, low \fesc\ is common, but the overall distribution also extends to very high \fesc. Both models include some galaxies with lower limits in O32, whose predicted \fesc\ values therefore also correspond to lower limits. Depending on how high the true \fesc\ for these galaxies are, 35-82\% of the ELG-O32 model galaxies and 33-41\% of the ELG-O32-$\beta$ galaxies have \fesc\ $\geq 0.2$. 

Because these galaxies are all spectroscopically confirmed, these samples could be biased toward galaxies with stronger emission lines and hence higher \fesc; the median EW(\oiii+H$\beta$) of the spectroscopic ELG-O32 sample is 1790 \AA\ \citep{fujimoto23, tang23, saxena23}, compared with 690 \AA\ for the photometric samples in the ELG-EW model, for example \citep{endsley21, endsley23, bouwens23}. In addition, these models may not underpredict \fesc\ for strong LCEs to the same extent as the ELG-EW model (see Figure~\ref{fig:highz_b} and Table~\ref{table:highz:metrics}). For the six galaxies with predicted \fesc\ $> 0.2$ in the ELG-O32 model, the ELG-EW predictions are indeed lower by $\Delta$\fesc=0.26 on average, which suggests that model differences also account for some of the higher \fesc\ values  compared to the ELG-EW distribution.

\begin{deluxetable*}{lllllll}
\tablecaption{Predictions for $z\gtrsim6$ Galaxies from the ELG-O32 Model}
\label{table:predictfujimoto}
\tablehead{
\colhead{Source ID} & \colhead{Alternate Name} & \colhead{$z_{\rm spec}$} & \colhead{\fesc} & \colhead{$f_{\rm esc, min}$} & \colhead{$f_{\rm esc, max}$} & \colhead{Reference}}
\startdata
CEERS-2 & CEERS1-3858 & 8.807 & $>$0.178 & $>$0.053 & $>$0.567 & \citet{fujimoto23} \\
CEERS-3 & CEERS1-3908 & 8.005 & $>$0.049 & $>$0.023 & $>$0.267 & \citet{fujimoto23} \\
CEERS-3 & CEERS1-3908 & 8.00 & $>$0.006 & $>$0 & $>$0.032 & \citet{tang23} \\
CEERS-7 & CEERS1-6059 & 7.993 & $>$0.119 & $>$0.038 & $>$0.453 & \citet{fujimoto23} \\
CEERS-20 & CEERS3-1748 & 7.769 & $>$0 & $>$0 & $>$0 & \citet{fujimoto23} \\
CEERS-23 & CEERS6-7603 & 8.881 & $>$0.008 & $>$0 & $>$0.041 & \citet{fujimoto23} \\
CEERS-23 & CEERS6-7603 & 8.881 & $>$0.007 & $>$0 & $>$0.038 & \citet{tang23} \\
CEERS-24 & CEERS6-7641 & 8.998 & $>$0.012 & $>$0 & $>$0.043 & \citet{fujimoto23} \\
CEERS-24 & CEERS6-7641 & 8.999 & $>$0.013 & $>$0 & $>$0.049 & \citet{tang23} \\
CEERS-44 & & 7.10 & $>$0.402 & $>$0.159 & $>$0.731 & \citet{tang23} \\
CEERS-407 & & 7.028 & 0.023 & 0.005 & 0.093 & \citet{tang23} \\
CEERS-498 & & 7.18 & $>$0.499 & $>$0.201 & $>$0.869 & \citet{tang23} \\
CEERS-499 & & 7.168 & $>$0 & $>$0 & $>$0 & \citet{tang23} \\
CEERS-698 & & 7.47 & 0.722 & 0.498 & 1 & \citet{tang23} \\
CEERS-1019 & & 8.678 & 1 & 0.644 & 1 & \citet{tang23} \\
CEERS-1025 & & 8.715 & 0.301 & 0.110 & 0.622 & \citet{tang23} \\
CEERS-1027 & & 7.819 & $>$0.650 & $>$0.453 & 1 & \citet{tang23} \\
CEERS-1038 & & 7.194 & $>$0.007 & $>$0 & $>$0.038 & \citet{tang23} \\
JADES-GS-z7-LA & & 7.278 & 0 & 0 & 0.012 & \citet{saxena23} \\
11027 & & 9.51 & 0.021 & 0.005 & 0.071 & \citet{williams23} \\
\enddata
\tablecomments{Predicted \fesc\ values from the ELG-O32 model. $z_{\rm spec}$ is the spectroscopic redshift. $f_{\rm esc, min}$ and $f_{\rm esc, max}$ represent the 15.9 and 84.1 percentiles of the model \fesc\ predictions. For galaxies with lower limits on O32, the \fesc\ predictions are also lower limits and are marked accordingly. The Reference column lists the publication used for the model input variables.} 
\end{deluxetable*}

\begin{deluxetable*}{llllll}
\tablecaption{Predictions for $z\gtrsim6$ Galaxies from the ELG-O32-$\beta$ Model}
\label{table:predictsaxenab}
\tablehead{
\colhead{Source ID} & \colhead{$z_{\rm spec}$} & \colhead{\fesc} & \colhead{$f_{\rm esc, min}$} & \colhead{$f_{\rm esc, max}$} & \colhead{Reference}}
\startdata
21842 & 7.982 & 0.09 & 0.03 & 0.41 & \citet{saxena24} \\
10013682 & 7.276 & 0.009 & 0.00 & 0.04 & \citet{saxena24} \\
16625 & 6.631 & 0.37 & 0.12 & 0.68 & \citet{saxena24} \\
18846 & 6.336 & 0.60 & 0.37 & 1.00 & \citet{saxena24} \\
19342 & 5.974 & 0.47 & 0.18 & 0.82 & \citet{saxena24} \\
9422 & 5.937 & 0.85 & 0.59 & 1.00 & \citet{saxena24} \\
6002 & 5.937 & 0.11 & 0.03 & 0.44 & \citet{saxena24} \\
12637 & 7.66 & 0.21 & 0.07 & 0.60 & \citet{saxena24} \\
15362 & 6.794 & 0.00 & 0.00 & 0.03 & \citet{saxena24} \\
13607 & 6.622 & 0.00 & 0.00 & 0.00 & \citet{saxena24} \\
14123 & 6.327 & 0.07 & 0.03 & 0.34 & \citet{saxena24} \\
58850 & 6.263 & 0.35 & 0.12 & 0.66 & \citet{saxena24} \\
17138 & 6.204 & 0.00 & 0.00 & 0.02 & \citet{saxena24} \\
9365 & 5.917 & 0.28 & 0.09 & 0.61 & \citet{saxena24} \\
11027 & 9.51 & 0.01 & 0.00 & 0.05 & \citet{williams23} \\
4590 & 8.495 & 0.01 & 0.00 & 0.05 & \citet{schaerer22b} \\
6355 & 7.664 & 0.06 & 0.03 & 0.35 & \citet{schaerer22b} \\
10612 & 7.66 & 0.17 & 0.05 & 0.53 & \citet{schaerer22b} \\
JADES-GS-z7-LA & 7.278 & 0.00 & 0.00 & 0.02 & \citet{saxena23} \\
10025 & 7.875 & 0.01 & 0.00 & 0.06 & \citet{mascia23} \\
100004 & 7.884 & $>$0.01 & $>$0.00 & $>$0.05 & \citet{mascia23} \\
10000 & 7.884 & 0.05 & 0.02 & 0.26 & \citet{mascia23} \\
10021 & 7.288 & 0.42 & 0.17 & 0.76 & \citet{mascia23} \\
100001 & 7.875 & 0.006 & 0.00 & 0.04 & \citet{mascia23} \\
100003 & 7.88 & 0.47 & 0.18 & 0.82 & \citet{mascia23} \\
100005 & 7.883 & 0.02 & 0.005 & 0.09 & \citet{mascia23} \\
150008 & 6.23 & $>$0.12 & $>$0.04 & $>$0.47 & \citet{mascia23} \\
\enddata
\tablecomments{Predicted \fesc\ values from the ELG-O32-$\beta$ model. $z_{\rm spec}$ is the spectroscopic redshift. $f_{\rm esc, min}$ and $f_{\rm esc, max}$ represent the 15.9 and 84.1 percentiles of the model \fesc\ predictions. For galaxies with lower limits on O32, the \fesc\ predictions are also lower limits and are marked accordingly. The Reference column lists the publication used for the model input variables.} 
\end{deluxetable*}

While we have generated initial predictions for $z\gtrsim6$ galaxies using our models, predictions for the largest samples are limited by a lack of spectroscopic information. We do find evidence of high \fesc\ $>0.2$ among smaller spectroscopic samples at $z\gtrsim6$, but these samples may be biased toward stronger emission-line galaxies. To obtain more accurate predictions, high-redshift studies should prioritize observations of nebular line ratios and galaxy sizes for larger samples. As {\it JWST} surveys continue, larger, more representative spectroscopic samples will improve estimates of the average \fesc\ at $z>6$. 

\subsection{Trends with UV Magnitude}
\label{sec:magtrends}

\begin{figure*}
\gridline{\fig{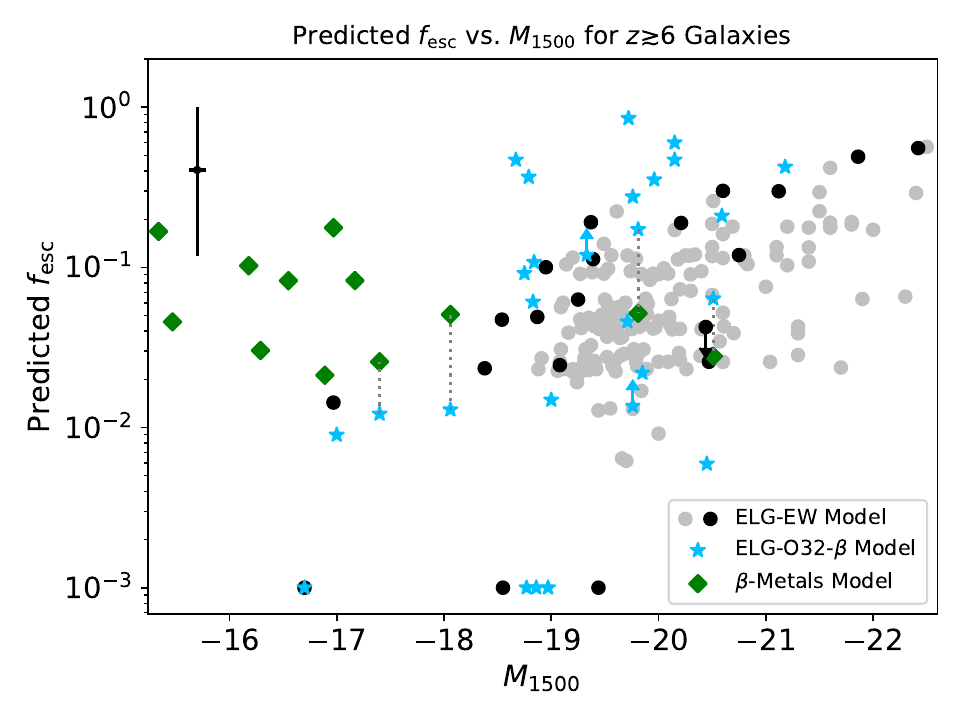}{0.5\textwidth}{(a)}
	\fig{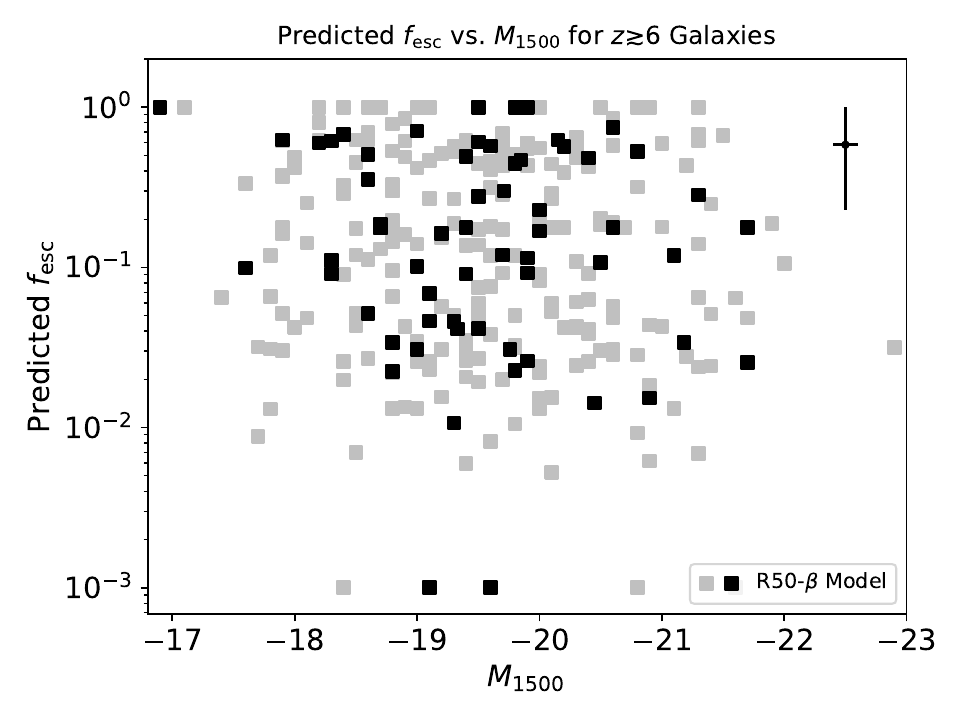}{0.5\textwidth}{(b)}}
\caption{Predicted \fesc\ as a function of $M_{\rm 1500}$ for $z\gtrsim6$ galaxies. Panel a shows the predicted \fesc\ from the ELG-EW model (circles) for galaxies with photometric redshifts (gray) and spectroscopic redshifts (black), from the ELG-O32-$\beta$ model (blue stars), and from the $\beta$-Metals model (green diamonds). Panel b shows the predicted \fesc\ from the R50-$\beta$ model (squares) with photometric (gray) and spectroscopic (black) redshifts. The crosses in the upper corners show the median uncertainties for each panel. Dark gray dotted lines connect the predictions for the same galaxy in different models. We see no strong trend between \fesc\ and $M_{\rm 1500}$, but many of the predictions suffer from a limited set of variables (ELG-EW and $\beta$-Metals models) or require extrapolation outside the LzLCS+ parameter space (R50-$\beta$ model).
\label{fig:endsleymag}}
\end{figure*}

We examine the predicted \fesc\ values as a function of magnitude in Figure~\ref{fig:endsleymag}. We plot predictions from the three models that apply to the largest sample sizes (ELG-EW, ELG-O32-$\beta$, and R50-$\beta$) and from the model that applies to the faintest galaxies ($\beta$-Metals; Table~\ref{table:predictatek}). We see no strong trend with magnitude, although the few brightest sources with $M_{\rm 1500} \lesssim -21.5$ do appear to have higher \fesc\ on average, with a median \fesc\ $=0.18$ vs.\ 0.07 for fainter sources. Aside from these rare sources, the ELG-EW, ELG-O32-$\beta$, and $\beta$-Metals samples tend to show \fesc\ within $\sim0.02-0.3$ across a wide range of magnitudes, from -16 to -22. The R50-$\beta$ model (Figure~\ref{fig:endsleymag}b) also shows a flat distribution of \fesc\ with magnitude, but it predicts many more galaxies at higher \fesc. However, as previously noted, 41\%\ of the galaxies in this model require extrapolating predictions to galaxies with extremely compact morphologies, which makes these predictions uncertain.

\begin{deluxetable*}{llllll}
\tablecaption{Predictions for $z\gtrsim6$ Galaxies from the $\beta$-Metals Model}
\label{table:predictatek}
\tablehead{
\colhead{Source ID} & \colhead{$z_{\rm spec}$} & \colhead{\fesc} & \colhead{$f_{\rm esc, min}$} & \colhead{$f_{\rm esc, max}$} & \colhead{Reference}}
\startdata
18924 & 7.7 & 0.05 & 0.02 & 0.30 & \citet{atek24} \\
16155 & 6.87 & 0.03 & 0.006 & 0.18 & \citet{atek24} \\
23920 & 6.00 & 0.10 & 0.03 & 0.46 & \citet{atek24} \\
12899 & 6.88 & 0.17 & 0.04 & 0.55 & \citet{atek24} \\
8613 & 6.38 & 0.18 & 0.04 & 0.57 & \citet{atek24} \\
23619 & 6.72 & 0.08 & 0.03 & 0.42 & \citet{atek24} \\
38335 & 6.23 & 0.02 & 0.00 & 0.11 & \citet{atek24} \\
27335 & 6.76 & 0.08 & 0.03 & 0.42 & \citet{atek24} \\
11027 & 9.51 & 0.03 & 0.005 & 0.15 & \citet{williams23} \\
4590 & 8.495 & 0.05 & 0.02 & 0.35 & \citet{schaerer22b} \\
6355 & 7.664 & 0.03 & 0.006 & 0.18 & \citet{schaerer22b} \\
10612 & 7.66 & 0.05 & 0.02 & 0.36 & \citet{schaerer22b} \\
\enddata
\tablecomments{Predicted \fesc\ values from the $\beta$-Metals model. $z_{\rm spec}$ is the spectroscopic redshift. $f_{\rm esc, min}$ and $f_{\rm esc, max}$ represent the 15.9 and 84.1 percentiles of the model \fesc\ predictions. The Reference column lists the publication used for the model input variables.} 
\end{deluxetable*}

Any trends with magnitude or lack thereof in the other models should also be taken with caution. All the models have a tendency to under-predict \fesc, but this tendency is most pronounced in the ELG-EW and $\beta$-Metals models (Figure~\ref{fig:highz_b}), which trace the extremes of the magnitude range in Figure~\ref{fig:endsleymag}. The more reliable ELG-O32-$\beta$ model, which better reproduces the LzLCS+ \fesc\ values, suggests that some of the galaxies in the $M_{\rm 1500} = -19$ to $-21$ magnitude range may have significantly high \fesc. The ELG-O32-$\beta$ sample does not include enough faint galaxies to reveal whether or not a trend between \fesc\ and magnitude exists, however. 

At the faintest magnitudes, the \citet{atek24} sample, included in the $\beta$-Metals model includes several extremely low-luminosity lensed galaxies observed as part of the {\it JWST} UNCOVER survey \citep{bezanson22, weaver24}. For these galaxies, the $\beta$-Metals model therefore extrapolates the LzLCS+ sample to an unobserved parameter space. The only statistically significant predictor variable in the $\beta$-Metals model is $\beta_{\rm 1550}$, and Figure~\ref{fig:endsleymag} illustrates that these galaxies' blue slopes suggest at least moderate \fesc\ (0.02-0.18; Table~\ref{table:predictatek}). Crucially, even these moderate \fesc\ values, which may be under-predicted, are still greater than what is required for such faint galaxies to dominate reionization given their high LyC photon production rates \citep{atek24}.

While moderate or high \fesc\ for faint, blue galaxies matches past predictions \citep[\eg,][]{chisholm22}, the high \fesc\ in the brighter sources in Figure~\ref{fig:endsleymag} may seem surprising at first glance. In all four models, the fitted coefficients (see the Appendix) show that a brighter $M_{\rm 1500}$ results in a higher \fesc\ {\it at fixed values of the other inputs (fixed E(B-V), fixed O32, etc).} This effect does not come from the association of higher observed UV luminosity with lower dust content. The data do not show such a correlation, and, because the models already include dust measurements, the model implies that brighter UV luminosities increase \fesc\ at {\it fixed} dust attenuation. In addition, we see a similar relationship with $M_*$, where higher $M_*$ is also associated with higher \fesc\ at fixed values of the other parameters. \citet{lin24} find a similar result using a different technique. Using the LzLCS+ data, they fit a logistic regression model for the probability of having LyC escape as a function of $M_{\rm 1500}$, $\beta_{\rm 1550}$, and O32 and likewise find a brighter $M_{\rm 1500}$ increases the probability of LyC escape for fixed values of the other variables. This scaling does not imply that higher UV luminosities correlate with \fesc\ in the LzLCS+ sample; if anything, the LzLCS+ indicates a slight trend in the opposite direction \citep{flury22b}. Instead, this scaling implies that a bright galaxy with LCE-like properties, such as high O32 and blue UV slope, is a more extreme object than a faint galaxy with identical properties. For instance, at fixed O32, the LzLCS+ galaxies with higher luminosities tend to have higher \fesc, but high O32 values are also uncommon among the bright galaxies as a whole, leading to an overall trend of decreasing \fesc\ with luminosity \citep[\eg,][]{flury22b}. 

The enhanced \fesc\ for bright LCE candidates affects the predictions in Figure~\ref{fig:endsleymag} by boosting the \fesc\ of a bright, dust-poor emission-line galaxy or compact galaxy relative to an otherwise similar faint galaxy. For instance, for the ELG-EW model, the brightest galaxies should have higher predicted \fesc\ only if they have equally low E(B-V) and equally high EW as the fainter galaxies. Indeed, the ELG-EW samples, mostly galaxies from \citet{endsley21, endsley23}, show no trend between E(B-V)$_{\rm UV}$ and $M_{\rm 1500}$ or between EW(\oiii+H$\beta$) and $M_{\rm 1500}$; with comparable dust and emission line strengths, the brighter galaxies therefore end up with higher predicted \fesc. \citet{endsley23} note that lower metallicities or higher \fesc\ among the fainter galaxies could suppress their EWs and account for the lack of an observed trend with $M_{\rm 1500}$. Hence, other properties, such as O32, might be necessary to distinguish faint galaxies with high \fesc. The EWs in the \citet{endsley21, endsley23} samples also come from photometry, which may be more uncertain than spectroscopic measurements \citep[\eg,][]{duan24}. 

The ELG-O32-$\beta$ samples have O32 measurements, but they similarly show no strong trend between $\beta_{\rm 1550}$ and $M_{\rm 1500}$ or between O32 and $M_{\rm 1500}$. Likewise, the \citet{morishita24} sample, which constitutes the bulk of the R50-$\beta$ model sample, exhibits a relatively flat trend between radius and $M_{\rm 1500}$.  Only the $\beta$-Metals samples, mostly from \citet{atek24}, show any trend between physical properties and $M_{\rm 1500}$, with the lowest metallicities and bluest slopes appearing among the faintest galaxies. 

The limitations of the models and the available $z\gtrsim6$ datasets make it difficult to discern any trends between $M_{\rm 1500}$ and \fesc. Limited measurements for galaxies at the brightest and faintest magnitudes restrict us to some of the less accurate Cox models. The models also require extrapolating to make predictions for galaxies outside the magnitude or radius range of the LzLCS+ sample. Better constraints on \fesc\ within the $z>6$ population will require larger spectroscopic samples across a wide range of galaxy magnitudes as well as better observational constraints on \fesc\ in a more diverse set of galaxies at lower redshifts.

\subsection{Notes on Individual Sources}
\label{sec:individ}

\subsubsection{Strong LCES}
\label{sec:strong}
Several galaxies in the ELG-O32 and ELG-O32-$\beta$ models appear to be extremely strong LCEs (Tables~\ref{table:predictfujimoto} and \ref{table:predictsaxenab}). One unusual galaxy, CEERS-1019, has a predicted \fesc\ $= 1$. Such a high \fesc\ would seemingly conflict with the presence of strong emission lines, since it should correspond to a complete absence of absorbing gas. However, given the uncertainty in the model predictions, the \fesc\ of CEERS-1019 could be as low as 0.6. Two other galaxies in the ELG-O32 model sample, CEERS-698 and CEERS-1027, have \fesc\ $\geq0.5$, and \fesc\ as high as 1 is consistent with their uncertainties. Interestingly, all three of these strong LCE candidates are strong \lya\ emitters, which suggests that they reside within ionized bubbles \citep{tang23}. Similarly, two galaxies in the ELG-O32-$\beta$ samples have predicted \fesc\ $>0.5$: 18846 and 9422 from JADES \citep{saxena24}. With O32$=70.6$, the ELG-O32-$\beta$ model infers \fesc\ $=0.85$ for 9422, much higher than the \fesc\ $=0.01$ value derived by \citet{saxena24} using the multivariate prediction method of \citet{choustikov24}; we discuss the differences between our predictions and those of other models in \S\ref{sec:compare}.

The true \fesc\ of some of these extreme galaxies may not be quite as high as the models predict, however. Because two of these strong LCEs, CEERS-1019 (\fesc\ $=1$) and CEERS-698 (\fesc\ $\sim0.7$), have brighter UV magnitudes than the LzLCS+ sample by 0.4-0.9, their high predicted \fesc\ could be unreliable. However, these same models successfully predicted the high \fesc\ of the $z\sim3$ galaxies Ion2 and J1316+2614 (Figure~\ref{fig:highz_b}), which are as bright as these $z>6$ galaxies. The LzLCS+ sample itself also contains only three LCEs with \fesc\ $>0.5$, which limits the models' ability to accurately determine predictive relationships in the high \fesc\ regime \citep{mascia23}. CEERS-1019 may differ from the LzLCS+ galaxies for additional reasons; it has unusually strong nitrogen emission and is a possible candidate for supermassive star formation \citep{marqueschaves24} or an AGN \citep{larson23}. The physical conditions occurring in extreme galaxies like CEERS-1019 likely are not present within the LzLCS+ sample. Empirical predictions based on $z=0.3$ galaxies would not be suitable for such objects. 

The predictions for the other strong LCE candidates are likely more reliable, provided their star-formation and nebular conditions are not significantly different from the LCEs at $z\sim3$. Galaxy 9422 has a higher O32 ratio than the LzLCS+, but the other strong leaker candidates fall within the LzLCS+ parameter space, so that the model is not extrapolating to an unobserved regime. Hence, our models suggest that several strong emission-line galaxies identified in the epoch of reionization may be strong LCEs with \fesc\ well above 0.2.

\subsubsection{JADES-GS-z7-LA}
\label{sec:jades}
\citet{saxena23} report the detection of a $z=7.3$ galaxy with strong \lya\ emission, JADES-GS-z7-LA, which is presumably located within an ionized bubble. With \fesclya$ = 0.96$, IGM absorption seems to have had little effect on the galaxy's \lya\ emission. We therefore choose to incorporate the measured EW(\lya) and \fesclya\ in our model predictions. We can apply most of the models in Table~\ref{table:highz:models} to JADES-GS-z7-LA, omitting only the TopThree, R50-$\beta$, and $\beta$-Metals models due to a lack of reported \sigsfr, radius, and 12+\logten(O/H). We compare the different \fesc\ predictions in Table~\ref{table:jades}. JADES-GS-z7-LA does fall outside of the LzLCS+ parameter space, which may limit the accuracy of the \fesc\ predictions; it is fainter than the LzLCS+ galaxies by 1.6 mag, and it has a higher EW(\lya) by 144.1 \AA.

The different Cox models disagree regarding the \fesc\ of JADES-GS-z7-LA, with predicted \fesc\ ranging from 0 to 0.40. The only nonzero predictions come from models that have information about UV dust attenuation and \lya, but even the extreme \lya\ properties of JADES-GS-z7-LA do not guarantee \fesc\ $>0.1$. The LAE-O32 and ELG-O32-$\beta$-\lya\ models include O32 and stellar mass as well as \lya\ and find \fesc\ $=0.017-0.085$. These two models are also among the best-performing models for the LzLCS+ sample (Table~\ref{table:highz:metrics}). JADES-GS-z7-LA is not devoid of dust, with E(B-V)$_{\rm UV}=0.10$, and its UV slope $\beta=-2.1$ is not extreme. These properties are consistent with the properties of $z\sim0.3$ galaxies with moderate \fesc; all the LzLCS+ galaxies with higher \fesc\ $ > 0.1$ have lower E(B-V)$_{\rm UV} < 0.1$, and a slope of $\beta_{\rm 1550}=-2.1$ matches the median value for the moderate LCEs with \fesc\ $=0.05-0.1$. 

Furthermore, as discussed above, given JADES-GS-z7-LA's low mass (\logten($M_*/$\Msol)=$7.15$), its O32 value of 8.8 may not be extreme. Even its strong \lya\ emission does not necessarily imply a low optical depth along the line of sight, since \lya\ photons can scatter. For instance, the non-leaking $z\sim0.3$ galaxy J1248+4259 \citep{izotov18b} has \fesc\ $\leq 0.013$ and EW(\lya)$=256$ \AA. One piece of evidence that favors a high \fesc\ for JADES-GS-z7-LA is the small offset between the \lya\ velocity and the systemic redshift \citep{saxena23}. With the low resolution of most of the LzLCS+ UV spectra, we cannot currently include this parameter in our models, and it may boost the predicted \fesc\ for JADES-GS-z7-LA. However, this low velocity offset could still be consistent with little to no LyC escape; its offset of $120\pm80$ \kmps\ also resembles the \lya\ profile of the non-leaker J1248+4259, whose red peak is offset from the systemic velocity by $<200$ \kmps. 

In agreement with the assessment of \citet{saxena23}, we thus find that the evidence for LyC escape in JADES-GS-z7-LA is ambiguous. The multivariate models that incorporate the most information suggest that it likely has a moderate \fesc\ $\sim0.017-0.085$ and may therefore not be solely responsible for producing its ionized bubble in the IGM \citep[\eg,][]{witstok24}.

\subsubsection{Galaxies with High O32}
\label{sec:higho32}
The case of JADES-GS-z7-LA highlights the fact that the Cox multivariate predictions can differ from single-variable estimates (e.g., based on \lya\ alone). Galaxies with high O32 provide another example of this result. For instance, \citet{williams23} present {\it JWST} observations of a lensed low-mass (\logten($M_*$/\Msol) = 7.7) galaxy, 11027, at $z=9.51$ and hypothesize that the galaxy has a high \fesc\ $> 0.1$ based on its high O32 ratio of 12 (\oiii$\lambda\lambda$5007+4959/\oii=16). However \citet{lin24} suggest that 11027 is not likely to be a strong LCE considering the combination of O32, UV magnitude, and UV slope. Our ELG-O32 and ELG-O32-$\beta$ Cox models (Tables~\ref{table:predictfujimoto}-\ref{table:predictsaxenab}) likewise predict \fesc\ of only 0.012-0.021. O32 alone is not sufficient to constrain \fesc, and high O32 values are common among the lowest mass galaxies in the LzLCS+ sample \citep{flury22b}, including weak LyC emitters. In contrast, the ELG-O32-$\beta$ model predicts that several galaxies with lower O32 than galaxy 11027 (e.g., 6355, 10000, 10612, 12637 with O32=6.3-10.6, \citealt{schaerer22b,mascia23,saxena24}) are actually more likely to be LCEs because of their higher luminosities and/or bluer UV slopes (Table~\ref{table:predictsaxenab}). The Cox predictions emphasize the fact that not all galaxies with high O32 have high \fesc\ \citep[\eg,][]{izotov18b, flury22b} and additional properties such as mass, UV luminosity, and dust extinction are important to consider. 

\begin{deluxetable*}{llll}
\tablecaption{\fesc\ Predictions for JADES-GS-z7-LA}
\label{table:jades}
\tablehead{
\colhead{Model} & \colhead{\fesc} & \colhead{$f_{\rm esc, min}$} & \colhead{$f_{\rm esc, max}$}}
\startdata
LAE & 0.400 & 0.142 & 0.736 \\ 
LAE-O32 & 0.085 & 0.028 & 0.325 \\ 
LAE-O32-nodust & 0 & 0 & 0.028 \\ 
ELG-EW & 0 & 0 & 0.026\\ 
ELG-O32 & 0 & 0 & 0.012 \\ 
ELG-O32-$\beta$ & 0 & 0 & 0.022 \\
ELG-O32-$\beta$-\lya\ & 0.017 & 0 & 0.059 \\ 
\enddata
\tablecomments{Predicted \fesc\ values for JADES-GS-z7-LA from different models. $f_{\rm esc, min}$ and $f_{\rm esc, max}$ represent the 15.9 and 84.1 percentiles of the model \fesc\ predictions.} 
\end{deluxetable*}

Multivariate models are an important tool for predicting \fesc\ in the epoch of reionization. Our results highlight the fact that multivariate models can give substantially different predictions from single variable estimates, and \fesc\ predictions should therefore incorporate as much information as possible.

\section{Comparison with Other Studies}
\label{sec:compare}
\subsection{Single-Variable Predictions}
\label{sec:compare:single}
\begin{figure*}
\plotone{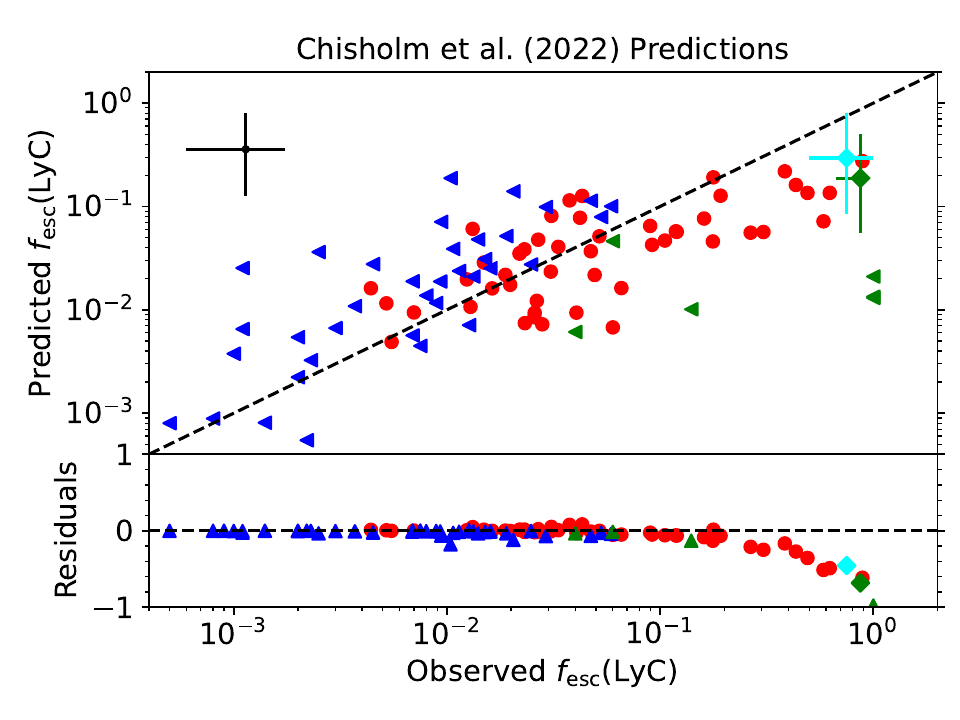}
\caption{The \fesc\ predictions for the LzLCS+ and $z\sim3$ galaxies from the \citet{chisholm22} model, which uses a single predictor variable: $\beta_{\rm 1550}$. Symbols are the same as in Figure~\ref{fig:highz_a}. The model reproduces the \fesc\ of LCEs but over-predicts the \fesc\ for the LzLCS+ non-detections and under-predicts \fesc\ for the strongest LCEs.
\label{fig:chisholmz3}}
\end{figure*}

Among the observable properties measured for the LzLCS+ sample, the UV slope $\beta_{\rm 1550}$ shows one of the strongest correlations with \fesc. Consequently, \citet{chisholm22} propose that $\beta_{\rm 1550}$ could predict \fesc\ for high-redshift galaxies and derive a relationship between $\beta_{\rm 1550}$ and \fesc\ based on the LzLCS+ dataset. In Figure~\ref{fig:chisholmz3}, we compare the predicted \fesc\ from this method with the observed \fesc\ for the LzLCS+ and for $z\sim3$ galaxies \citep{vanzella16, bassett19, marqueschaves22}. We list the goodness-of-fit statistics for the LzLCS+ with the \citet{chisholm22} model in Table~\ref{table:literature}. The \citet{chisholm22} model's high $R^2=0.45$ and low RMS$=0.43$ are comparable to the best-performing Cox models in Tables~\ref{table:highz:metrics} and \ref{table:literature}, which demonstrates that $\beta_{\rm 1550}$ alone can indeed predict \fesc\ for LCEs reasonably well. Consistent with this result, Paper I finds that, of the variables accessible at high redshift, $\beta_{\rm 1550}$ is the most important variable to include in the Cox models.

Despite its success for LCEs, however, the \citet{chisholm22} model has more difficulty in constraining \fesc\ in the LzLCS+ non-detections, as indicated by its lower concordance $C=0.76$. This concordance is lower than all the Cox models in Tables~\ref{table:highz:metrics} and \ref{table:literature}, which have $C=0.77-0.83$. The difference between the predictions for detections and non-detections is apparent in Figure~\ref{fig:chisholmz3}. For LzLCS+ galaxies with \fesc\ detections or upper limits between $0.01-0.1$, the one-to-one relation between the \citet{chisholm22} model predictions and the observations runs right between the LzLCS+ detections. However, nearly all the non-detections are above this line, indicating that the model is systematically over-predicting their \fesc. In contrast, for one of the better performing Cox models, such as the ELG-O32 model in Figure~\ref{fig:highz_b}b, both detections and non-detections in this same \fesc\ range fall on either side of the one-to-one relation, indicating that predictions for detections and non-detections are comparable. 

Figure~\ref{fig:chisholmz3} also applies the \citet{chisholm22} model to two $z\sim3$ LCEs: Ion2 and J1316+2614. The model correctly identifies both galaxies as LCEs but under-predicts their \fesc, although Ion2's observed \fesc\ is consistent with the stated uncertainties in the \citet{chisholm22} predictions. Comparing the Ion2 predictions with those from the Cox models discussed in \S\ref{sec:z3}, we find that the \citet{chisholm22} single-variable $\beta_{\rm 1550}$ model is one of the least accurate at predicting Ion2's \fesc, with a predicted \fesc\ $=0.29$, corresponding to RMS=$0.40$ dex. As discussed in \citet{chisholm22}, $\beta_{\rm 1550}$ only tracks the loss of LyC photons due to dust. Scatter below and above the \citet{chisholm22} $\beta_{\rm 1550}$-\fesc\ relation results from variations in the absorbing \hi\ column. Other properties, such as O32, \sigsfr, and UV magnitude, may better track the \hi\ component of LyC absorption for galaxies like Ion2.

\begin{figure*}
\gridline{\fig{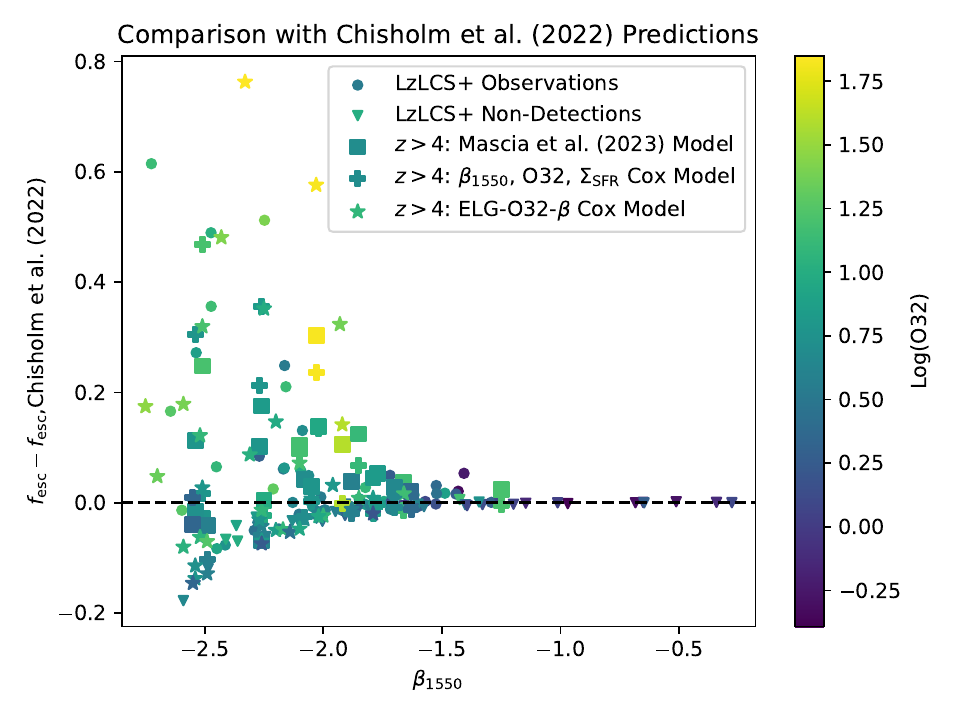}{0.5\textwidth}{(a)}
	\fig{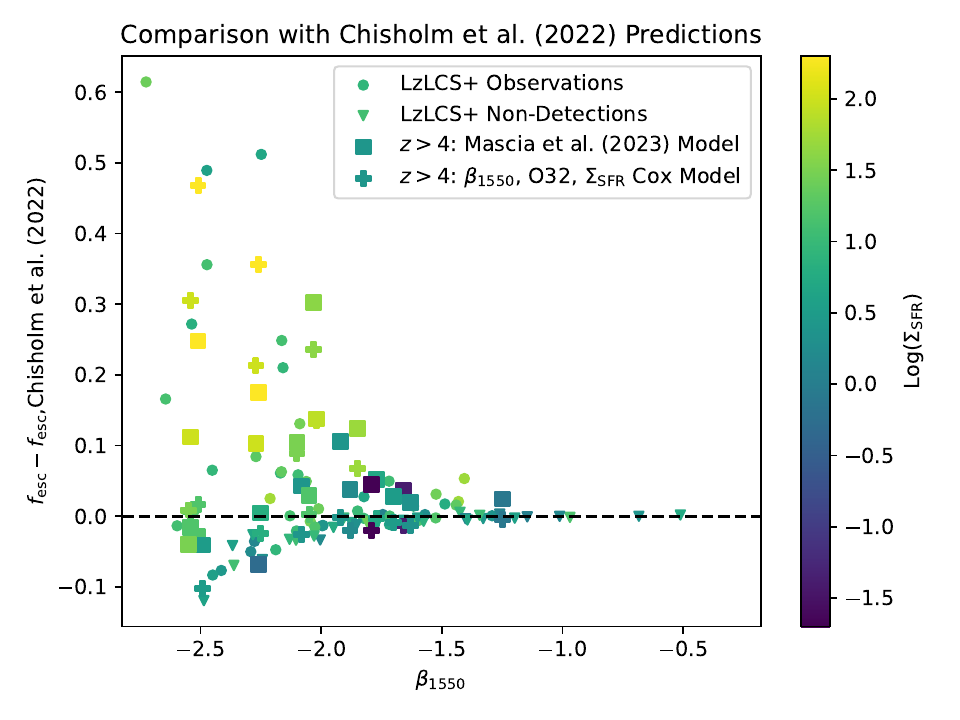}{0.5\textwidth}{(b)}
	}
\caption{Circles (detections) and triangles (upper limits) show the difference between LzLCS+ observations and \citet{chisholm22} predicted \fesc\ as a function of observed $\beta_{\rm 1550}$. Other symbols show the difference between multivariate model predictions for $z>4$ samples and the \citet{chisholm22} single-variable predictions as a function of $\beta_{\rm 1550}$. Squares compare the \fesc\ predictions from \citet{mascia23}, using $\beta_{\rm 1550}$, O32, and $r_{\rm 50,NUV}$. Crosses show \fesc\ predictions for the same $z>4$ \citet{mascia23} sample for a Cox model using $\beta_{\rm 1550}$, O32, and \sigsfr. Stars show the ELG-O32-$\beta$ Cox Model predictions for $z>4$ galaxies from \citet{williams23}, \citet{schaerer22b}, \citet{mascia23}, and \citet{saxena24}, where the input variables are $\beta_{\rm 1550}$, O32, and $M_{\rm 1500}$. We color-code galaxies by their observed O32 ratio in panel (a) and by \sigsfr\ in panel (b). At blue UV slopes, the LzLCS+ observations and high-redshift model predictions can differ significantly from the \citet{chisholm22} model predictions, depending on the galaxies' other properties, such as O32 and \sigsfr.
\label{fig:chisholm_predict}}
\end{figure*}

Figure~\ref{fig:chisholm_predict} illustrates how including these other variables affects predictions at $z>4$. We plot the difference between the observed LzLCS+ \fesc\ and the \citet{chisholm22} predictions as a function of $\beta_{\rm 1550}$.  Galaxies are color-coded by their O32 ratio (Figure~\ref{fig:chisholm_predict}a) or \sigsfr\ (Figure~\ref{fig:chisholm_predict}b). For the reddest $\beta_{\rm 1550}$, all the model predictions and observations agree that galaxies have little to no LyC escape. However, at blue UV slopes, the disagreement increases. The \citet{chisholm22} model assigns all galaxies a single value of \fesc, whereas the LzLCS+ observations show a range of \fesc\ and deviate from the predictions by as much as $\Delta$\fesc\ $=0.2$ lower or $\Delta$\fesc\ $=0.6$ higher at the bluest slopes. For the LzLCS+ galaxies, this difference is associated with other differences in the galaxies' properties; galaxies with higher O32 or higher \sigsfr\ tend to be offset to higher \fesc, and galaxies with lower \fesc\ than the \citet{chisholm22} predictions tend to have lower O32 or lower \sigsfr. 

The multivariate model predictions reflect these trends with other properties and assign galaxies a higher or lower \fesc\ at fixed $\beta_{\rm 1550}$ accordingly. In Figure~\ref{fig:chisholm_predict}, we also plot the difference between the predicted \fesc\ for $z>4$ galaxies from multivariate models vs.\ the single-variable \citet{chisholm22} model. All these multivariate models use $\beta_{\rm 1550}$ in addition to two or more other variables. The \citet{mascia23} model (squares) predicts \fesc\ using a linear fit to $\beta_{\rm 1550}$, \logten(O32) and \logten($r_{\rm 50,NUV}$) for their sample of $z>4$ JWST-GLASS galaxies. We show predictions from two Cox models in crosses and stars: a Cox model using the three top-ranked, high-redshift accessible variables from Paper I ($\beta_{\rm 1550}$, O32, and \sigsfr) for the \citet{mascia23} sample and the ELG-O32-$\beta$ Model (\S\S\ref{sec:z3}-\ref{sec:z6}), which uses $\beta_{\rm 1550}$, O32, and $M_{\rm 1500}$ as inputs to predict \fesc\ for $z>4$ galaxies from \citet{williams23}, \citet{schaerer22b}, \citet{mascia23}, and \citet{saxena24}. Like the LzLCS+ observations, the multivariate $z>4$ predictions can significantly deviate from the \citet{chisholm22} predictions, especially at blue UV slopes. High-redshift surveys relying only on $\beta_{\rm 1550}$ could therefore miss some of the physical differences among blue galaxies that might indicate variations in \fesc. Properties such as O32 and \sigsfr\ are sensitive to feedback in galaxies and may complement $\beta_{\rm 1550}$ measurements by tracing a galaxy's ability to carve low column density \hi\ channels. 

\begin{deluxetable*}{lllll}
\tablecaption{Goodness-of-Fit Statistics for Literature vs.\ Cox Models}
\label{table:literature}
\tablehead{
\colhead{Model} & \colhead{$R^2$} & \colhead{$R^2_{\rm adj}$} & \colhead{RMS} & \colhead{$C$}}
\startdata
\citet{chisholm22} & 0.45 & 0.43 & 0.43 & 0.76 \\
\citet{choustikov24} & -1.10 & -1.40 & 0.83 & 0.57 \\ 
\citet{mascia23} & 0.57 & 0.54 & 0.38 & 0.81 \\ 
JWST Cox Model (Paper I) & 0.29 & 0.14 & 0.47 & 0.83 \\ 
Cox Model with $\beta_{\rm 1550}$, \sigsfr, O32 & 0.34 & 0.29 & 0.46 & 0.83\\
Cox Model with \citet{choustikov24} Variables & 0.28 & 0.16 & 0.48 & 0.82\\ 
Cox Model with \citet{mascia23} Variables & 0.40 & 0.35 & 0.44 & 0.82 \\ 
\enddata
\tablecomments{A comparison of the goodness-of-fit statistics for \fesc\ predictions for the LzLCS+ sample from the \citet{chisholm22}, \citet{choustikov24}, and \citet{mascia23} literature models and various Cox models. We list statistics for the JWST Model from Paper I and a model limited to its three top-ranked variables. We also list statistics for Cox models run using the same sets of variables as in the \citet{choustikov24} and \citet{mascia23} models.} 
\end{deluxetable*}

\subsection{Multivariate Predictions}
\label{sec:compare:multi}

Recently, \citet{choustikov24} and \citet{mascia23} have developed multivariate linear regression models to predict \fesc\ from a set of observable variables. \citet{choustikov24} derive their model for \fesc\ from synthetic spectra of $z>4$ galaxies from the SPHINX cosmological radiation hydrodynamics simulation. \citet{mascia23} base their predictions on the LzLCS+ sample but adopt the upper limit in \fesc\ as the observed value for non-detections. Here, we use galaxies at low and high redshift to compare the predictions of the Cox models with the predictions from these literature models. 

\begin{figure*}
\gridline{\fig{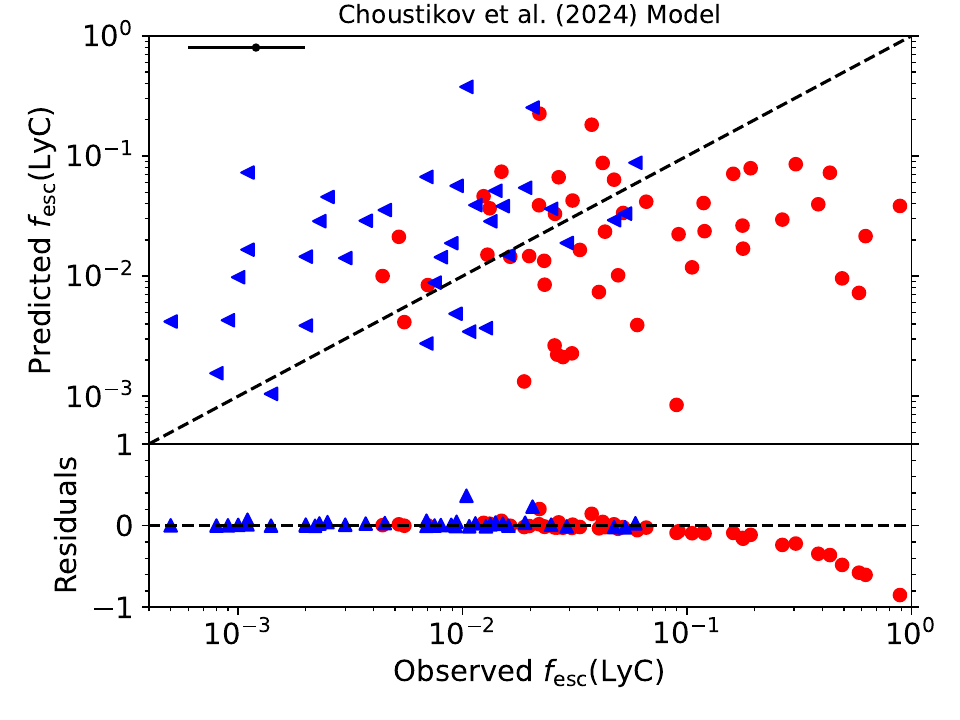}{0.5\textwidth}{(a)}
	\fig{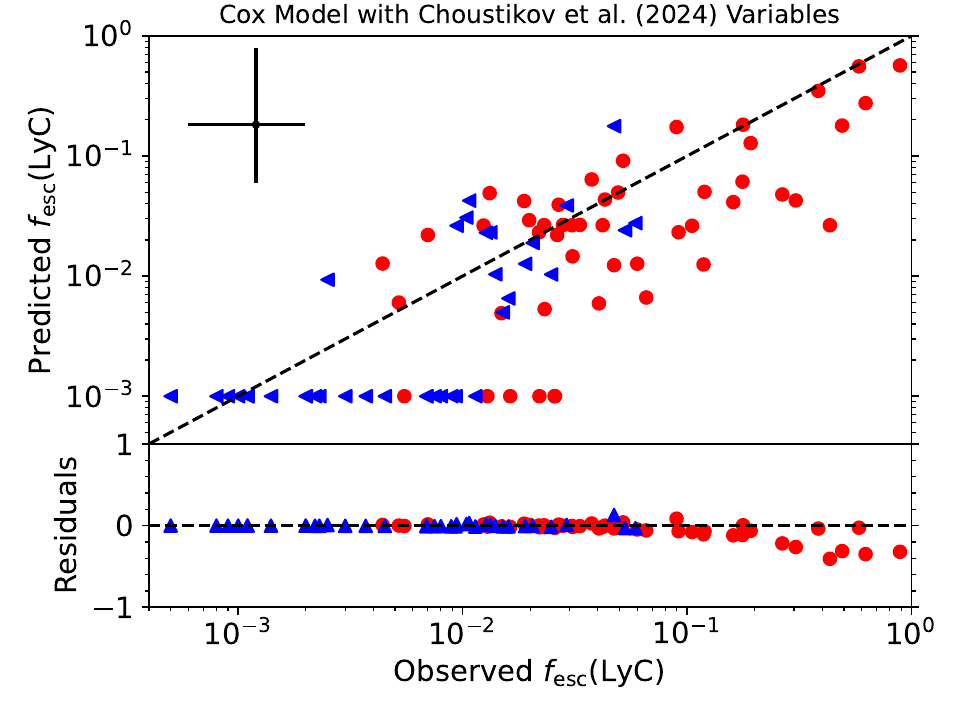}{0.5\textwidth}{(b)}
	}
\caption{(a) The \fesc\ predictions from the \citet{choustikov24} literature model compared with the LzLCS+ observations. (b) The \fesc\ predictions from a Cox model run using the same variables as the \citet{choustikov24} model: $\beta_{\rm 1550}$, E(B-V)$_{\rm neb}$, $L$(H$\beta$), $M_{\rm 1500}$, R23, and O32. Symbols are the same as in Figure~\ref{fig:highz_a}. The \citet{choustikov24} model does not reproduce the \fesc\ observations from the LzLCS+, but a Cox model using the same input variables does recover the observed \fesc.
\label{fig:choustikov}}
\end{figure*}

In Figure~\ref{fig:choustikov}a, we show the results of applying the \citet{choustikov24} model to the LzLCS+ sample, and we list the corresponding goodness-of-fit statistics in Table~\ref{table:literature}. The \citet{choustikov24} model fails to predict the observed \fesc\ in the LzLCS+ sample. This result does not depend on our method of calculating \fesc; we find a similarly poor fit if we substitute the \fesc\ derived from H$\beta$, as used in the \citet{choustikov24} simulations. In Figure~\ref{fig:choustikov}b, we show \fesc\ predictions from a Cox model run using the same set of variables ($\beta_{\rm 1550}$, E(B-V)$_{\rm neb}$, $L$(H$\beta$), $M_{\rm 1500}$, R23$ = $(\oiii~$\lambda\lambda$5007,4959+\oii$\lambda$3727)/H$\beta$, and O32) as in the \citet{choustikov24} linear regression mode, and we list the goodness-of-fit metrics in Table~\ref{table:literature}. 

This model performs comparably to the other Cox models in Table~\ref{table:highz:metrics}, which shows that the reason for the \citet{choustikov24} model's difficulty is not its set of variables. Rather, in the \citet{choustikov24} model, \fesc\ anti-correlates with O32, whereas the LzLCS+ data and the Cox model using the \citet{choustikov24} variable set indicate a correlation between \fesc\ and O32. The compact, lower-mass LCEs in the LzLCS do not have counterparts in the SPHINX simulations; these compact LCEs often have high O32 ratios $\gtrsim10$ \citep{flury22b}, whereas the SPHINX galaxies with O32 $ > 10$ typically have lower values of \fesc. Hence, while we find that high O32 increases \fesc\ in our multivariate predictions, \citet{choustikov24} find the opposite, that at fixed $\beta$ and E(B-V), galaxies with older ages and lower O32 have higher \fesc. Radiative feedback with a turbulent gas structure may allow \fesc\ from high O32 galaxies at earlier ages than predicted by cosmological simulations \citep[\eg,][]{kakiichi21,kimm19,choustikov24}. However, further studies of $z\sim3$ galaxies are also necessary to test whether the LzLCS population and the LzLCS-derived Cox models correctly describe the high-redshift population.

Like our Cox model predictions in \S\ref{sec:z6}, \citet{choustikov24} predict a low \fesc\ $=0.03$ for JADES-GS-z7-LA, due to its moderate UV slope and dust content. However, because of the different dependence of \fesc\ on O32, the \citet{choustikov24} model predicts a dramatically lower \fesc\ for CEERS-44, CEERS-698, CEERS-1019, and CEERS-1027 (\fesc\ $=0.006-0.1$) and for JADES 9422 (\fesc\ $=0.01$; \citealt{saxena24}) than our model predictions (\fesc\ $>0.4$ to 1). \citet{choustikov24} point out that their predicted \fesc\ values are consistent with \fesclya\ $ >$ \fesc\ as observed locally \citep[\eg,][]{flury22b}, but IGM effects may complicate any comparison with \fesclya. As noted above (\S\ref{sec:z6}), the Cox model predictions for some of these galaxies require extrapolating outside the LzLCS+ parameter space, and their \fesc\ may not truly be as extreme as this model predicts. Still, these galaxies remain plausible candidates for high \fesc\ given the observed trends seen at $z\sim0.3$ and $z\sim3$ (\S\ref{sec:z3}). LyC observations of galaxies with similar properties are necessary to better constrain their \fesc.

\begin{figure*}
\gridline{\fig{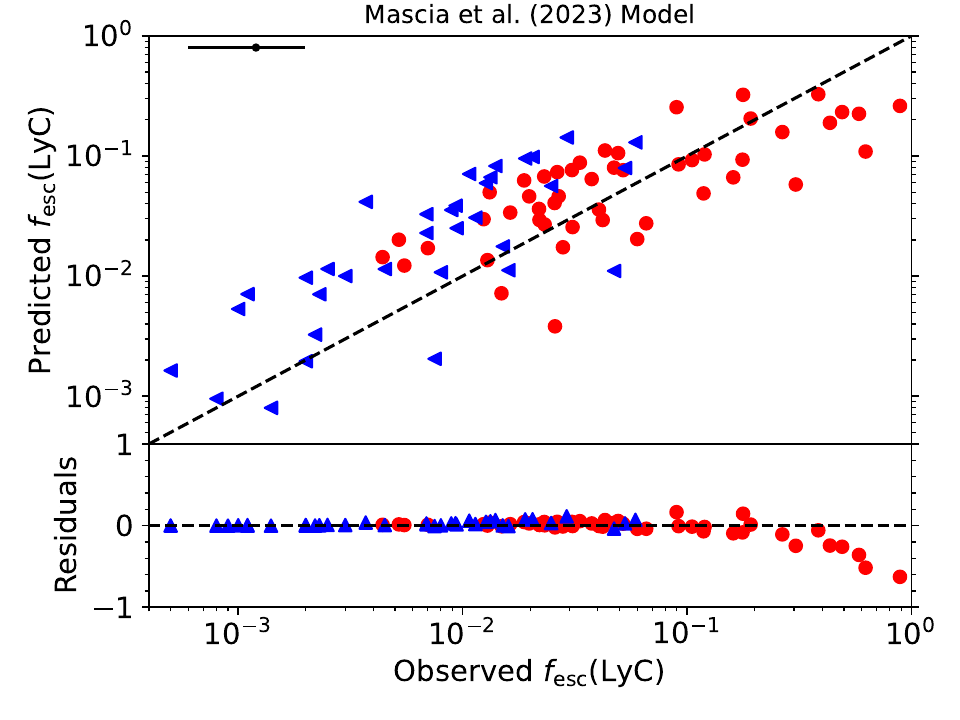}{0.5\textwidth}{(a)}
	\fig{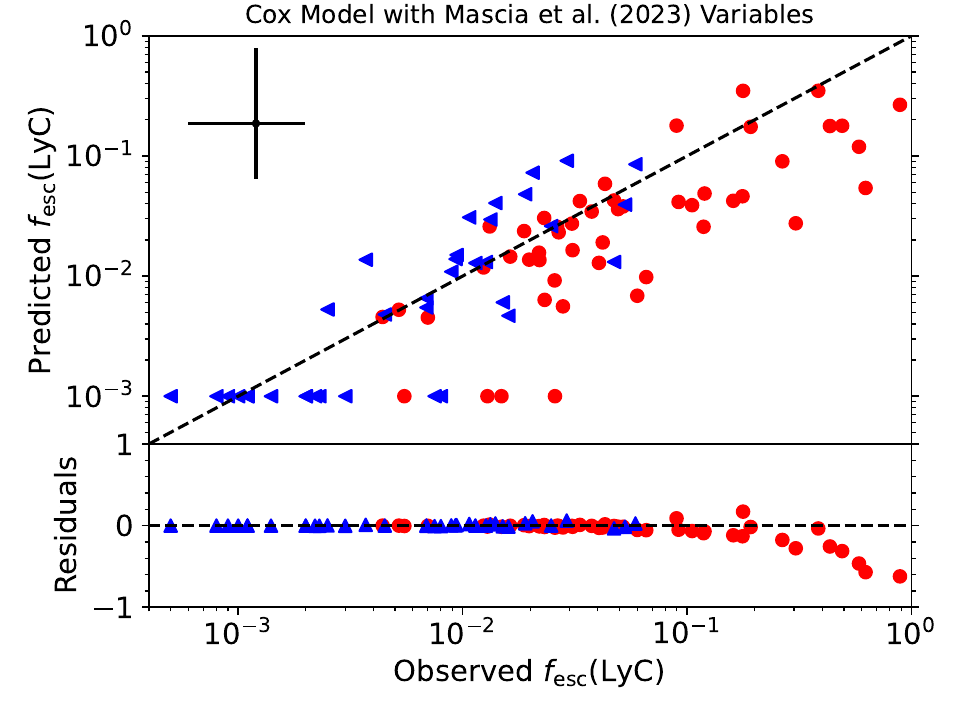}{0.5\textwidth}{(b)}
	}
\caption{(a) The \fesc\ predictions from the \citet{mascia23} literature model compared with the LzLCS+ observations. (b) The \fesc\ predictions from a Cox model run using the same variables as the \citet{mascia23} model: $\beta_{\rm 1550}$, $r_{\rm 50,NUV}$, and O32. Symbols are the same as in Figure~\ref{fig:highz_a}. Both models perform similarly, but the Cox model does slightly better at predicting \fesc\ in weak LCEs and non-detections.
\label{fig:mascia}}
\end{figure*}

In contrast to the results for the \citet{choustikov24} model, the \citet{mascia23} model (Figure~\ref{fig:mascia}a and Table~\ref{table:literature}) reproduces \fesc\ for the LzLCS+ galaxies reasonably well. This finding is unsurprising, since the \citet{mascia23} model is in fact derived from the LzLCS+ dataset. A Cox model run using the same variables ($\beta_{\rm 1550}$, $r_{\rm 50,NUV}$, and O32) performs comparably to the \citet{mascia23} model (Figure~\ref{fig:mascia}b and Table~\ref{table:literature}). The largest difference occurs for weak LCEs (\fesc\ $< 0.05$) and non-detections. In these cases, the \citet{mascia23} model tends to systematically over-predict \fesc, whereas by incorporating information from non-detections, Cox model predictions are more evenly distributed above and below the observed values. However, we note that the over-prediction of \fesc\ in the \citet{mascia23} model is small, $\Delta$\fesc\ of only a few percent. 

Despite its derivation from the same dataset, the Cox and \citet{mascia23} models give different predictions at high redshift. The \citet{mascia23} model predicts \fesc\ $=0.19$ for Ion2, the only $z\sim3$ LCE with the required measurements, whereas the Cox models more accurately match its observed \fesc\ $>0.5$ (\S\ref{sec:z3}). This difference suggests that galaxy luminosity, incorporated in the Cox models in either \sigsfr\ or in $M_{\rm 1500}$ may be important in reproducing \fesc\ for the strongest LCEs. 

\begin{figure*}
\gridline{\fig{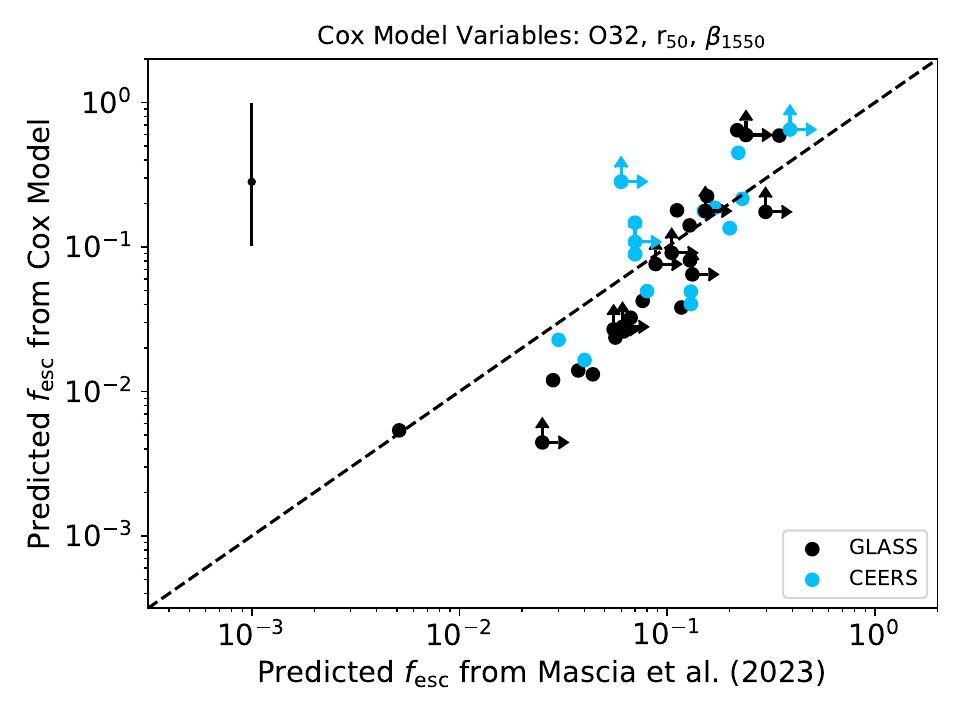}{0.5\textwidth}{(a)}
	\fig{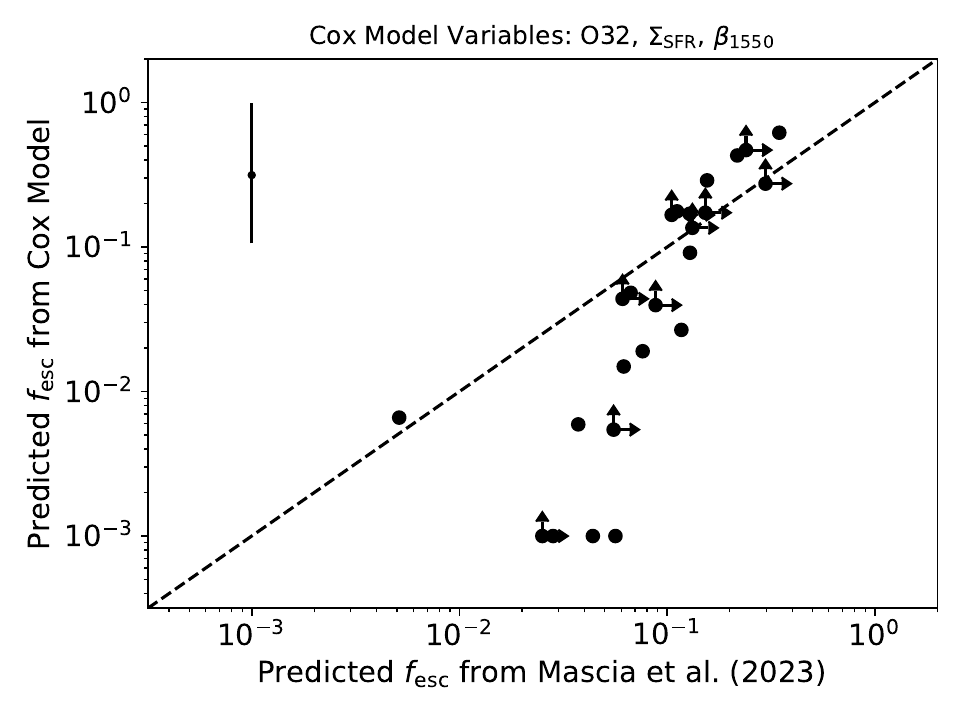}{0.5\textwidth}{(b)}
	}
\gridline{\fig{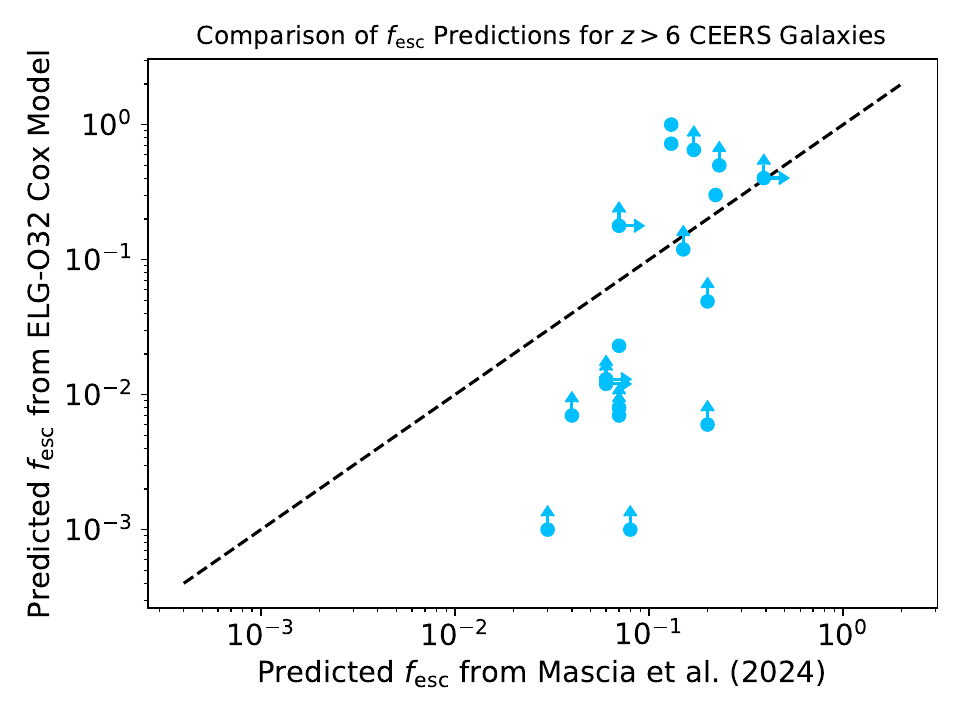}{0.5\textwidth}{(c)}
	\fig{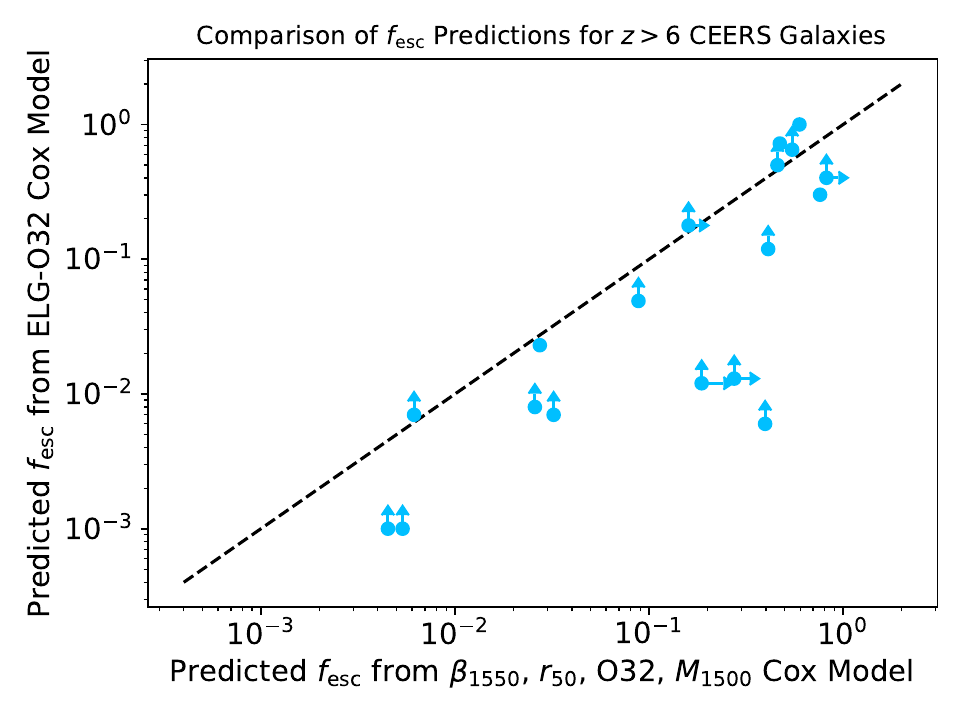}{0.5\textwidth}{(d)}
	}
\caption{A comparison of \fesc\ predictions for high-redshift galaxies from Cox models and the \citet{mascia23} model. (a) A Cox model using $\beta_{\rm 1550}$, $r_{\rm 50,NUV}$, and O32 gives similar predicted \fesc\ as the \citet{mascia23} model for $z>4$ galaxies from the GLASS-JWST (black) and CEERS (blue) surveys \citep{mascia23, mascia24}. (b) If \sigsfr\ is substituted for $r_{\rm 50,NUV}$ in the Cox model, the Cox model \fesc\ predictions for the GLASS galaxies deviate from those of the \citet{mascia23} model. (c) For the CEERS galaxies, the \fesc\ predictions from the ELG-O32 Cox model disagree with the \citet{mascia24} predictions. (d) A Cox model using $M_{\rm 1500}$ in addition to the \citet{mascia23} variables begins to agree more closely with the ELG-O32 Cox model predictions. Adding a measurement of galaxy luminosity in the form of \sigsfr\ or $M_{\rm 1500}$ can significantly affect the \fesc\ predictions.
\label{fig:mascia_highz}}
\end{figure*}

In Figure~\ref{fig:mascia_highz}, we compare the model predictions for GLASS-JWST galaxies \citep{treu22} at $z=4-8$ from \citet{mascia23} with predictions from Cox models using similar combinations of variables. For a Cox model using the same variables as in \citet{mascia23}, the predictions from \citet{mascia23} and the corresponding Cox model track each other (Figure~\ref{fig:mascia_highz}a), aside from the aforementioned tendency of the \citet{mascia23} model to give higher predictions at low \fesc. However, the \citet{mascia23} and Cox models disagree more strongly when \sigsfr\ is used as a variable in the Cox model instead of $r_{\rm 50,NUV}$ (Figure~\ref{fig:mascia_highz}b). The inclusion of \sigsfr\ shifts some galaxies with compact sizes but weak star formation to much lower \fesc. We conclude that including some measure of luminosity or SFR may be important to accurately identify non-leakers and extreme LCEs like Ion2. 

\citet{mascia24} also apply their multivariate model to galaxies from CEERS and find significantly different results than the predictions from our ELG-O32 Cox model for these galaxies (\S\ref{sec:z6}, Table~\ref{table:predictfujimoto}). We compare these predictions in Figure~\ref{fig:mascia_highz}c. Again, the disagreement primarily arises from the inclusion of luminosity in the ELG-O32 Cox model but not the \citet{mascia23} model. As with the GLASS sample, when we run a Cox model with the same input variables ($\beta_{\rm 1550}$, $r_{\rm 50,NUV}$, and O32), our predictions agree closely with the \citet{mascia24} predictions (Figure~\ref{fig:mascia_highz}a). Adding one additional variable, $M_{\rm 1500}$, as a measure of luminosity begins to bring the predicted \fesc\ into agreement with the ELG-O32 predictions (Figure~\ref{fig:mascia_highz}d). The remaining disagreement between the \citet{mascia24} and ELG-O32 Cox model predictions is due to the use of radius as a variable in the \citet{mascia23} model and different estimates of dust content and ionization from different publications \citep{mascia24, fujimoto23, tang23}. 

Observational and theoretical studies agree that \fesc\ depends on multiple physical parameters. However, studies have not yet reached a consensus as to which parameters matter and how. Predictions for \fesc\ in $z\gtrsim6$ galaxies from this work, \citet{choustikov24}, \citet{mascia23} and \citet{mascia24} disagree because of different adopted scalings with O32 and with luminosity. Observationally testing and distinguishing between these predictions with larger samples at $z<6$ will be critical to reliably predict \fesc\ in the epoch of reionization.

\section{Implications for Reionization}
\label{sec:discuss}

As demonstrated in \S\ref{sec:z3}, the empirical Cox models derived from the LzLCS+ show promise as diagnostics of \fesc\ at high redshift. However, with the limited measurements at $z>6$ available so far, the models do not decisively show which galaxy populations dominate reionization (\S\ref{sec:magtrends}). Because a combination of properties regulates a galaxy's optical depth, we need estimates of many factors to accurately predict \fesc. Dust attenuation, galaxy morphology, ionization, and UV luminosity all play a role in promoting LyC escape. Based primarily on dust attenuation, we would expect faint galaxies to be strong LCEs \citep[\eg,][]{chisholm22, atek24}. Nevertheless, at a fixed value of $\beta_{\rm 1550}$, \fesc\ still shows considerable spread in the LzLCS+ and in multivariate predictions (Figure~\ref{fig:chisholm_predict}), spanning a range of up to $\sim0.7$ in \fesc. Without estimates of other properties (e.g., \sigsfr, O32), we therefore cannot determine whether faint galaxies have \fesc\ above or below the average LzLCS+ value for the same UV color. 

At the same time, we cannot yet rule out the contribution of more luminous galaxies. Several galaxies with $M_{\rm 1500} \leq -20$ (e.g., CEERS 498, 1027, 698, 1019; JADES 18846; GLASS-JWST 100003, 10021) have the low dust content and high ionization suggestive of extreme \fesc\ $\geq 0.5$. The LzLCS+ data seem to suggest that for the same O32 ratio and $\beta_{\rm 1500}$, a more luminous galaxy will have higher \fesc\ (see also \citealt{lin24}). This effect could possibly be an observational bias, as the LyC flux should be easier to detect for more luminous objects. However, we note that at fixed O32, brighter LzLCS+ galaxies show higher ratios of 900 \AA\ to 1100 \AA\ flux, which suggests that their high \fesc\ is genuine. Moreover, the trend of higher \fesc\ with luminosity only appears in galaxies that have properties associated with LyC escape, such as high O32. Overall, the brighter galaxies in the LzLCS+ have fewer LyC detections \citep{flury22b}, which suggests that the detections are not biased toward the brightest galaxies. Furthermore, including UV luminosity in the multivariate models appears necessary to reproduce the \fesc\ of Ion2 (\S\ref{sec:z3} and \S\ref{sec:compare}), which suggests that the increase of \fesc\ with luminosity may be a real phenomenon.

 In the multivariate models, $M_{\rm 1500}$ might help to break the degeneracy between ionization parameter and optical depth in galaxies with high O32. For instance, low-metallicity galaxies tend to have both low luminosities and inherently high ionization parameters \citep[\eg,][]{nagao06, dopita06}, such that they may be more likely to have high O32 even at high optical depth. Alternatively, LCEs with higher UV luminosities could be able to ionize gas over a wider opening angle, so that we are statistically more likely to observe a high \fesc\ line of sight.

Of course, to evaluate a galaxy population's influence on reionization, we must also consider how many ionizing photons are produced and the total LyC input into the IGM, not just \fesc.  Unfortunately, the Cox models do not do as well at predicting the ionizing to non-ionizing UV flux ratio or the total LyC luminosity (Paper I). Galaxies with high O32 and high nebular EWs may possess elevated $\xi_{\rm ion}$ values \citep[\eg,][]{schaerer16, tang19, maseda20, naidu22}, such that at fixed \fesc, these galaxies will provide more LyC photons to the IGM. These same properties scale with \fesc\ in the multivariate Cox models, and our identified candidate strong LCEs at $z>6$ generally have O32 $> 10$ and/or EW(\oiii+H$\beta$) $> 1500$ \AA. High \fesc\ may therefore be coupled with high $\xi_{\rm ion}$ \citep[\eg,][]{schaerer16, naidu22}. 

However, ionizing photon production may vary with other parameters as well, such as galaxy luminosity \citep[\eg,][]{bouwens16, finkelstein19}. \citet{fujimoto23} estimate $\xi_{\rm ion}$ for spectroscopically confirmed galaxies at $z\sim8-9$ and find that $\xi_{\rm ion}$ is two times higher at $M_{\rm UV}\sim-19.5$ compared to $M_{\rm UV}\sim-21.5$. Although this enhanced efficiency is not enough to compensate for the six times fainter UV luminosity at $M_{\rm UV}\sim-19.5$ vs.\ -21.5, these fainter galaxies are also approximately 40 times more numerous \citep{bouwens15}. \citet{atek24} infer similarly high $\xi_{\rm ion}$ for even fainter ($M_{\rm UV}\sim-17$ to -15) and even more numerous galaxies, which only need \fesc $< 0.05$ to drive reionization. For the faintest galaxies we consider ($M_{\rm 1500} > -17.5$), the median of the Cox model predicted \fesc\ values is near this threshold, with median \fesc $\sim 0.04$. However, these \fesc\ estimates require refinement with additional parameters (e.g., O32, \sigsfr). If \fesc\ does not vary strongly with luminosity, as suggested by our preliminary, limited models, the fainter population would dominate reionization due to their higher $\xi_{\rm ion}$ and greater numbers. 

Assessing the main contributors to reionization requires progress on several fronts. At $z>6$, we need estimates of nebular and morphological parameters in addition to $\beta_{\rm 1550}$ and $M_{\rm 1500}$ for galaxy samples spanning a wide range of luminosities. Parameters such as O32 and \sigsfr\ have some of the greatest effects on the \fesc\ prediction accuracy for both the LzLCS+ (Paper I) and galaxies at $z\sim3$ (\S\ref{sec:z3}). The limited magnitude range of the LzLCS+ reference sample ($M_{\rm 1500} \sim-18.5$ to $\sim-21.5$) also introduces uncertainty. Our \fesc\ estimates for spectroscopic samples within this magnitude range are likely reasonable and suggest moderately high median \fesc\ $\sim0.04-0.05$, with some galaxies reaching \fesc\ as high as 0.6-0.7 (see Figure~\ref{fig:endsleymag} and predictions in Table~\ref{table:predictfujimoto} for the ELG-O32 model and Table~\ref{table:predictsaxenab} for the ELG-O32-$\beta$ model). However, our \fesc\ estimates for fainter and brighter galaxy populations rely on extrapolation and are less trustworthy. To confirm these estimates, we need to explore \fesc\ and its dependence on galaxy properties across a wider magnitude range at $z<6$. Lastly, observations of the IGM and galaxy population at $z>6$ will provide a further test of \fesc\ predictions. An accurate model of \fesc\ should reproduce both the timing and topology of reionization with the $z>6$ galaxy population. While much progress is needed to both improve and confirm \fesc\ estimates, our current \fesc\ predictions imply that plausible contributors to reionization appear at all magnitude ranges, and star-forming galaxies with the required levels of LyC escape do indeed exist at $z>6$. 

\section{Summary}
\label{sec:conclusions}
Quantifying the LyC escape fraction (\fesc) of galaxies is critical to understand the sources of cosmic reionization. We have developed a flexible tool for predicting \fesc\ empirically using combinations of observable variables available in the $z\sim0.3$ LzLCS+ reference sample (Paper I). We generate multivariate diagnostics for \fesc\ with the Cox proportional hazards model \citep{cox72}, a survival analysis technique that appropriately treats data with upper limits. We test Cox models developed from the $z\sim0.3$ galaxies on \fesc\ observations at $z\sim3$, and we apply the models to several samples of $z\gtrsim6$ galaxies to predict \fesc\ for galaxies in the epoch of reionization. The Appendix provides all the information necessary to use the multivariate Cox models to predict \fesc\ for other samples of galaxies, including samples at $z\sim3$ and samples at $z\gtrsim6$, where the LyC is unobservable.\footnote{We have also developed a python script, available at \url{https://github.com/sflury/LyCsurv}, that allows the user to generate new Cox proportional hazards models to predict \fesc\ using custom combinations of variables for any given galaxy population. Version 0.1.0 of the code is archived in Zenodo \citep{lycsurv}.}

We summarize our main findings below.
\begin{enumerate}

\item  The models successfully reproduce the observed \fesc\ values for the high-redshift $z\sim3$ galaxies and often have a lower RMS for the $z\sim3$ samples than for the $z\sim0.3$ galaxies. The success of these models suggests that low-redshift and high-redshift LCEs may share similar properties. (\S\ref{sec:z3})

\item The best-performing models for the $z\sim3$ galaxies include the dust attenuation inferred from the UV SED (E(B-V)$_{\rm UV}$) or the UV slope $\beta_{\rm 1550}$, plus either O32 or a morphological measurement ($r_{\rm 50,NUV}$ or \sigsfr). However, this result is tentative; each model is tested on a different subset of high-redshift galaxies, which makes comparing the models difficult. To determine which variable combinations best predict \fesc\ at high redshift, we require larger $z\sim3$ samples of LyC-emitters with the full suite of input variables. (\S\ref{sec:z3})  

\item We generate new Cox models based on the LzLCS+ observations, which incorporate variables measured for $z\gtrsim6$ samples. One model, using $M_{\rm 1500}$, $M_*$, E(B-V)$_{\rm UV}$ and EW(\oiii+H$\beta$) as input variables, applies to 180 galaxies from \citet{endsley21, endsley23, bouwens23, tang23, fujimoto23, saxena23} and predicts a median \fesc\ $=0.047$ for this combined sample. Of these 180 galaxies, 27\%\ have \fesc\ $>0.1$ and only 6\%\ have \fesc\ $>0.2$. However, this set of variables results in predictions that tend to underestimate \fesc, and the galaxy samples mostly consist of photometric measurements. A second model, using $M_{\rm 1500}$, $M_*$, $\beta_{\rm 1550}$, and $r_{\rm 50,NUV}$ as input variables, applies to 278 galaxies, mostly with photometric redshifts, from \citet{morishita24} and \citet{mascia23}. This model predicts a higher median \fesc\ $=0.14$, with 56\%\ and 39\%\ of the galaxies having \fesc\ $>0.1$ and \fesc\ $>0.2$, respectively. Many of these galaxies are more compact than the LzLCS+ sample, and this model therefore requires extrapolating beyond the LzLCS+ parameter space. (\S\ref{sec:z6samples})

\item Smaller samples of spectroscopically-confirmed $z\gtrsim6$ galaxies have higher predicted \fesc, likely because they tend to include stronger emission-line galaxies. We use a model with $\beta_{\rm 1550}$, $M_{\rm 1500}$, and O32 to predict \fesc\ for 27 spectroscopically-confirmed galaxies \citep{williams23, schaerer22b, mascia23, saxena23, saxena24} and find that 33-41\%\ of these galaxies have \fesc\ $>0.2$. For a smaller sample of 17 galaxies \citep{williams23, tang23, fujimoto23, saxena23}, using $M_*$ as an additional input variable and E(B-V)$_{\rm UV}$ instead of $\beta_{\rm 1550}$, we find that 46-85\%\ of the galaxies have \fesc\ $>0.2$. These models identify five galaxies at $z>5.9$ that may have a line-of-sight \fesc\ as high as 0.5-1: CEERS-698, CEERS-1019, CEERS-1027 from \citet{tang23} and JADES 18846 and JADES 9422 from \citet{saxena24}. (\S\S\ref{sec:z6samples}, \ref{sec:individ}).

\item Currently, the predicted \fesc\ values for different galaxy samples show no strong trend with UV magnitude across the range of $M_{\rm 1500}=-16$ to $-22$. However, many of these models are limited by a lack of spectroscopic information. The models suggest that low-luminosity galaxies ($M_{\rm 1500} > -18$) have at least moderate \fesc\ $\sim0.05$. If this \fesc\ is coupled with a high ionizing photon production efficiency, such faint galaxies could substantially contribute to reionization \citep[\eg,][]{atek24,simmonds24}. Additional measurements of variables such as O32 and \sigsfr\ for more $z\gtrsim6$ galaxies could enable more accurate \fesc\ predictions for both bright and faint galaxies and might increase the numbers of suspected strong LCEs. (\S\ref{sec:magtrends})

\item The multivariate predictions for $z\gtrsim6$ galaxies can differ strongly from \fesc\ predictions based on a single variable. We predict \fesc\ for the strong \lya\ Emitter, JADES-GS-z7-LA, discovered at $z=7.278$ \citep{saxena23}. Despite its exceedingly high \fesclya, which would seem to imply a high \fesc, the two Cox models that include the most available information (dust, \lya, O32, and luminosity) predict that JADES-GS-z7-LA is a moderate LCE with \fesc\ $=0.017-0.085$. Another example is the galaxy 11027 from \citet{williams23}, which has O32 $=12$, similar to the $z\gtrsim6$ galaxies with predicted \fesc\ $\geq0.6$. Yet, models incorporating other information about the galaxy, such as its mass (\logten($M_*$/\Msol)=7.7), UV magnitude ($M_{\rm 1500}=-17.4$), and dust attenuation ($\beta_{\rm 1550}=-2$) predict \fesc\ $\leq0.02$. However, we caution that some $z\gtrsim6$ galaxies, including these ones, fall outside the parameter space covered by the $z\sim0.3$ galaxy sample, which make make our \fesc\ predictions less reliable. (\S\ref{sec:individ})

\item We compare our model predictions to single variable predictions \citep{chisholm22} and to other multivariate \fesc\ predictions derived from simulations \citep{choustikov24} or observations \citep{mascia23, mascia24}. The Cox models more accurately predict \fesc\ in non-detections than the $\beta_{\rm 1550}$ model from \citet{chisholm22}, and the inclusion of feedback-sensitive variables such as O32 and \sigsfr\ may better trace the effect of absorption from \hi\ in LCEs with low dust content. Turning to multivariate models, the LzLCS+ sample implies a correlation between O32 and \fesc, rather than the anti-correlation adopted in the \citet{choustikov24} model. As a result, we find different predicted \fesc\ for high-redshift galaxies, and the \citet{choustikov24} model fails to reproduce the \fesc\ of the LzLCS+ galaxies. In contrast, our predictions agree with the \citet{mascia23} model, when we use their same set of input variables. However, our high-redshift \fesc\ predictions differ strongly when we add UV luminosity or \sigsfr\ as input variables.

\end{enumerate}

LyC-emitting galaxies show similar physical characteristics at both low and high redshift, which suggests that the same processes may govern LyC escape across cosmic time. Because a variety of factors can influence LyC escape, accurately predicting \fesc\ requires information about galaxy luminosities, dust, nebular properties, and ionization. Direct measurements of LyC at low redshift, combined with multivariate statistical models for \fesc, can give insight into the possible \fesc\ of galaxies in the epoch of reionization. Nevertheless, additional observations are necessary before these techniques can reach their full potential. At low redshift, LyC observations that push to new parameter space, especially fainter and brighter UV magnitudes, will aid the application of low-redshift results to high-redshift samples. Measurements of relevant galaxy properties, especially in the rest-frame optical, for larger samples of $z\sim3$ LCEs will better test the relative performance of these models at high redshift. Based on our current results, models using O32 and \sigsfr\ appear promising but are not yet applicable to large samples at $z>6$. By measuring these properties for larger, representative galaxy samples at $z>6$, high-redshift surveys can connect galaxies in the epoch of reionization with their LyC-emitting counterparts at low redshift. 

\begin{acknowledgments}
\nolinenumbers
We thank Aayush Saxena for providing information about the dust attenuation in JADES-GS-z7-LA, and we thank the anonymous referee for feedback that improved the paper's readability. AEJ and SRF acknowledge support from NASA through grant HST-GO-15626. STScI is operated by AURA under NASA contract NAS-5-26555. ASL acknowledges support from Knut and Alice Wallenberg Foundation. MT acknowledges support from the NWO grant 0.16.VIDI.189.162 (``ODIN”).
\end{acknowledgments}

\appendix
In Tables~\ref{atab:U}-\ref{atab:Saxena_fesc}, we provide the information necessary to predict \fesc\ using the models discussed in this paper, and we give an example of how to use these models below. We have also developed a python script, available on GitHub\footnote{\texttt{LyCsurv} codebase: \url{https://github.com/sflury/LyCsurv}.}, that allows the user to generate and apply new Cox proportional hazards models to predict \fesc\ using a desired set of observed variables (version 0.1.0 of the code is archived in Zenodo; \citealt{lycsurv}).

Here, we provide the parameters for the main models presented in Paper I and in this work. The data for the fiducial model from Paper I appears in Table~\ref{atab:U}. The best-performing model for the LzLCS+ sample is the fiducial model modified to use EW(\hi,abs), and these model parameters appear in Table~\ref{atab:T}. In Table~\ref{atab:JWSTbeta}, we list the parameters for the Paper I JWST model, which only uses variables accessible at $z>6$. Two alternative JWST models use only the top-ranked accessible variables: $\beta_{\rm 1550}$, \sigsfr, and either O32 or \neiii/\oii. We provide the data for these models in Tables~\ref{atab:trioJWST2} and \ref{atab:trioJWST1}. Finally, Tables~\ref{atab:Ion2}-\ref{atab:Atek2} present the parameters for the models used for the $z\sim3$ and $z\gtrsim6$ \fesc\ predictions discussed in Sections \S\S\ref{sec:z3}-\ref{sec:z6}. 

Each table first lists the goodness-of-fit statistics for the LzLCS+ galaxies as a measure of the model performance. It then provides the fitted coefficients ($b_i$) for each included variable and the reference values ${\bar x_i}$, which are the mean of the LzLCS+ $x_i$ values, where $x_i$ is the input variable ($\beta_{\rm 1550}$, \logten(O32), etc). Finally, the table lists the baseline cumulative hazard function, HF$_0$, calculated by the {\tt lifelines} Cox Model fitting routine for each of the observed \fesc\ values for LCEs in the LzLCS+. Together, these parameters predict \fesc\ for a given galaxy with a set of input variables $x$ as follows (see Section \S\ref{sec:cox} for more detail). 

To predict \fesc\ for a galaxy, we first calculate the partial hazards function ph($x$) for its set of variables:
\begin{equation}
{\rm ph}(x) = \exp[\sum_{i=1}^{n} b_i(x_i - {\bar x_i})].
\end{equation}
We use the coefficients $b_i$ given in the model table, the galaxy's observed values $x_i$ for each variable, and the mean values ${\bar x_i}$ also given in the model table. For example, for the galaxy Ion2, we have observed values of E(B-V)$_{\rm UV}=0$, \logten(O32)=0.851, and \logten(\sigsfr/(\Msol\ yr$^{-1}$ kpc$^{-2}$))=2.03. Using $b_i$ (-10.083, 1.240, 1.369) and ${\bar x_i}$ (0.158, 0.521, 0.705) from Table~\ref{atab:Ion2}, we calculate
\begin{equation}
{\rm ph}(x) = \exp[-10.083*(0-0.158)+1.240*(0.851-0.521)+1.369*(2.03-0.705)] = 45.43.
\end{equation}
We then use this value of ph($x)$ to scale the baseline cumulative hazard function HF$_0$(\fesc) and calculate the Survival Function $S$(\fesc), which represents the probability that the true escape fraction, $f_{\rm esc, true}$, is lower than each tabulated \fesc:
\begin{equation}
S(f_{\rm esc}) = \exp[-HF_0(f_{\rm esc})*{\rm ph}(x)].
\end{equation}
From Table~\ref{atab:Ion2}, for \fesc\ $=0.8890$, HF$_0$ is 0.003949 and $S$(0.8890) is therefore $\exp[-0.003949*45.43]=0.8358$. For \fesc\ $=0.6247$, $HF_0=0.008253$ and $S(0.6247)=\exp[-0.008253*45.43]=0.6873$, and so on, for each value of \fesc. Thus, according to the ``TopThree model, there is a 68.73\%\ probability that Ion2 has a measured \fesc\ $<0.6247$ and an 83.58\%\ probability that Ion2's \fesc\ $<0.8890$. To find the median predicted \fesc, we calculate $S$ for each value of \fesc\ in the table and find the value where $S$ reaches 0.5; \fesc\ is predicted to be above this value 50\%\ of the time and below this value 50\%\ of the time. For Ion2, we see that $S(0.5838) = 0.5629$ and $S(0.4911)=0.4577$. We linearly interpolate to find where $S$(\fesc)$=0.5$, finding that $S(0.53)\sim0.5$. The predicted \fesc\ of Ion2 is then \fesc\ $=0.53$. Similarly, to determine the uncertainties in the predicted \fesc, we find where $S$(\fesc) reaches 0.159 and 0.841. According to the model, \fesc\ will be between these \fesc\ values 68\%\ of the time. For some galaxies, $S$(\fesc) is always $>0.5$ for the tabulated \fesc\ values, indicating a $>$50\%\ probability that \fesc\ is smaller than the smallest tabulated \fesc. This situation corresponds to an arbitrarily small predicted \fesc, \fesc\ $\sim0$. Conversely, if $S$(\fesc) is always $<0.5$, \fesc\ is arbitrarily large and \fesc\ $\sim1$. 

\begin{deluxetable*}{llll}
\tablecaption{Fiducial Model (Paper I)}
\label{atab:U}
\tablehead{\multicolumn{4}{c}{{\it Statistics}}}
\startdata
$R^2$ & $R^2_{\rm adj}$ & RMS & $C$ \\
0.60 & 0.49 & 0.36 & 0.89 \\
\hline
\multicolumn{4}{c}{{\it Model Parameters}} \\
\multicolumn{2}{l}{Variable} & $b_i$ & ${\bar x_i}$ \\
\hline
\multicolumn{2}{l}{\logten($M_*$/\Msol)} & 0.626 & 8.932 \\
\multicolumn{2}{l}{$M_{\rm 1500}$} & -0.462 & -19.84 \\
\multicolumn{2}{l}{\logten(EW(H$\beta$)/\AA)} & -3.682e-02 & 1.954 \\
\multicolumn{2}{l}{E(B-V)$_{\rm neb}$} & 6.820 & 0.145 \\
\multicolumn{2}{l}{12+\logten(O/H)} & 1.462 & 8.112 \\
\multicolumn{2}{l}{\logten(\oiii~$\lambda5007$/\oii~$\lambda3727$)} & 2.651 & 0.521 \\
\multicolumn{2}{l}{\logten($\Sigma_{\rm SFR}$/(\Msol yr$^{-1}$ kpc$^{-2}$)} & 1.834 & 0.705 \\
\multicolumn{2}{l}{E(B-V)$_{\rm UV}$} & -14.897 & 0.158 \\
\multicolumn{2}{l}{\fesclya} & 7.002 & 0.220 \\
\multicolumn{2}{l}{EW(LIS)(\AA)\tablenotemark{a}} & -0.264 & 1.060 \\
\hline
\multicolumn{4}{c}{{\it Baseline Cumulative Hazard}} \\
\fesc & $H_0$(\fesc) & \fesc (cont.) & $H_0$ (cont.) \\
\hline
0.8890 & 0.001038 & 0.0376 & 0.099723 \\
0.6247 & 0.002243 & 0.0333 & 0.110434 \\
0.5838 & 0.003495 & 0.0309 & 0.121567 \\
0.4911 & 0.004907 & 0.0307 & 0.132833 \\
0.4333 & 0.006444 & 0.0280 & 0.146030 \\
0.3845 & 0.008002 & 0.0268 & 0.160176 \\
0.3052 & 0.009680 & 0.0264 & 0.175696 \\
0.2659 & 0.011478 & 0.0258 & 0.191559 \\
0.1921 & 0.013460 & 0.0257 & 0.208518 \\
0.1777 & 0.015602 & 0.0232 & 0.226165 \\
0.1767 & 0.017962 & 0.0231 & 0.244148 \\
0.1607 & 0.020514 & 0.0220 & 0.264128 \\
0.1197 & 0.023136 & 0.0219 & 0.284535 \\
0.1185 & 0.025995 & 0.0198 & 0.310698 \\
0.1053 & 0.028889 & 0.0188 & 0.349019 \\
0.0917 & 0.031854 & 0.0163 & 0.390900 \\
0.0898 & 0.035544 & 0.0149 & 0.436669 \\
0.0658 & 0.041090 & 0.0132 & 0.489950 \\
0.0600 & 0.046719 & 0.0129 & 0.552766 \\
0.0519 & 0.052798 & 0.0124 & 0.618132 \\
0.0493 & 0.059108 & 0.0070 & 0.765657 \\
0.0473 & 0.065551 & 0.0055 & 1.052480 \\
0.0431 & 0.073313 & 0.0052 & 1.396596 \\
0.0421 & 0.081962 & 0.0044 & 1.962125 \\
0.0405 & 0.090790 & & \\
\enddata
\tablenotetext{a}{For this variable, a positive EW denotes net absorption.}
\end{deluxetable*}

\begin{deluxetable*}{llll}
\tablecaption{Fiducial Model with EW(\hi,abs) (Paper I)}
\label{atab:T}
\tablehead{\multicolumn{4}{c}{{\it Statistics}}}
\startdata
$R^2$ & $R^2_{\rm adj}$ & RMS & $C$ \\
0.69 & 0.60 & 0.31 & 0.91 \\
\hline
\multicolumn{4}{c}{{\it Model Parameters}} \\
\multicolumn{2}{l}{Variable} & $b_i$ & ${\bar x_i}$ \\
\hline
\multicolumn{2}{l}{\logten($M_*$/\Msol)} & 0.454 & 8.932 \\
\multicolumn{2}{l}{$M_{\rm 1500}$} & -1.370 & -19.84 \\
\multicolumn{2}{l}{\logten(EW(H$\beta$)/\AA)} & -2.756 & 1.954 \\
\multicolumn{2}{l}{E(B-V)$_{\rm neb}$} & 7.127 & 0.145 \\
\multicolumn{2}{l}{12+\logten(O/H)} & 1.729 & 8.112 \\
\multicolumn{2}{l}{\logten(\oiii~$\lambda5007$/\oii~$\lambda3727$)} & 5.170 & 0.521 \\
\multicolumn{2}{l}{\logten($\Sigma_{\rm SFR}$/(\Msol yr$^{-1}$ kpc$^{-2}$)} & 0.475 & 0.705 \\
\multicolumn{2}{l}{E(B-V)$_{\rm UV}$} & -21.358 & 0.158 \\
\multicolumn{2}{l}{\fesclya} & 3.988 & 0.220 \\
\multicolumn{2}{l}{EW(\hi,abs)(\AA)\tablenotemark{a}} & -1.487 & 2.468 \\
\hline
\multicolumn{4}{c}{{\it Baseline Cumulative Hazard}} \\
\fesc & $H_0$(\fesc) & \fesc (cont.) & $H_0$ (cont.) \\
\hline
0.8890 & 0.000270 & 0.0376 & 0.084116 \\
0.6247 & 0.000750 & 0.0333 & 0.093819 \\
0.5838 & 0.001276 & 0.0309 & 0.104034 \\
0.4911 & 0.001888 & 0.0307 & 0.114298 \\
0.4333 & 0.002637 & 0.0280 & 0.126069 \\
0.3845 & 0.003402 & 0.0268 & 0.139434 \\
0.3052 & 0.004460 & 0.0264 & 0.154484 \\
0.2659 & 0.005616 & 0.0258 & 0.170455 \\
0.1921 & 0.006786 & 0.0257 & 0.186968 \\
0.1777 & 0.008170 & 0.0232 & 0.203982 \\
0.1767 & 0.009731 & 0.0231 & 0.221288 \\
0.1607 & 0.011393 & 0.0220 & 0.240703 \\
0.1197 & 0.013207 & 0.0219 & 0.260881 \\
0.1185 & 0.015226 & 0.0198 & 0.285998 \\
0.1053 & 0.017322 & 0.0188 & 0.319845 \\
0.0917 & 0.019459 & 0.0163 & 0.367457 \\
0.0898 & 0.023429 & 0.0149 & 0.419286 \\
0.0658 & 0.028468 & 0.0132 & 0.479264 \\
0.0600 & 0.033620 & 0.0129 & 0.554142 \\
0.0519 & 0.039120 & 0.0124 & 0.633334 \\
0.0493 & 0.045164 & 0.0070 & 0.793057 \\
0.0473 & 0.051386 & 0.0055 & 1.175799 \\
0.0431 & 0.058404 & 0.0052 & 1.623712 \\
0.0421 & 0.066728 & 0.0044 & 2.096537 \\
0.0405 & 0.075377 & & \\
\enddata
\tablenotetext{a}{For this variable, a positive EW denotes net absorption.}
\end{deluxetable*}

\begin{deluxetable*}{llll}
\tablecaption{Full JWST Model (Paper I)}
\label{atab:JWSTbeta}
\tablehead{\multicolumn{4}{c}{{\it Statistics}}}
\startdata
$R^2$ & $R^2_{\rm adj}$ & RMS & $C$ \\
0.29 & 0.14 & 0.47 & 0.83 \\
\hline
\multicolumn{4}{c}{{\it Model Parameters}} \\
\multicolumn{2}{l}{Variable} & $b_i$ & ${\bar x_i}$ \\
\hline
\multicolumn{2}{l}{\logten($M_*$/\Msol)} & 6.877e-02 & 8.932 \\
\multicolumn{2}{l}{$M_{\rm 1500}$} & -0.397 & -19.84 \\
\multicolumn{2}{l}{\logten(EW(H$\beta$)/\AA)} & -1.181 & 1.954 \\
\multicolumn{2}{l}{E(B-V)$_{\rm neb}$} & 0.710 & 0.145 \\
\multicolumn{2}{l}{12+\logten(O/H)} & -2.750e-02 & 8.112 \\
\multicolumn{2}{l}{\logten(\oiii~$\lambda5007$/\oii~$\lambda3727$)} & 2.740 & 0.521 \\
\multicolumn{2}{l}{\logten($\Sigma_{\rm SFR}$/(\Msol yr$^{-1}$ kpc$^{-2}$)} & 1.082 & 0.705 \\
\multicolumn{2}{l}{$\beta_{\rm 1550}$} & -2.166 & -1.810 \\
\hline
\multicolumn{4}{c}{{\it Baseline Cumulative Hazard}} \\
\fesc & HF$_0$(\fesc) & \fesc (cont.) & HF$_0$ (cont.) \\
\hline
0.8890 & 0.003071 & 0.0376 & 0.196676 \\
0.6247 & 0.006718 & 0.0333 & 0.211196 \\
0.5838 & 0.010477 & 0.0309 & 0.226775 \\
0.4911 & 0.014362 & 0.0307 & 0.242624 \\
0.4333 & 0.018463 & 0.0280 & 0.261375 \\
0.3845 & 0.022643 & 0.0268 & 0.280335 \\
0.3052 & 0.027504 & 0.0264 & 0.301007 \\
0.2659 & 0.032501 & 0.0258 & 0.322308 \\
0.1921 & 0.037624 & 0.0257 & 0.343655 \\
0.1777 & 0.043075 & 0.0232 & 0.365982 \\
0.1767 & 0.049150 & 0.0231 & 0.388662 \\
0.1607 & 0.055522 & 0.0220 & 0.413036 \\
0.1197 & 0.062102 & 0.0219 & 0.437812 \\
0.1185 & 0.069015 & 0.0198 & 0.466136 \\
0.1053 & 0.076045 & 0.0188 & 0.496786 \\
0.0917 & 0.083300 & 0.0163 & 0.529638 \\
0.0898 & 0.090759 & 0.0149 & 0.565906 \\
0.0658 & 0.100730 & 0.0132 & 0.611820 \\
0.0600 & 0.110801 & 0.0129 & 0.672871 \\
0.0519 & 0.121542 & 0.0124 & 0.739342 \\
0.0493 & 0.132930 & 0.0070 & 0.872244 \\
0.0473 & 0.144577 & 0.0055 & 1.060466 \\
0.0431 & 0.156904 & 0.0052 & 1.258225 \\
0.0421 & 0.169902 & 0.0044 & 1.535025 \\
0.0405 & 0.183203 & & \\
\enddata
\end{deluxetable*}

\begin{deluxetable*}{llll}
\tablecaption{Limited JWST Model (Paper I): $\beta_{\rm 1550}$, \logten($\Sigma_{\rm SFR}$), \logten(O32)}
\label{atab:trioJWST2}
\tablehead{\multicolumn{4}{c}{{\it Statistics}}}
\startdata
$R^2$ & $R^2_{\rm adj}$ & RMS & $C$ \\
0.34 & 0.29 & 0.46 & 0.83 \\
\hline
\multicolumn{4}{c}{{\it Model Parameters}} \\
\multicolumn{2}{l}{Variable} & $b_i$ & ${\bar x_i}$ \\
\hline
\multicolumn{2}{l}{\logten(\oiii~$\lambda5007$/\oii~$\lambda3727$)} & 0.996 & 0.521 \\
\multicolumn{2}{l}{\logten($\Sigma_{\rm SFR}$/(\Msol yr$^{-1}$ kpc$^{-2}$)} & 1.404 & 0.705 \\
\multicolumn{2}{l}{$\beta_{\rm 1550}$} & -2.274 & -1.810 \\
\hline
\multicolumn{4}{c}{{\it Baseline Cumulative Hazard}} \\
\fesc & HF$_0$(\fesc) & \fesc (cont.) & HF$_0$ (cont.) \\
\hline
0.8890 & 0.003311 & 0.0376 & 0.196361 \\
0.6247 & 0.007118 & 0.0333 & 0.210168 \\
0.5838 & 0.011019 & 0.0309 & 0.225189 \\
0.4911 & 0.015019 & 0.0307 & 0.240530 \\
0.4333 & 0.019271 & 0.0280 & 0.259278 \\
0.3845 & 0.023685 & 0.0268 & 0.278291 \\
0.3052 & 0.028643 & 0.0264 & 0.298441 \\
0.2659 & 0.033702 & 0.0258 & 0.319131 \\
0.1921 & 0.038913 & 0.0257 & 0.339855 \\
0.1777 & 0.044432 & 0.0232 & 0.361703 \\
0.1767 & 0.050682 & 0.0231 & 0.383929 \\
0.1607 & 0.057192 & 0.0220 & 0.407993 \\
0.1197 & 0.063925 & 0.0219 & 0.432411 \\
0.1185 & 0.070990 & 0.0198 & 0.460300 \\
0.1053 & 0.078219 & 0.0188 & 0.491690 \\
0.0917 & 0.085704 & 0.0163 & 0.525259 \\
0.0898 & 0.093495 & 0.0149 & 0.563415 \\
0.0658 & 0.103034 & 0.0132 & 0.615517 \\
0.0600 & 0.112666 & 0.0129 & 0.678401 \\
0.0519 & 0.123204 & 0.0124 & 0.747133 \\
0.0493 & 0.134221 & 0.0070 & 0.893528 \\
0.0473 & 0.145450 & 0.0055 & 1.105639 \\
0.0431 & 0.157431 & 0.0052 & 1.324779 \\
0.0421 & 0.170192 & 0.0044 & 1.623314 \\
0.0405 & 0.183168 & & \\
\enddata
\end{deluxetable*}

\begin{deluxetable*}{llll}
\tablecaption{Limited JWST Model (Paper I): $\beta_{\rm 1550}$, \logten($\Sigma_{\rm SFR}$), \logten(\neiii/\oii)}
\label{atab:trioJWST1}
\tablehead{\multicolumn{4}{c}{{\it Statistics}}}
\startdata
$R^2$ & $R^2_{\rm adj}$ & RMS & $C$ \\
0.40 & 0.35 & 0.44 & 0.83 \\
\hline
\multicolumn{4}{c}{{\it Model Parameters}} \\
\multicolumn{2}{l}{Variable} & $b_i$ & ${\bar x_i}$ \\
\hline
\multicolumn{2}{l}{\logten(\neiii~$\lambda$3869/\oii~$\lambda3727$)} & 1.315 & -0.532 \\
\multicolumn{2}{l}{\logten($\Sigma_{\rm SFR}$/(\Msol yr$^{-1}$ kpc$^{-2}$)} & 1.343 & 0.705 \\
\multicolumn{2}{l}{$\beta_{\rm 1550}$} & -2.192 & -1.810 \\
\hline
\multicolumn{4}{c}{{\it Baseline Cumulative Hazard}} \\
\fesc & HF$_0$(\fesc) & \fesc (cont.) & HF$_0$ (cont.) \\
\hline
0.8890 & 0.003276 & 0.0376 & 0.196231 \\
0.6247 & 0.007022 & 0.0333 & 0.210160 \\
0.5838 & 0.010866 & 0.0309 & 0.225310 \\
0.4911 & 0.014827 & 0.0307 & 0.240797 \\
0.4333 & 0.019033 & 0.0280 & 0.260131 \\
0.3845 & 0.023395 & 0.0268 & 0.279810 \\
0.3052 & 0.028307 & 0.0264 & 0.300595 \\
0.2659 & 0.033303 & 0.0258 & 0.322009 \\
0.1921 & 0.038458 & 0.0257 & 0.343462 \\
0.1777 & 0.043933 & 0.0232 & 0.366090 \\
0.1767 & 0.050163 & 0.0231 & 0.389078 \\
0.1607 & 0.056638 & 0.0220 & 0.413937 \\
0.1197 & 0.063307 & 0.0219 & 0.439215 \\
0.1185 & 0.070313 & 0.0198 & 0.467915 \\
0.1053 & 0.077462 & 0.0188 & 0.500143 \\
0.0917 & 0.084871 & 0.0163 & 0.535109 \\
0.0898 & 0.092566 & 0.0149 & 0.574662 \\
0.0658 & 0.102235 & 0.0132 & 0.628551 \\
0.0600 & 0.111988 & 0.0129 & 0.693361 \\
0.0519 & 0.122604 & 0.0124 & 0.764890 \\
0.0493 & 0.133634 & 0.0070 & 0.918966 \\
0.0473 & 0.144908 & 0.0055 & 1.139392 \\
0.0431 & 0.156907 & 0.0052 & 1.368115 \\
0.0421 & 0.169810 & 0.0044 & 1.677980 \\
0.0405 & 0.182933 & & \\
\enddata
\end{deluxetable*}

\begin{deluxetable*}{llll}
\tablecaption{TopThree Model}
\label{atab:Ion2}
\tablehead{\multicolumn{4}{c}{{\it Statistics}}}
\startdata
$R^2$ & $R^2_{\rm adj}$ & RMS & $C$ \\
0.38 & 0.34 & 0.44 & 0.82 \\
\hline
\multicolumn{4}{c}{{\it Model Parameters}} \\
\multicolumn{2}{l}{Variable} & $b_i$ & ${\bar x_i}$ \\
\hline
\multicolumn{2}{l}{E(B-V)$_{\rm UV}$} & -10.083 & 0.158 \\
\multicolumn{2}{l}{\logten(\oiii~$\lambda5007$/\oii~$\lambda3727$)} & 1.240 & 0.521 \\
\multicolumn{2}{l}{\logten($\Sigma_{\rm SFR}$/(\Msol yr$^{-1}$ kpc$^{-2}$)} & 1.369 & 0.705 \\
\hline
\multicolumn{4}{c}{{\it Baseline Cumulative Hazard}} \\
\fesc & HF$_0$(\fesc) & \fesc (cont.) & HF$_0$ (cont.) \\
\hline
0.8890 & 0.003949 & 0.0376 & 0.206394 \\
0.6247 & 0.008253 & 0.0333 & 0.220201 \\
0.5838 & 0.012651 & 0.0309 & 0.234665 \\
0.4911 & 0.017203 & 0.0307 & 0.249373 \\
0.4333 & 0.022038 & 0.0280 & 0.268514 \\
0.3845 & 0.027025 & 0.0268 & 0.287870 \\
0.3052 & 0.032320 & 0.0264 & 0.308337 \\
0.2659 & 0.037741 & 0.0258 & 0.329388 \\
0.1921 & 0.043347 & 0.0257 & 0.350502 \\
0.1777 & 0.049274 & 0.0232 & 0.372931 \\
0.1767 & 0.055904 & 0.0231 & 0.395743 \\
0.1607 & 0.062835 & 0.0220 & 0.420512 \\
0.1197 & 0.069951 & 0.0219 & 0.445571 \\
0.1185 & 0.077408 & 0.0198 & 0.473320 \\
0.1053 & 0.085092 & 0.0188 & 0.505102 \\
0.0917 & 0.093180 & 0.0163 & 0.539307 \\
0.0898 & 0.101614 & 0.0149 & 0.578156 \\
0.0658 & 0.111233 & 0.0132 & 0.630685 \\
0.0600 & 0.120940 & 0.0129 & 0.688622 \\
0.0519 & 0.131675 & 0.0124 & 0.754340 \\
0.0493 & 0.142953 & 0.0070 & 0.895800 \\
0.0473 & 0.154429 & 0.0055 & 1.107382 \\
0.0431 & 0.166817 & 0.0052 & 1.327823 \\
0.0421 & 0.179768 & 0.0044 & 1.631787 \\
0.0405 & 0.192981 & & \\
\enddata
\end{deluxetable*}

\begin{deluxetable*}{llll}
\tablecaption{LAE Model}
\label{atab:Steidel}
\tablehead{\multicolumn{4}{c}{{\it Statistics}}}
\startdata
$R^2$ & $R^2_{\rm adj}$ & RMS & $C$ \\
0.21 & 0.15 & 0.50 & 0.81 \\
\hline
\multicolumn{4}{c}{{\it Model Parameters}} \\
\multicolumn{2}{l}{Variable} & $b_i$ & ${\bar x_i}$ \\
\hline
\multicolumn{2}{l}{$M_{\rm 1500}$} & -1.022 & -19.84 \\
\multicolumn{2}{l}{E(B-V)$_{\rm UV}$} & -13.943 & 0.157 \\
\multicolumn{2}{l}{EW(\lya)(\AA)\tablenotemark{a}} & 1.716e-02 & 68.16 \\
\hline
\multicolumn{4}{c}{{\it Baseline Cumulative Hazard}} \\
\fesc & $H_0$(\fesc) & \fesc (cont.) & $H_0$ (cont.) \\
\hline
0.8890 & 0.004221 & 0.0376 & 0.217054 \\
0.6247 & 0.008679 & 0.0333 & 0.232258 \\
0.5838 & 0.013292 & 0.0309 & 0.247919 \\
0.4911 & 0.018280 & 0.0307 & 0.264020 \\
0.4333 & 0.023421 & 0.0280 & 0.281165 \\
0.3845 & 0.028705 & 0.0268 & 0.298428 \\
0.3052 & 0.034055 & 0.0264 & 0.316930 \\
0.2659 & 0.039708 & 0.0258 & 0.335720 \\
0.1921 & 0.046128 & 0.0257 & 0.355020 \\
0.1777 & 0.053219 & 0.0232 & 0.375092 \\
0.1767 & 0.060496 & 0.0231 & 0.395379 \\
0.1607 & 0.068146 & 0.0220 & 0.416328 \\
0.1197 & 0.075984 & 0.0219 & 0.437797 \\
0.1185 & 0.084457 & 0.0198 & 0.461351 \\
0.1053 & 0.093132 & 0.0188 & 0.489491 \\
0.0917 & 0.102126 & 0.0163 & 0.518083 \\
0.0898 & 0.111330 & 0.0149 & 0.549562 \\
0.0658 & 0.120646 & 0.0132 & 0.584802 \\
0.0600 & 0.130097 & 0.0129 & 0.623000 \\
0.0519 & 0.140548 & 0.0124 & 0.671296 \\
0.0493 & 0.151240 & 0.0070 & 0.788813 \\
0.0473 & 0.162539 & 0.0055 & 0.952357 \\
0.0431 & 0.175252 & 0.0052 & 1.126629 \\
0.0421 & 0.188749 & 0.0044 & 1.368313 \\
0.0405 & 0.202829 & & \\
\enddata
\tablenotetext{a}{Positive EW(\lya) denotes net emission.}
\end{deluxetable*}

\begin{deluxetable*}{llll}
\tablecaption{LAE-O32 Model}
\label{atab:Fletcher}
\tablehead{\multicolumn{4}{c}{{\it Statistics}}}
\startdata
$R^2$ & $R^2_{\rm adj}$ & RMS & $C$ \\
0.37 & 0.29 & 0.45 & 0.82 \\
\hline
\multicolumn{4}{c}{{\it Model Parameters}} \\
\multicolumn{2}{l}{Variable} & $b_i$ & ${\bar x_i}$ \\
\hline
\multicolumn{2}{l}{$M_{\rm 1500}$} & -0.935 & -19.84 \\
\multicolumn{2}{l}{\logten($M_*$/\Msol)} & 0.961 & 8.920 \\
\multicolumn{2}{l}{E(B-V)$_{\rm UV}$} & -15.987 & 0.157 \\
\multicolumn{2}{l}{\logten(\oiii~$\lambda5007$/\oii~$\lambda3727$)} & 1.863 & 0.523 \\
\multicolumn{2}{l}{EW(\lya)(\AA)\tablenotemark{a}} & 1.506e-02 & 68.16 \\
\hline
\multicolumn{4}{c}{{\it Baseline Cumulative Hazard}} \\
\fesc & $H_0$(\fesc) & \fesc (cont.) & $H_0$ (cont.) \\
\hline
0.8890 & 0.002782 & 0.0376 & 0.186577 \\
0.6247 & 0.005695 & 0.0333 & 0.200373 \\
0.5838 & 0.008706 & 0.0309 & 0.214414 \\
0.4911 & 0.012297 & 0.0307 & 0.228667 \\
0.4333 & 0.016140 & 0.0280 & 0.245277 \\
0.3845 & 0.020018 & 0.0268 & 0.261970 \\
0.3052 & 0.024057 & 0.0264 & 0.280210 \\
0.2659 & 0.028335 & 0.0258 & 0.298707 \\
0.1921 & 0.033372 & 0.0257 & 0.317460 \\
0.1777 & 0.038847 & 0.0232 & 0.337641 \\
0.1767 & 0.044733 & 0.0231 & 0.358145 \\
0.1607 & 0.051327 & 0.0220 & 0.379255 \\
0.1197 & 0.058081 & 0.0219 & 0.400762 \\
0.1185 & 0.065680 & 0.0198 & 0.424827 \\
0.1053 & 0.073441 & 0.0188 & 0.451867 \\
0.0917 & 0.081748 & 0.0163 & 0.479495 \\
0.0898 & 0.090210 & 0.0149 & 0.510421 \\
0.0658 & 0.099141 & 0.0132 & 0.545046 \\
0.0600 & 0.108225 & 0.0129 & 0.583697 \\
0.0519 & 0.117912 & 0.0124 & 0.633505 \\
0.0493 & 0.127906 & 0.0070 & 0.745027 \\
0.0473 & 0.138188 & 0.0055 & 0.916397 \\
0.0431 & 0.149428 & 0.0052 & 1.095363 \\
0.0421 & 0.161445 & 0.0044 & 1.343768 \\
0.0405 & 0.173974 & & \\
\enddata
\tablenotetext{a}{Positive EW(\lya) denotes net emission.}
\end{deluxetable*}

\begin{deluxetable*}{llll}
\tablecaption{LAE-O32-nodust Model}
\label{atab:Nakajima}
\tablehead{\multicolumn{4}{c}{{\it Statistics}}}
\startdata
$R^2$ & $R^2_{\rm adj}$ & RMS & $C$ \\
0.02 & -0.06 & 0.59 & 0.77 \\
\hline
\multicolumn{4}{c}{{\it Model Parameters}} \\
\multicolumn{2}{l}{Variable} & $b_i$ & ${\bar x_i}$ \\
\hline
\multicolumn{2}{l}{$M_{\rm 1500}$} & -0.913 & -19.84 \\
\multicolumn{2}{l}{\logten(\oiii~$\lambda5007$/\oii~$\lambda3727$)} & 3.475 & 0.523 \\
\multicolumn{2}{l}{EW(\lya)(\AA)\tablenotemark{a}} & 2.219e-03 & 68.16 \\
\hline
\multicolumn{4}{c}{{\it Baseline Cumulative Hazard}} \\
\fesc & $H_0$(\fesc) & \fesc (cont.) & $H_0$ (cont.) \\
\hline
0.8890 & 0.006247 & 0.0376 & 0.266926 \\
0.6247 & 0.012954 & 0.0333 & 0.282683 \\
0.5838 & 0.019834 & 0.0309 & 0.299106 \\
0.4911 & 0.027167 & 0.0307 & 0.315676 \\
0.4333 & 0.034710 & 0.0280 & 0.333150 \\
0.3845 & 0.042320 & 0.0268 & 0.350663 \\
0.3052 & 0.050342 & 0.0264 & 0.369016 \\
0.2659 & 0.058559 & 0.0258 & 0.388124 \\
0.1921 & 0.067137 & 0.0257 & 0.407508 \\
0.1777 & 0.076178 & 0.0232 & 0.427799 \\
0.1767 & 0.085515 & 0.0231 & 0.448402 \\
0.1607 & 0.095217 & 0.0220 & 0.469654 \\
0.1197 & 0.104993 & 0.0219 & 0.491134 \\
0.1185 & 0.115165 & 0.0198 & 0.513681 \\
0.1053 & 0.125425 & 0.0188 & 0.539040 \\
0.0917 & 0.135985 & 0.0163 & 0.565361 \\
0.0898 & 0.146749 & 0.0149 & 0.593704 \\
0.0658 & 0.158378 & 0.0132 & 0.626154 \\
0.0600 & 0.170137 & 0.0129 & 0.661094 \\
0.0519 & 0.182291 & 0.0124 & 0.708036 \\
0.0493 & 0.194771 & 0.0070 & 0.783818 \\
0.0473 & 0.208242 & 0.0055 & 0.882875 \\
0.0431 & 0.222481 & 0.0052 & 0.989315 \\
0.0421 & 0.237057 & 0.0044 & 1.116652 \\
0.0405 & 0.251873 & & \\
\enddata
\tablenotetext{a}{Positive EW(\lya) denotes net emission.}
\end{deluxetable*}

\begin{deluxetable*}{llll}
\tablecaption{ELG-EW Model}
\label{atab:EndFletch}
\tablehead{\multicolumn{4}{c}{{\it Statistics}}}
\startdata
$R^2$ & $R^2_{\rm adj}$ & RMS & $C$ \\
0.14 & 0.05 & 0.53 & 0.79 \\
\hline
\multicolumn{4}{c}{{\it Model Parameters}} \\
\multicolumn{2}{l}{Variable} & $b_i$ & ${\bar x_i}$ \\
\hline
\multicolumn{2}{l}{$M_{\rm 1500}$} & -0.248 & -19.84 \\
\multicolumn{2}{l}{\logten($M_*$/\Msol)} & 0.771 & 8.920 \\
\multicolumn{2}{l}{E(B-V)$_{\rm UV}$} & -12.149 & 0.157 \\
\multicolumn{2}{l}{\logten(EW(\oiii$\lambda\lambda$5007,4959+H$\beta$)/\AA)} & 1.893 & 2.725 \\
\hline
\multicolumn{4}{c}{{\it Baseline Cumulative Hazard}} \\
\fesc & HF$_0$(\fesc) & \fesc (cont.) & HF$_0$ (cont.) \\
\hline
0.8890 & 0.005687 & 0.0376 & 0.240170 \\
0.6247 & 0.011541 & 0.0333 & 0.254842 \\
0.5838 & 0.017533 & 0.0309 & 0.269815 \\
0.4911 & 0.023684 & 0.0307 & 0.285298 \\
0.4333 & 0.030206 & 0.0280 & 0.302262 \\
0.3845 & 0.036819 & 0.0268 & 0.319275 \\
0.3052 & 0.043720 & 0.0264 & 0.337274 \\
0.2659 & 0.050851 & 0.0258 & 0.355343 \\
0.1921 & 0.058194 & 0.0257 & 0.373607 \\
0.1777 & 0.065762 & 0.0232 & 0.392994 \\
0.1767 & 0.073806 & 0.0231 & 0.412665 \\
0.1607 & 0.082393 & 0.0220 & 0.433109 \\
0.1197 & 0.091141 & 0.0219 & 0.453932 \\
0.1185 & 0.100320 & 0.0198 & 0.477037 \\
0.1053 & 0.109807 & 0.0188 & 0.502072 \\
0.0917 & 0.119757 & 0.0163 & 0.527919 \\
0.0898 & 0.129963 & 0.0149 & 0.559078 \\
0.0658 & 0.140592 & 0.0132 & 0.596941 \\
0.0600 & 0.151376 & 0.0129 & 0.637311 \\
0.0519 & 0.162692 & 0.0124 & 0.679393 \\
0.0493 & 0.174373 & 0.0070 & 0.783432 \\
0.0473 & 0.186132 & 0.0055 & 0.931661 \\
0.0431 & 0.198855 & 0.0052 & 1.084933 \\
0.0421 & 0.212176 & 0.0044 & 1.283606 \\
0.0405 & 0.226118 & & \\
\enddata
\end{deluxetable*}

\begin{deluxetable*}{llll}
\tablecaption{ELG-O32 Model}
\label{atab:Fujimoto}
\tablehead{\multicolumn{4}{c}{{\it Statistics}}}
\startdata
$R^2$ & $R^2_{\rm adj}$ & RMS & $C$ \\
0.42 & 0.36 & 0.44 & 0.79 \\
\hline
\multicolumn{4}{c}{{\it Model Parameters}} \\
\multicolumn{2}{l}{Variable} & $b_i$ & ${\bar x_i}$ \\
\hline
\multicolumn{2}{l}{$M_{\rm 1500}$} & -0.727 & -19.84 \\
\multicolumn{2}{l}{\logten($M_*$/\Msol)} & 0.693 & 8.920 \\
\multicolumn{2}{l}{E(B-V)$_{\rm UV}$} & -10.230 & 0.157 \\
\multicolumn{2}{l}{\logten(\oiii~$\lambda5007$/\oii~$\lambda3727$)} & 3.810 & 0.523 \\
\hline
\multicolumn{4}{c}{{\it Baseline Cumulative Hazard}} \\
\fesc & HF$_0$(\fesc) & \fesc (cont.) & HF$_0$ (cont.) \\
\hline
0.8890 & 0.003619 & 0.0376 & 0.205717 \\
0.6247 & 0.007534 & 0.0333 & 0.219975 \\
0.5838 & 0.011580 & 0.0309 & 0.234524 \\
0.4911 & 0.015999 & 0.0307 & 0.249271 \\
0.4333 & 0.020713 & 0.0280 & 0.266818 \\
0.3845 & 0.025461 & 0.0268 & 0.284396 \\
0.3052 & 0.030795 & 0.0264 & 0.303606 \\
0.2659 & 0.036338 & 0.0258 & 0.323166 \\
0.1921 & 0.042111 & 0.0257 & 0.342907 \\
0.1777 & 0.048239 & 0.0232 & 0.363696 \\
0.1767 & 0.055076 & 0.0231 & 0.384742 \\
0.1607 & 0.062454 & 0.0220 & 0.406670 \\
0.1197 & 0.069944 & 0.0219 & 0.428908 \\
0.1185 & 0.077866 & 0.0198 & 0.453903 \\
0.1053 & 0.085926 & 0.0188 & 0.481290 \\
0.0917 & 0.094440 & 0.0163 & 0.509903 \\
0.0898 & 0.103111 & 0.0149 & 0.541931 \\
0.0658 & 0.112911 & 0.0132 & 0.579921 \\
0.0600 & 0.122814 & 0.0129 & 0.622891 \\
0.0519 & 0.133160 & 0.0124 & 0.670930 \\
0.0493 & 0.144055 & 0.0070 & 0.779178 \\
0.0473 & 0.155155 & 0.0055 & 0.930883 \\
0.0431 & 0.167143 & 0.0052 & 1.089376 \\
0.0421 & 0.179615 & 0.0044 & 1.293047 \\
0.0405 & 0.192628 & & \\
\enddata
\end{deluxetable*}

\begin{deluxetable*}{llll}
\tablecaption{ELG-O32-$\beta$ Model}
\label{atab:Saxenab}
\tablehead{\multicolumn{4}{c}{{\it Statistics}}}
\startdata
$R^2$ & $R^2_{\rm adj}$ & RMS & $C$ \\
0.30 & 0.25 & 0.48 & 0.79 \\
\hline
\multicolumn{4}{c}{{\it Model Parameters}} \\
\multicolumn{2}{l}{Variable} & $b_i$ & ${\bar x_i}$ \\
\hline
\multicolumn{2}{l}{$M_{\rm 1500}$} & -0.852 & -19.84 \\
\multicolumn{2}{l}{$\beta_{\rm 1550}$} & -1.479 & -1.819 \\
\multicolumn{2}{l}{\logten(\oiii~$\lambda5007$/\oii~$\lambda3727$)} & 3.307 & 0.523 \\
\hline
\multicolumn{4}{c}{{\it Baseline Cumulative Hazard}} \\
\fesc & $H_0$(\fesc) & \fesc (cont.) & $H_0$ (cont.) \\
\hline
0.8890 & 0.004039 & 0.0376 & 0.219993 \\
0.6247 & 0.008739 & 0.0333 & 0.234945 \\
0.5838 & 0.013629 & 0.0309 & 0.250650 \\
0.4911 & 0.018793 & 0.0307 & 0.266648 \\
0.4333 & 0.024186 & 0.0280 & 0.284011 \\
0.3845 & 0.029664 & 0.0268 & 0.301400 \\
0.3052 & 0.035829 & 0.0264 & 0.320107 \\
0.2659 & 0.042177 & 0.0258 & 0.339289 \\
0.1921 & 0.048750 & 0.0257 & 0.358640 \\
0.1777 & 0.055873 & 0.0232 & 0.378715 \\
0.1767 & 0.063520 & 0.0231 & 0.398969 \\
0.1607 & 0.071496 & 0.0220 & 0.420108 \\
0.1197 & 0.079577 & 0.0219 & 0.441571 \\
0.1185 & 0.088029 & 0.0198 & 0.465104 \\
0.1053 & 0.096590 & 0.0188 & 0.491366 \\
0.0917 & 0.105423 & 0.0163 & 0.518661 \\
0.0898 & 0.114447 & 0.0149 & 0.548480 \\
0.0658 & 0.124389 & 0.0132 & 0.584009 \\
0.0600 & 0.134414 & 0.0129 & 0.625072 \\
0.0519 & 0.144961 & 0.0124 & 0.671057 \\
0.0493 & 0.155928 & 0.0070 & 0.780254 \\
0.0473 & 0.167396 & 0.0055 & 0.920758 \\
0.0431 & 0.179884 & 0.0052 & 1.068231 \\
0.0421 & 0.192936 & 0.0044 & 1.252412 \\
0.0405 & 0.206403 & & \\
\enddata
\end{deluxetable*}

\begin{deluxetable*}{llll}
\tablecaption{ELG-O32-$\beta$-Ly$\alpha$ Model}
\label{atab:Saxena_fesc}
\tablehead{\multicolumn{4}{c}{{\it Statistics}}}
\startdata
$R^2$ & $R^2_{\rm adj}$ & RMS & $C$ \\
0.40 & 0.32 & 0.45 & 0.81 \\
\hline
\multicolumn{4}{c}{{\it Model Parameters}} \\
\multicolumn{2}{l}{Variable} & $b_i$ & ${\bar x_i}$ \\
\hline
\multicolumn{2}{l}{$M_{\rm 1500}$} & -0.488 & -19.84 \\
\multicolumn{2}{l}{\logten($M_*$/\Msol)} & 0.969 & 8.920 \\
\multicolumn{2}{l}{$\beta_{\rm 1550}$} & -1.863 & -1.819 \\
\multicolumn{2}{l}{\logten(\oiii~$\lambda5007$/\oii~$\lambda3727$)} & 3.138 & 0.523 \\
\multicolumn{2}{l}{\fesclya} & 2.323 & 0.221 \\
\hline
\multicolumn{4}{c}{{\it Baseline Cumulative Hazard}} \\
\fesc & $H_0$(\fesc) & \fesc (cont.) & $H_0$ (cont.) \\
\hline
0.8890 & 0.003503 & 0.0376 & 0.204488 \\
0.6247 & 0.007231 & 0.0333 & 0.219681 \\
0.5838 & 0.011119 & 0.0309 & 0.235210 \\
0.4911 & 0.015539 & 0.0307 & 0.250950 \\
0.4333 & 0.020235 & 0.0280 & 0.268220 \\
0.3845 & 0.024968 & 0.0268 & 0.285541 \\
0.3052 & 0.030584 & 0.0264 & 0.303822 \\
0.2659 & 0.036388 & 0.0258 & 0.322431 \\
0.1921 & 0.042636 & 0.0257 & 0.341630 \\
0.1777 & 0.049220 & 0.0232 & 0.361799 \\
0.1767 & 0.056445 & 0.0231 & 0.382292 \\
0.1607 & 0.064118 & 0.0220 & 0.403271 \\
0.1197 & 0.071902 & 0.0219 & 0.424872 \\
0.1185 & 0.080101 & 0.0198 & 0.450274 \\
0.1053 & 0.088375 & 0.0188 & 0.481809 \\
0.0917 & 0.096898 & 0.0163 & 0.514436 \\
0.0898 & 0.105662 & 0.0149 & 0.550082 \\
0.0658 & 0.115446 & 0.0132 & 0.591342 \\
0.0600 & 0.125364 & 0.0129 & 0.640608 \\
0.0519 & 0.135529 & 0.0124 & 0.694320 \\
0.0493 & 0.145923 & 0.0070 & 0.809258 \\
0.0473 & 0.156598 & 0.0055 & 0.964455 \\
0.0431 & 0.167861 & 0.0052 & 1.129894 \\
0.0421 & 0.179836 & 0.0044 & 1.333051 \\
0.0405 & 0.192127 & & \\
\enddata
\end{deluxetable*}

\begin{deluxetable*}{llll}
\tablecaption{R50-$\beta$ Model}
\label{atab:Morishita}
\tablehead{\multicolumn{4}{c}{{\it Statistics}}}
\startdata
$R^2$ & $R^2_{\rm adj}$ & RMS & $C$ \\
0.36 & 0.30 & 0.44 & 0.84 \\
\hline
\multicolumn{4}{c}{{\it Model Parameters}} \\
\multicolumn{2}{l}{Variable} & $b_i$ & ${\bar x_i}$ \\
\hline
\multicolumn{2}{l}{$M_{\rm 1500}$} & -0.196 & -19.84 \\
\multicolumn{2}{l}{\logten($M_*$/\Msol)} & 0.361 & 8.932 \\
\multicolumn{2}{l}{$\beta_{\rm 1550}$} & -2.766 & -1.810 \\
\multicolumn{2}{l}{$\beta_{\rm 1550}$} & -4.472 & -0.210 \\
\hline
\multicolumn{4}{c}{{\it Baseline Cumulative Hazard}} \\
\fesc & $H_0$(\fesc) & \fesc (cont.) & $H_0$ (cont.) \\
\hline
0.8890 & 0.003630 & 0.0376 & 0.196594 \\
0.6247 & 0.007542 & 0.0333 & 0.209834 \\
0.5838 & 0.011518 & 0.0309 & 0.223769 \\
0.4911 & 0.015572 & 0.0307 & 0.237960 \\
0.4333 & 0.019859 & 0.0280 & 0.253960 \\
0.3845 & 0.024345 & 0.0268 & 0.270600 \\
0.3052 & 0.029476 & 0.0264 & 0.288151 \\
0.2659 & 0.034731 & 0.0258 & 0.306004 \\
0.1921 & 0.040111 & 0.0257 & 0.323881 \\
0.1777 & 0.045787 & 0.0232 & 0.342762 \\
0.1767 & 0.052227 & 0.0231 & 0.362072 \\
0.1607 & 0.059035 & 0.0220 & 0.382566 \\
0.1197 & 0.066366 & 0.0219 & 0.403879 \\
0.1185 & 0.074039 & 0.0198 & 0.431991 \\
0.1053 & 0.081864 & 0.0188 & 0.463839 \\
0.0917 & 0.089852 & 0.0163 & 0.497097 \\
0.0898 & 0.098085 & 0.0149 & 0.533738 \\
0.0658 & 0.107216 & 0.0132 & 0.583785 \\
0.0600 & 0.116484 & 0.0129 & 0.645270 \\
0.0519 & 0.126573 & 0.0124 & 0.709809 \\
0.0493 & 0.137188 & 0.0070 & 0.848036 \\
0.0473 & 0.147942 & 0.0055 & 1.054349 \\
0.0431 & 0.159479 & 0.0052 & 1.269787 \\
0.0421 & 0.171621 & 0.0044 & 1.562856 \\
0.0405 & 0.184018 & & \\
\enddata
\end{deluxetable*}

\begin{deluxetable*}{llll}
\tablecaption{$\beta$-Metals Model}
\label{atab:Atek2}
\tablehead{\multicolumn{4}{c}{{\it Statistics}}}
\startdata
$R^2$ & $R^2_{\rm adj}$ & RMS & $C$ \\
0.11 & 0.02 & 0.55 & 0.77 \\
\hline
\multicolumn{4}{c}{{\it Model Parameters}} \\
\multicolumn{2}{l}{Variable} & $b_i$ & ${\bar x_i}$ \\
\hline
\multicolumn{2}{l}{$M_{\rm 1500}$} & -7.760e-02 & -19.84 \\
\multicolumn{2}{l}{\logten($M_*$/\Msol)} & 0.397 & 8.920 \\
\multicolumn{2}{l}{$\beta_{\rm 1550}$} & -2.227 & -1.819 \\
\multicolumn{2}{l}{12+\logten(O/H)} & -1.225 & 8.111 \\
\hline
\multicolumn{4}{c}{{\it Baseline Cumulative Hazard}} \\
\fesc & HF$_0$(\fesc) & \fesc (cont.) & HF$_0$ (cont.) \\
\hline
0.8890 & 0.006818 & 0.0376 & 0.272610 \\
0.6247 & 0.013970 & 0.0333 & 0.288649 \\
0.5838 & 0.021388 & 0.0309 & 0.305054 \\
0.4911 & 0.029052 & 0.0307 & 0.322206 \\
0.4333 & 0.037085 & 0.0280 & 0.341201 \\
0.3845 & 0.045286 & 0.0268 & 0.360368 \\
0.3052 & 0.054005 & 0.0264 & 0.380230 \\
0.2659 & 0.062894 & 0.0258 & 0.400381 \\
0.1921 & 0.071991 & 0.0257 & 0.420683 \\
0.1777 & 0.081252 & 0.0232 & 0.441694 \\
0.1767 & 0.091036 & 0.0231 & 0.463044 \\
0.1607 & 0.101098 & 0.0220 & 0.485047 \\
0.1197 & 0.111387 & 0.0219 & 0.507680 \\
0.1185 & 0.121976 & 0.0198 & 0.533235 \\
0.1053 & 0.132764 & 0.0188 & 0.560019 \\
0.0917 & 0.143767 & 0.0163 & 0.587757 \\
0.0898 & 0.154960 & 0.0149 & 0.618128 \\
0.0658 & 0.166527 & 0.0132 & 0.652861 \\
0.0600 & 0.178210 & 0.0129 & 0.691653 \\
0.0519 & 0.190433 & 0.0124 & 0.732358 \\
0.0493 & 0.202887 & 0.0070 & 0.824911 \\
0.0473 & 0.215452 & 0.0055 & 0.936173 \\
0.0431 & 0.228888 & 0.0052 & 1.051480 \\
0.0421 & 0.243133 & 0.0044 & 1.195328 \\
0.0405 & 0.257825 & & \\
\enddata
\end{deluxetable*}

\bibliography{latestrefs}
\bibliographystyle{aasjournal}

\end{document}